\documentclass[journal=jctcce,manuscript=articlel]{achemso}
\usepackage{graphicx,dcolumn,bm,xcolor,microtype,multirow,amsmath,amssymb,amsfonts,physics,float,lscape,soul,rotating,longtable}
\usepackage[version=4]{mhchem}

\usepackage[colorlinks = true,
            linkcolor = blue,
            urlcolor  = black,
            citecolor = blue,
            anchorcolor = black]
            {hyperref}

\usepackage{amsmath}
\usepackage{newtxtext,newtxmath}

\usepackage[normalem]{ulem}

\definecolor{darkgreen}{RGB}{0, 180, 0}

\newcommand{\mc}{\multicolumn}

\newcommand{\ra}{\rightarrow}
\newcommand{\eg}{\textit{e.g.}}
\newcommand{\ie}{\textit{i.e.}}

\newcommand{\Td}{\%T_1}

\newcommand{\CASPT}{CASPT2}
\newcommand{\NEV}{NEVPT2}
\newcommand{\PNEV}{PC-NEVPT2}

\newcommand{\AD}{ADC(2)}
\newcommand{\AT}{ADC(3)}
\newcommand{\CCD}{CC2}
\newcommand{\CCSD}{CCSD}
\newcommand{\STEOM}{STEOM-CCSD}
\newcommand{\CCT}{CC3}

\newcommand{\CCSDT}{CCSDT}
\newcommand{\CCSDTQ}{CCSDTQ}
\newcommand{\CCSDTQP}{CCSDTQP}

\newcommand{\SCI}{SCI}

\newcommand{\FCI}{FCI}

\newcommand{\Pop}{6-31+G(d)}
\newcommand{\AVDZ}{aVDZ}
\newcommand{\AVTZ}{aVTZ}

\newcommand{\AVQZ}{aVQZ}

\newcommand{\AVPZ}{aV5Z}

\newcommand{\MaxP}{Max($+$)}
\newcommand{\MaxN}{Max($-$)}

\newcommand{\pis}{\pi^\star}

\newcommand{\Val}{\mathrm{V}}

\newcommand{\LCPQ}{Laboratoire de Chimie et Physique Quantiques, CNRS et Universit\'e Toulouse III - Paul Sabatier, 118 route de Narbonne, 31062 Toulouse, France}
\newcommand{\CEISAM}{Universit\'e de Nantes, CNRS,  CEISAM UMR 6230, F-44000 Nantes, France}
\newcommand{\Pisa}{Dipartimento di Chimica e Chimica Industriale, University of Pisa, Via Moruzzi 3, 56124 Pisa, Italy}

\title{A Mountaineering Strategy to Excited States: Highly-Accurate Energies and Benchmarks for Medium Size Molecules}

\author{Pierre-Fran{\c c}ois Loos}
	\email{loos@irsamc.ups-tlse.fr}
	\affiliation[LCPQ, Toulouse]{\LCPQ}
\author{Filippo Lipparini}
	\affiliation[DC, Pisa]{\Pisa}
	\email{filippo.lipparini@unipi.it}
\author{Martial Boggio-Pasqua}
	\affiliation[LCPQ, Toulouse]{\LCPQ}
\author{Anthony Scemama}
	\affiliation[LCPQ, Toulouse]{\LCPQ}
\author{Denis Jacquemin}
	\email{Denis.Jacquemin@univ-nantes.fr}
	\affiliation[UN, Nantes]{\CEISAM}

\begin{document}

\begin{tocentry}
\begin{center}
\includegraphics[scale=.27]{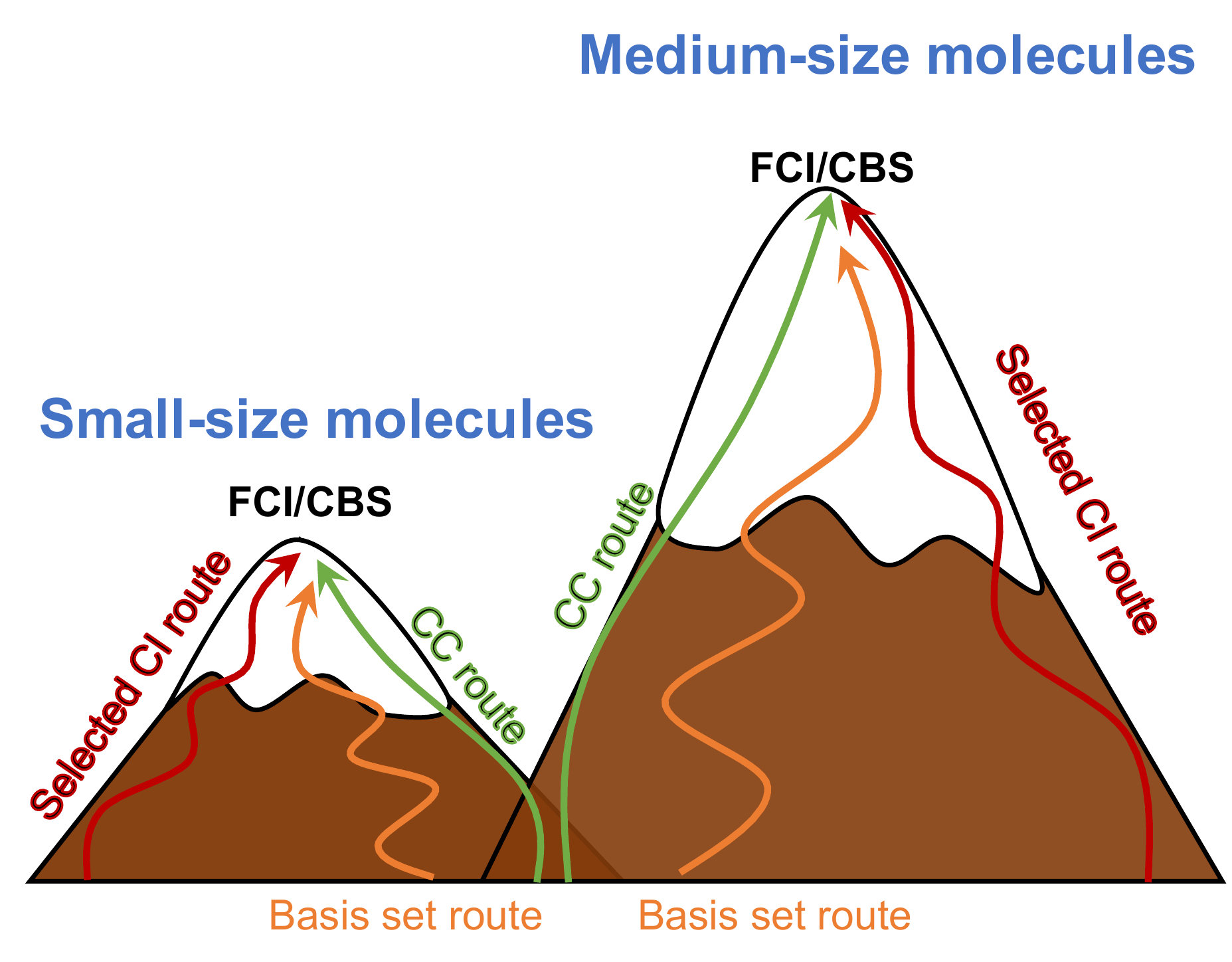}
\end{center}
\end{tocentry}

\begin{abstract}
Following our previous work focussing on compounds containing up to 3 non-hydrogen atoms [\emph{J. Chem. Theory Comput.} {\bfseries 14} (2018) 4360--4379], we present here highly-accurate vertical transition energies
obtained for 27 molecules encompassing 4, 5, and 6 non-hydrogen atoms: acetone, acrolein, benzene, butadiene, cyanoacetylene, cyanoformaldehyde, cyanogen, cyclopentadiene, cyclopropenone, cyclopropenethione,
diacetylene, furan, glyoxal, imidazole, isobutene, methylenecyclopropene, propynal, pyrazine, pyridazine, pyridine, pyrimidine, pyrrole, tetrazine, thioacetone, thiophene, thiopropynal, and triazine. To obtain these energies, we
use equation-of-motion coupled cluster theory up to the highest technically possible excitation order for these systems (CC3, EOM-CCSDT, and EOM-CCSDTQ), selected configuration interaction (SCI) calculations (with tens
of millions of determinants in the reference space), as well as the multiconfigurational $n$-electron valence state perturbation theory (NEVPT2) method. All these approaches are applied in combination with diffuse-containing
atomic basis sets. For all transitions, we report at least CC3/\emph{aug}-cc-pVQZ vertical excitation energies as well as CC3/\emph{aug}-cc-pVTZ oscillator strengths for each dipole-allowed transition. We show that CC3
almost systematically delivers transition energies in agreement with higher-level methods with a typical deviation of $\pm 0.04$ eV, except for transitions with a dominant double excitation character where the error is much larger.
The present contribution gathers a large, diverse and accurate set of more than 200 highly-accurate transition energies for states of various natures (valence, Rydberg, singlet, triplet, $n \ra \pis$, $\pi \ra \pis$, \ldots).
We use this series of theoretical best estimates to benchmark a series of popular methods for excited state calculations: CIS(D), ADC(2), CC2, STEOM-CCSD, EOM-CCSD, CCSDR(3), CCSDT-3,  CC3, as well
as {\NEV}.  The results of these benchmarks are compared to the available literature data.
\end{abstract}
\clearpage

%
%
\section{Introduction}

Accurately describing transition energies between the electronic ground state (GS) and excited states (ES) remains an important challenge in quantum chemistry. When dealing with large compounds in complex environments,
one is typically limited to the use of time-dependent density-functional theory (TD-DFT), \cite{Cas95,Ulr12b,Ada13a} a successful yet far from flawless approach. In particular, to perform TD-DFT calculations, one must
choose an ``appropriate'' exchange-correlation functional, which is difficult yet primordial as the impact of the exchange-correlation functional is exacerbated within TD-DFT as compared to DFT. \cite{Lau13} Such selection
can, of course, rely on the intrinsic features of the various exchange-correlation functional families, \eg, it is well-known that range-separated hybrids provide a more physically-sound description of long-range charge-transfer
transitions than semi-local exchange-correlation functionals. \cite{Dre04,Pea08} However, to obtain a quantitative assessment of the accuracy that can be expected from TD-DFT calculations, benchmarks cannot be avoided.
This is why so many assessments of TD-DFT performance for various ES properties are available. \cite{Lau13}

While several of these benchmarks rely on experimental data as reference (typically band shapes \cite{Die04,Die04b,Avi13,Cha13,Lat15b,Mun15,Vaz15,San16b} or 0-0 energies
\cite{Die04b,Goe10a,Jac12d,Chi13b,Win13,Fan14b,Jac14a,Jac15b,Loo19b}), using theoretical best estimates (TBE) based on state-of-the-art computational methods \cite{Sch08,Sau09,Sil10b,Sil10c,Sch17,Loo18a}
are advantageous as they allow comparisons on a perfectly equal footing (same geometry, vertical transitions, no environmental effects, etc).  In such a case, the challenge is in fact to obtain accurate TBE, as the
needed top-notch theoretical models generally come with a dreadful scaling with system size and, in addition, typically require large atomic basis sets to deliver transition energies close to the complete basis set (CBS) limit. \cite{Gin19}

More than 20 years ago, Serrano-Andr\`es, Roos, and collaborators compiled an impressive series of reference transition energies for several typical conjugated organic molecules (butadiene, furan, pyrrole, tetrazine, \ldots).
\cite{Ful92,Ser93,Ser93b,Ser93c,Lor95b,Mer96,Mer96b,Roo96,Ser96b} To this end, they relied on experimental GS geometries and the complete-active-space second-order perturbation theory ({\CASPT}) approach with the largest
active spaces and basis sets one could dream of at the time.  These {\CASPT} values were later used to assess the performance of TD-DFT combined with various exchange-correlation functionals, \cite{Toz99b,Bur02} and remained for
a long time the best theoretical references available on the market. However, beyond comparisons with experiments, which are always challenging when computing vertical transition energies, \cite{San16b} there was no approach
available at that time to ascertain the accuracy of these transition energies. Nowadays, it is of common knowledge that CASPT2 has the tendency of underestimating vertical excitation energies in organic molecules when IPEA shift 
is not included. It is also known that the use of a standard value of 0.25 au for this IPEA shift may lead to overestimating of the transition energies making the use of this shift questionable.\cite{Zob17}

A decade ago, Thiel and coworkers defined TBE for 104 singlet and 63 triplet valence ES in 28 small and medium conjugated CNOH organic molecules. \cite{Sch08,Sil10b,Sil10c}  These TBE were computed on MP2/6-31G(d) structures
with several levels of theory, notably {\CASPT} and various coupled cluster (CC) variants ({\CCD}, {\CCSD}, and {\CCT}). Interestingly, while the default theoretical protocol used by Thiel and coworkers to define their
first series of TBE was {\CASPT}, \cite{Sch08} the vast majority of their most recent TBE (the so-called ``TBE-2'' in Ref.~\citenum{Sil10c}) were determined at the {\CCT} level of theory with the \emph{aug}-cc-pVTZ (aVTZ) basis set,
often using a basis set extrapolation technique.  More specifically, CC3/TZVP values were corrected for basis set incompleteness errors by the difference between the {\CCD}/{\AVTZ} and {\CCD}/TZVP results.  \cite{Sil10b,Sil10c}
Many works have exploited Thiel's TBE for assessing low-order methods,
\cite{Sil08,Goe09,Jac09c,Roh09,Sau09,Jac10c,Jac10g,Sil10,Mar11,Jac11a,Hui11,Del11,Tra11,Pev12,Dom13,Dem13,Sch13b,Voi14,Har14,Yan14b,Sau15,Pie15,Taj16,Mai16,Ris17,Dut18,Hel19,Haa20} highlighting further
their value for the electronic structure community. In contrast, the number of extensions/improvements of this original set remains quite limited. For example, K\'ann\'ar and Szalay computed, in 2014, {\CCSDT}/TZVP reference energies
for 17 singlet states of six molecules. \cite{Kan14} Three years later, the same authors reported 46 {\CCSDT}/{\AVTZ} transition energies in small compounds containing two or three non-hydrogen atoms (ethylene, acetylene, formaldehyde,
formaldimine, and formamide). \cite{Kan17}

Following the same philosophy, two years ago, we reported a set of 106 transition energies for which it was technically possible to reach the full configuration interaction (FCI) limit by performing high-order CC (up to {\CCSDTQP}) and selected
CI (SCI) calculations on {\CCT}/{\AVTZ} GS structures. \cite{Loo18a}  We exploited these TBE to benchmark many ES methods. \cite{Loo18a}  Among our conclusions, we found that {\CCSDTQ} yields near-{\FCI} quality excitation energies,
whereas we could not  detect any significant differences between {\CCT} and {\CCSDT} transition energies, both being very accurate with mean absolute errors (MAE) as small as $0.03$ eV compared to {\FCI}.

Although these conclusions agree well with earlier studies, \cite{Wat13,Kan14,Kan17} they obviously only hold for single excitations, \ie, transitions with $\Td$ in the  $80$--$100\%$ range.   Therefore, we also recently proposed a set of
20 TBE for transitions exhibiting a significant double-excitation character (\ie, with $\Td$ typically below $80\%$). \cite{Loo19c} Unsurprisingly, our results clearly evidenced that the error in CC methods is intimately related to the $\Td$ value.
For example, for the ES with a significant yet not dominant double excitation character [such as the infamous $A_g$ ES of butadiene ($\Td = 75\%$)] CC methods including triples deliver rather accurate estimates (MAE of $0.11$ eV with {\CCT}
and $0.06$ eV with {\CCSDT}), surprisingly outperforming second-order multi-reference schemes such as {\CASPT} or the generally robust $n$-electron valence state perturbation theory ({\NEV}). In contrast, for ES with a dominant double
excitation character, \eg, the low-lying $(n,n) \ra (\pis,\pis)$ excitation in nitrosomethane ($\Td = 2\%$), single-reference methods (not including quadruples) have been found to be unsuitable with MAEs of $0.86$ and $0.42$ eV for {\CCT}
and {\CCSDT}, respectively.  In this case, multiconfigurational methods are in practice required to obtain accurate results. \cite{Loo19c}

A clear limit of our 2018 work \cite{Loo18a} was the size of the compounds put together in our set. These were limited to $1$--$3$ non-hydrogen atoms, hence introducing a potential ``chemical'' bias. Therefore, we have decided, in the
present contribution, to consider larger molecules with organic compounds encompassing 4, 5, and 6 non-hydrogen atoms.  For such systems, performing {\CCSDTQ} calculations with large one-electron basis sets is elusive. Moreover,
the convergence of the {\SCI} energy with respect to the number of determinants is obviously slower for these larger compounds, hence extrapolating to the {\FCI} limit with an error of $\sim 0.01$ eV is rarely achievable in practice.
Consequently, the ``brute-force'' determination of {\FCI}/CBS estimates, as in our earlier work, \cite{Loo18a} is definitely out of reach here. Anticipating this problem, we have recently investigated bootstrap CBS extrapolation techniques.
\cite{Loo18a,Loo19c} In particular, we have demonstrated that, following an ONIOM-like scheme, \cite{Chu15} one can very accurately estimate such limit by correcting high-level values obtained in a small basis by the difference between
{\CCT} results obtained in a larger basis and in the same small basis.\cite{Loo18a} We globally follow such strategy here. In addition, we also perform {\NEV} calculations in an effort to check the consistency of our estimates. This is particularly
critical for ES with intermediate $\Td$ values. Using this protocol, we define a set of more than 200 \emph{aug}-cc-pVQZ reference transition energies, most being within $\pm 0.03$ eV of the {\FCI} limit. These reference energies are obtained
on {\CCT}/{\AVTZ} geometries and additional basis set corrections (up to quadruple-$\zeta$ at least) are also provided for {\CCT}. Together with the results obtained in our two earlier works, \cite{Loo18a,Loo19c} the present TBE will
hopefully contribute to climb a rung higher on the ES accuracy ladder.

%
%
\section{Computational Details}
\label{sec-met}

Unless otherwise stated, all transition energies are computed in the frozen-core approximation (with a large core for the sulfur atoms). Pople's {\Pop} and Dunning's \emph{aug}-cc-pVXZ (X $=$ D, T, Q, and 5) atomic basis sets are systematically
employed in our excited-state calculations. In the following, we employ the aVXZ shorthand notations for these diffuse-containing basis sets. We note that an alternative family of more compact diffuse basis sets (such as \emph{jun}-cc-pVTZ) have 
been proposed by Truhlar and coworkers.\cite{Pap11} Such variants could be better suited to reach CBS-quality transition energies at a smaller computational cost. As we intend to provide benchmark values here, we nevertheless stick to the 
original Dunning's bases, which are directly available in almost any quantum chemistry codes. Various statistical quantities are reported in the remaining of this paper: the mean signed error (MSE),
mean absolute error (MAE), root mean square error (RMSE), standard deviation of the errors (SDE), as well as the positive [\MaxP] and negative [\MaxN] maximum errors. Here, we globally follow the same procedure as in Ref.~\citenum{Loo18a},
so that we only briefly outline the various theoretical methods that we have employed in the subsections below.

\subsection{Geometries}

The molecules considered herein are displayed in Scheme \ref{Sch-1}.
Consistently with our previous work, \cite{Loo18a} we systematically use {\CCT}/{\AVTZ} GS geometries obtained without applying the frozen-core approximation.  The cartesian coordinates (in bohr) of each compound can be found in the
Supporting Information (SI). Several structures have been extracted from previous contributions, \cite{Bud17,Jac18a,Bre18a} whereas the missing structures were optimized using DALTON \cite{dalton} and/or CFOUR, \cite{cfour} applying
default parameters in both cases.

\begin{scheme}[htp]
  \begin{centering}
  	\includegraphics[scale=0.8,viewport=2cm 9.5cm 19cm 27.5cm,clip]{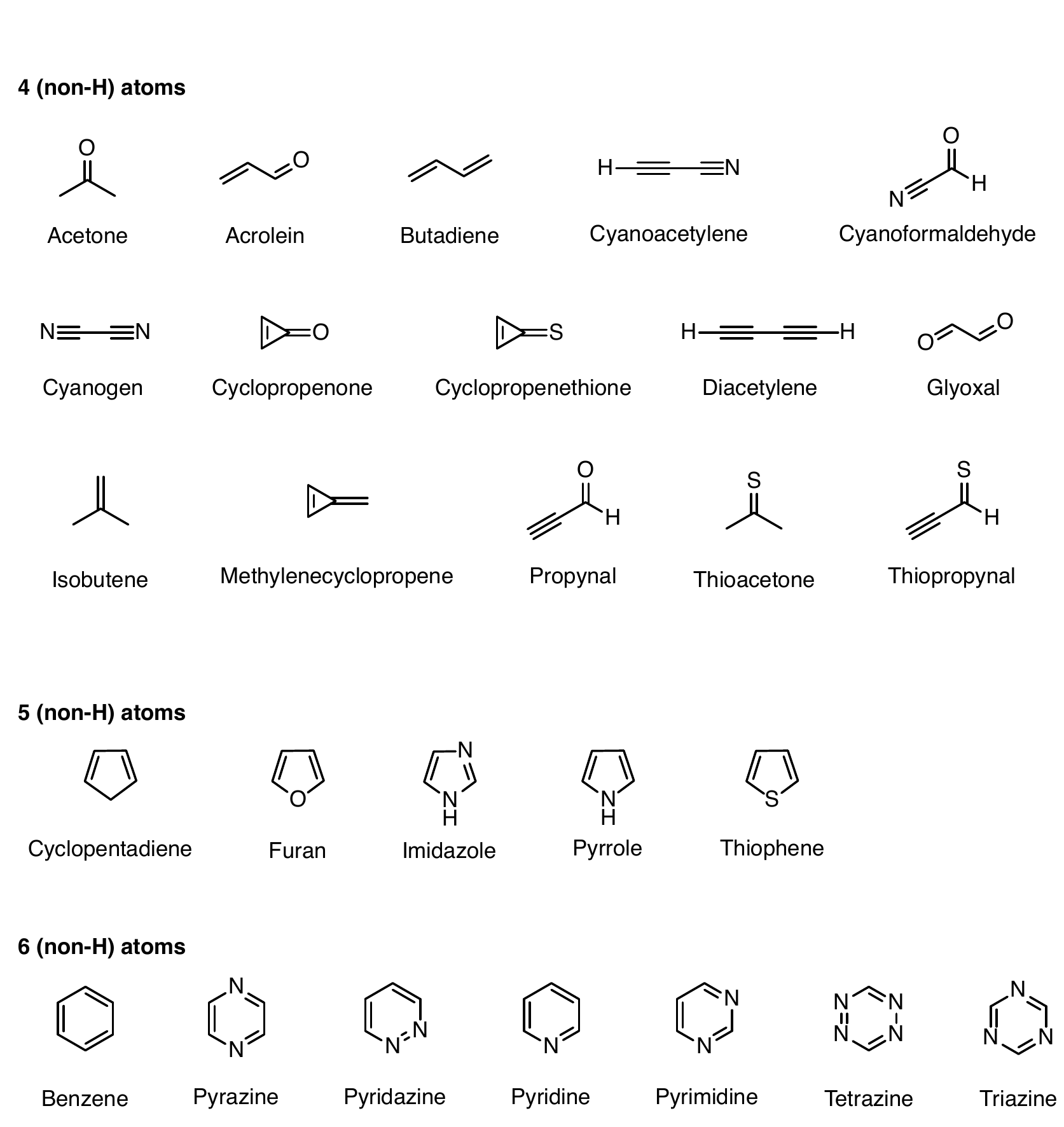}
  \end{centering}	
  \caption{Representation of the considered compounds.}
  \label{Sch-1}
\end{scheme}

\subsection{Selected Configuration Interaction}

Because SCI methods are less widespread than the other methods mentioned in the Introduction, we shall detail further their main features.
All the SCI calculations have been performed in the frozen-core approximation with the latest version of QUANTUM PACKAGE \cite{Gar19} using the Configuration Interaction using a Perturbative Selection made Iteratively (CIPSI) algorithm to select the
most important determinants in the FCI space. Instead of generating all possible excited determinants like a conventional CI calculation, the iterative CIPSI algorithm performs a sparse exploration of the FCI space via a selection of the most relevant
determinants using a second-order perturbative criterion. At each iteration, the variational (or reference) space is enlarged with new determinants. CIPSI can be seen as a deterministic version of the FCIQMC algorithm developed by Alavi and
coworkers. \cite{Boo09} We refer the interested reader to Ref.~\citenum{Gar19} where our implementation of the CIPSI algorithm is detailed.

Excited-state calculations are performed within a state-averaged formalism which means that the CIPSI algorithm select determinants simultaneously for the GS and ES. Therefore, all electronic states share the same set of determinants with different CI coefficients.
Our implementation of the CIPSI algorithm for ES is detailed in Ref.~\citenum{Sce19}. For each system, a preliminary SCI calculation is performed using Hartree-Fock orbitals in order to generate SCI wavefunctions with at least 5,000,000 determinants.
State-averaged natural orbitals are then computed based on this wavefunction, and a new, larger SCI calculation is performed with this new set of orbitals. This has the advantage to produce a smoother and faster convergence of the SCI energy towards the FCI limit.
For the largest systems, an additional iteration is sometimes required in order to obtain better quality natural orbitals and hence well-converged calculations.

The total SCI energy is defined as the sum of the (zeroth-order) variational energy (computed via diagonalization of the CI matrix in the reference space) and a second-order perturbative correction which takes into account the external determinants, \ie,
the determinants which do not belong to the variational space but are linked to the reference space via a non-zero matrix element. The magnitude of this second-order correction, $E^{(2)}$,  provides a qualitative idea of the ``distance'' to the FCI limit.
For maximum efficiency, the total SCI energy is linearly extrapolated to $E^{(2)} = 0$ (which effectively corresponds to the FCI limit) using the two largest SCI wavefunctions. These extrapolated total energies (simply labeled as FCI in the remaining of the paper)
are then used to compute vertical excitation energies. Although it is not possible to provide a theoretically-sound error bar, we estimate the extrapolation error by the difference in excitation energy between the largest SCI wavefunction and its corresponding
extrapolated value. We believe that it provides a very safe estimate of the extrapolation error. Additional information about the SCI wavefunctions and excitation energies as well as their extrapolated values can be found in the SI.

\subsection{NEVPT2}

The {\NEV} calculations have been performed with MOLPRO \cite{molpro} within the partially-contracted scheme ({\PNEV}), which is theoretically superior to its strongly-contracted version due to the larger number of perturbers and greater flexibility. \cite{Ang01,Ang01b,Ang02}
These NEVPT2 calculations are performed on top of a state-averaged complete-active-space self-consistent field calculation always including at least the ground state with the excited state of interest
Active spaces carefully chosen and tailored for the desired transitions
have been selected. The definition of the active space considered for each system as well as the number of states in the state-averaged calculation is provided in the SI.

\subsection{Other wavefunction calculations}

For the other levels of theory, we apply a variety of programs, namely, CFOUR,\cite{cfour}  DALTON,\cite{dalton} GAUSSIAN,\cite{Gaussian16} ORCA,\cite{Nee12} MRCC,\cite{Rol13,mrcc}  and Q-CHEM. \cite{Sha15} CFOUR is used for
 {\CCT}, \cite{Chr95b,Koc97}  CCSDT-3, \cite{Wat96,Pro10}  {\CCSDT} \cite{Nog87} and  {\CCSDTQ}\cite{Kuc91}; Dalton for  {\CCD}, \cite{Chr95,Hat00}  {\CCSD},\cite{Pur82}  CCSDR(3), \cite{Chr96b} and  {\CCT} \cite{Chr95b,Koc97}; Gaussian
for CIS(D); \cite{Hea94,Hea95}  ORCA for the similarity-transformed (ST) equation-of-motion (EOM) CCSD ({\STEOM})\cite{Noo97,Dut18}; MRCC for  {\CCSDT} \cite{Nog87} and  {\CCSDTQ}; \cite{Kuc91} and Q-Chem for {\AD}. \cite{Dre15}
Default program settings were applied. We note that for {\STEOM} we only report states that are characterized by an active character percentage of $98\%$ or larger. In all the software mentioned above, point group symmetry was
systematically employed to reduce the computational effort. It should be noted that we do not perform "GS" CC calculations in a specific symmetry to deduce ES energies. All the CC results reported below correspond to excited-state calculations 
within the EOM or linear-response (LR) formalisms, both delivering strictly identical results for transition energies. These formalisms are also applied to get the triplet ES energies directly from the closed-shell singlet GS. In other words, all our 
calculations systematically consider a restricted closed-shell ground state. Finally, the reported CC3 oscillator strengths have been determined within the  LR formalism.

%
%
\section{Main results}
\label{sec-res}

In the following, we present results obtained for molecules containing four, five, and six (non-hydrogen) atoms. In all cases, we test several atomic basis sets and push the CC excitation order as high as technically possible.
Given that the {\SCI} energy converges rather slowly for these  systems, we provide an estimated error bar for these extrapolated {\FCI} values (\emph{vide supra}). In most cases, these extrapolated FCI reference data
are used as a ``safety net'' to demonstrate the overall consistency of the various approaches rather than as definitive reference values (see next Section). As a further consistency check, we also report {\NEV}/{\AVTZ} excitation
energies for all states.  We underline that, except when specifically discussed, all ES present a dominant single-excitation character (see also next Section), so that we do not expect serious CC breakdowns. This is especially
true for triplet ES that are known to be characterized by very large $\Td$ values in the vast majority of the cases.  \cite{Sch08} Consequently, we concentrate most of our computational effort on the obtention of accurate transition
energies for singlet states. To assign the different ES, we use literature data, as well as the usual criteria, \ie, relative energies, spatial and spin symmetries, compositions from the underlying molecular orbitals, and oscillator strengths.
This allows clear-cut assignments for the vast majority of the cases. There are however some state/method combinations for which strong mixing between ES of the same symmetry makes unambiguous assignments almost impossible.

\subsection{Molecules with four non-hydrogen atoms}

\subsubsection{Cyanoacetylene, cyanogen, and diacetylene}

\begin{sidewaystable}[htp]
\caption{\small Vertical transition energies (in eV) of cyanoacetylene, cyanogen, and diacetylene. All states have a valence $\pi \ra \pis$ character.}
\label{Table-1}
  \begin{footnotesize}
\begin{tabular}{l|p{.5cm}p{1.0cm}p{1.2cm}p{1.4cm}|p{.5cm}p{1.0cm}p{1.2cm}p{1.4cm}|p{.5cm}p{1.0cm}p{1.2cm}|p{.5cm}|p{.5cm}|p{.6cm}p{.6cm}}
\hline
		 \mc{14}{c}{Cyanoacetylene}\\
		& \mc{4}{c}{\Pop} & \mc{4}{c}{\AVDZ}& \mc{3}{c}{\AVTZ} &  \mc{1}{c}{\AVQZ} &  \mc{1}{c}{\AVPZ} &  \mc{2}{c}{Litt.}\\
State 	& {\CCT} & {\CCSDT} & {\CCSDTQ}  & {\FCI} & {\CCT} & {\CCSDT} & {\CCSDTQ}  & {\FCI}& {\CCT} & {\CCSDT}   & {\NEV} & {\CCT} & {\CCT}& Th.$^a$ & Exp.$^b$   \\
\hline
$^1\Sigma^-$ 			&6.02&6.04&6.02&6.02$\pm$0.01	&5.92&5.92&5.91&5.84$\pm$0.09	&5.80&5.81&5.78&	5.79	&5.79	&5.46&4.77\\
$^1\Delta$ 			&6.29&6.31&6.29&6.28$\pm$0.01	&6.17&6.19&6.17&6.14$\pm$0.05	&6.08&6.09&6.10&	6.06 &6.06	&5.81&5.48\\
$^3\Sigma^+$			&4.44&4.45&	   &4.45$\pm$0.03	&4.43&4.43&	   &4.41$\pm$0.06	&4.45&4.44&4.45&	4.46 &4.47	&&\\
$^3\Delta$ 			&5.35&5.34&       &5.32$\pm$0.03	&5.28&5.27&	   &5.20$\pm$0.08	&5.22&5.21&5.19&	5.22 &5.22	&&\\
$^1A''$[F]$^c$			&3.70&3.72&3.70&3.67$\pm$0.03	&3.60&3.62&3.60&3.59$\pm$0.02	&3.54&3.56&3.50&	3.54	&		&&\\
\hline
		 \mc{14}{c}{Cyanogen}\\
		& \mc{4}{c}{\Pop} & \mc{4}{c}{\AVDZ}& \mc{3}{c}{\AVTZ} &  \mc{1}{c}{\AVQZ} &  \mc{1}{c}{\AVPZ} & \mc{1}{c}{Litt.}\\
State 	& {\CCT} & {\CCSDT} & {\CCSDTQ}  & {\FCI} & {\CCT} & {\CCSDT} & {\CCSDTQ}  & {\FCI}& {\CCT} & {\CCSDT}   & {\NEV}& {\CCT} & {\CCT}& Exp.$^d$ \\
\hline
$^1\Sigma_u^-$ 		&6.62&6.63&6.62&6.58$\pm$0.03	&6.52&6.52&6.51&6.44$\pm$0.08	&6.39&6.40&6.32&	6.38	&6.38	&5.63\\
$^1\Delta_u$ 			&6.88&6.89&6.88&6.87$\pm$0.02	&6.77&6.78&6.77&6.74$\pm$0.04	&6.66&6.67&6.66&	6.64	&6.64	&5.99\\
$^3\Sigma_u^+$		&4.92&4.92&4.94&4.91$\pm$0.06	&4.89&4.89&	   &4.87$\pm$0.07	&4.90&4.89&4.88&	4.91	&4.91	&4.13\\
$^1\Sigma_u^-$[F]$^c$	&5.27&5.28&5.26&5.31$\pm$0.05	&5.19&5.20&5.18&5.26$\pm$0.09	&5.06&5.07&4.97&	5.05	&5.05	&	\\
\hline
		 \mc{14}{c}{Diacetylene}\\
		& \mc{4}{c}{\Pop} & \mc{4}{c}{\AVDZ}& \mc{3}{c}{\AVTZ} &  \mc{1}{c}{\AVQZ} &  \mc{1}{c}{\AVPZ} & \mc{1}{c}{Litt.}\\
State 	& {\CCT} & {\CCSDT} & {\CCSDTQ}  & {\FCI} & {\CCT} & {\CCSDT} & {\CCSDTQ}  & {\FCI}& {\CCT} & {\CCSDT}   & {\NEV}& {\CCT} & {\CCT}&  Exp.$^e$ \\
\hline
$^1\Sigma_u^-$ 		&5.57&5.58&5.56&5.52$\pm$0.06	&5.44&5.45&5.43&5.47$\pm$0.02	&5.34&5.35&5.33&	5.33	&5.33	&4.81\\
$^1\Delta_u$ 			&5.83&5.85&	   &5.84$\pm$0.01	&5.69&5.70&5.69&5.69$\pm$0.02	&5.61&5.62&5.61&	5.60	&5.60	&5.06\\
$^3\Sigma_u^+$		&4.07&4.08&4.09&4.04$\pm$0.07	&4.06&4.06&	   &4.07$\pm$0.04	&4.08&	  &4.08&	4.10	&4.11	&2.7  \\
$^3\Delta_u$ 			&4.93&4.93&4.92&4.94$\pm$0.01	&4.86&4.85&	   &4.85$\pm$0.02	&4.80&4.79&4.78&	4.80	&4.80	&3.21\\
\hline
 \end{tabular}
  \end{footnotesize}
\vspace{-0.2 cm}
\begin{flushleft}
\begin{footnotesize}
$^a${{\CASPT} results from Ref.~\citenum{Luo08};}
$^b${Experimental 0-0 energies from Refs.~\citenum{Job66a} and \citenum{Job66b} (vacuum UV experiments);}
$^c${Vertical fluorescence energy of the lowest excited state;}
$^d${Experimental 0-0 energies from Refs.~\citenum{Cal63} ($^3\Sigma_u^+$), \citenum{Bel69} ($^1\Sigma_u^-$), and \citenum{Fis72} ($^1\Delta_u$), all analyzing vacuum electronic spectra;}
$^e${Experimental 0-0 energies from Ref.~\citenum{Hai79} (singlet ES, vacuum UV experiment) and Ref.~\citenum{All84} (triplet ES, EELS). In the latter contribution, the $2.7$ eV value for the $^3\Sigma_u^+$
state is the onset, whereas an estimate of the vertical energy ($4.2 \pm 0.2$ eV) is given for the $^3\Delta_u$ state.}
\end{footnotesize}
\end{flushleft}
\end{sidewaystable}

The ES of these three closely related linear molecules containing two triple bonds have been quite rarely theoretically investigated, \cite{Fis03,Pat06,Luo08,Loo18b,Loo19a} though (rather old) experimental measurements of their
0-0 energies are available for several ES. \cite{Cal63,Job66a,Job66b,Bel69,Fis72,Har77,Hai79,All84}  Our main results are collected in Tables \ref{Table-1} and S1. We consider only low-lying valence  $\pi \ra \pis$ transitions, which
are all characterized by a strongly dominant single excitation nature ($\Td > 90\%$, \emph{vide infra}). For cyanoacetylene, the {\FCI}/{\Pop} estimates come with small error bars, and one notices an excellent agreement between
these values and their {\CCSDTQ} counterparts, a statement holding for the Dunning double-$\zeta$ basis set results for which the {\FCI} uncertainties are however larger.  Using the  {\CCSDTQ}  values as references, it appears
that the previously obtained {\CASPT} estimates\cite{Luo08} are, as expected, too low and that the {\CCT} transition energies are slightly more accurate than their CCSDT counterparts, although all CC estimates of Table \ref{Table-1}
come, for a given basis set, in a very tight energetic window. There is also a very neat agreement between the CC/{\AVTZ} and {\NEV}/{\AVTZ}. All these facts provide strong evidences that the CC estimates can be fully trusted for
these three linear systems.  The basis set effects are quite significant for the valence ES of cyanoacetylene with successive drops of the transition energies by approximately $0.10$ eV, when going from {\Pop} to {\AVDZ}, and from
{\AVDZ} to {\AVTZ}. The lowest triplet state appears less basis set sensitive, though.  As expected, extending further the basis set size (to quadruple- and quintuple-$\zeta$) leaves the results pretty much unchanged. The same
observation holds when adding a second set of diffuse functions, or when correlating the core electrons (see the SI).  Obviously, both cyanogen and diacetylene yield very similar trends, with limited methodological effects and quite
large basis set effects, except for the $^1\Sigma_g^+ \ra {} ^3\Sigma_u^+$ transitions. We note that all CC3 and CCSDT values are, at worst, within $\pm 0.02$ eV of the {\FCI} window, \ie, all methods presented in Table \ref{Table-1}
provide very consistent estimates. For all the states reported in this Table, the average absolute deviation between {\NEV}/{\AVTZ} and {\CCT}/{\AVTZ} ({\CCSDT}/{\AVTZ}) is as small as $0.02$ ($0.03$) eV, the lowest absorption and
emission energies of cyanogen being the only two cases showing significant deviations. As a final note, all our vertical absorption (emission) energies are significantly larger (smaller) than the experimentally measured 0-0 energies,
as they should. We refer the interested reader to previous works, \cite{Fis03,Loo19a} for comparisons between theoretical ({\CASPT} and {\CCT}) and experimental 0-0 energies for these three compounds.

\subsubsection{Cyclopropenone, cyclopropenethione, and methylenecyclopropene}

These three related compounds present a three-membered $sp^2$ carbon cycle conjugated to an external $\pi$ bond. While the ES of methylenecyclopropene have regularly been investigated with theoretical tools in the past,
\cite{Mer96,Roo96,Car10b,Lea12,Gua13,Dad14,Gua14,Sch17,Bud17} the only investigations of vertical transitions we could find for the two other derivatives are a detailed {\CASPT} study of Serrano-Andr\'es and
coworkers in 2002, \cite{Ser02} and a more recent work reporting the three lowest-lying singlet states of cyclopropenone at the {\CASPT}/6-31G level.\cite{Liu14b}

\begin{table}[htp]
\caption{\small Vertical transition energies (in eV) for cyclopropenone, cyclopropenethione, and methylenecyclopropene.}
\label{Table-2}
  \begin{footnotesize}
\begin{tabular}{l|p{.5cm}p{.9cm}p{1.1cm}p{1.4cm}|p{.5cm}p{1.1cm}|p{.5cm}p{.9cm}p{1.2cm}|p{.6cm}p{.6cm}p{.6cm}}
\hline
		 \mc{12}{c}{Cyclopropenone}\\
		& \mc{4}{c}{\Pop} & \mc{2}{c}{\AVDZ}& \mc{3}{c}{\AVTZ}  &  \mc{2}{c}{Litt.}\\
State 	& {\CCT} & {\CCSDT} & {\CCSDTQ}  & {\FCI} & {\CCT}  &  {\CCSDT}& {\CCT} & {\CCSDT}  & {\NEV}  & Th.$^a$ & Exp.$^b$  \\
\hline
$^1B_1 (n \ra \pis)$		&4.32&4.34&4.36& 4.38$\pm$0.02		&4.22&4.23	&4.21&4.24&4.04	&4.25&4.13	\\
$^1A_2 (n \ra \pis)$		&5.68&5.65&5.65& 5.64$\pm$0.06		&5.59&5.56	&5.57&5.55&5.85	&5.59&5.5		\\
$^1B_2 (n \ra 3s)$		&6.39&6.38&6.41&					&6.21&6.19	&6.32&6.31&6.51	&6.90&6.22	\\
$^1B_2 (\pi \ra \pis$)		&6.70&6.67&6.68&					&6.56&6.54	&6.54&6.53&6.82	&5.96&6.1		\\
$^1B_2 (n \ra 3p)$		&6.92&6.91&6.94&					&6.88&6.86	&6.96&6.95&7.07	&7.24&6.88	\\
$^1A_1 (n \ra 3p)$		&7.00&7.00&7.03&					&6.88&6.87	&7.00&6.99&7.28	&7.28&		\\
$^1A_1 (\pi \ra \pis)$		&8.51&8.49&8.51&					&8.32&8.29	&8.28&8.26&8.19	&7.80&$\sim$8.1	\\
$^3B_1 (n \ra \pis)$		&4.02&4.03&	   & 4.00$\pm$0.07		&3.90&3.92	&3.91&3.93&3.51	&4.05&		\\
$^3B_2 (\pi \ra \pis)$		&4.92&4.92&	   & 4.95$\pm$0.00		&4.90&4.89	&4.89&4.88&5.10	&4.81&		\\
$^3A_2 (n \ra \pis)$		&5.48&5.44&	   &					&5.38&5.35	&5.37&5.35&5.60	&5.56&		\\
$^3A_1 (\pi \ra \pis)$		&6.89&6.88&	   & 					&6.79&6.78	&6.83&6.79&7.16	&6.98&		\\
\hline
		 \mc{12}{c}{Cyclopropenethione}\\
		& \mc{4}{c}{\Pop} & \mc{2}{c}{\AVDZ}& \mc{3}{c}{\AVTZ}  &  \mc{2}{c}{Litt.}\\
State 	& {\CCT} & {\CCSDT} & {\CCSDTQ}  & {\FCI} & {\CCT}  & {\CCSDT}& {\CCT} & {\CCSDT} & {\NEV}  & Th.$^a$ \\
\hline
$^1A_2 (n \ra \pis)$		&3.46&3.44&3.44& 3.45$\pm$0.01		&3.47&3.45	&3.43&3.41&3.52	&3.23	&	\\
$^1B_1 (n \ra \pis)$		&3.45&3.44&3.45& 3.44$\pm$0.05		&3.42&3.42	&3.43&3.44&3.50	&3.47	&	\\
$^1B_2 (\pi \ra \pis)$		&4.67&4.64&4.62& 4.59$\pm$0.09		&4.66&4.64	&4.64&4.62&4.77	&4.34	&	\\
$^1B_2 (n \ra 3s)$		&5.26&5.24&5.27&					&5.23&5.21	&5.34&5.31&5.35	&4.98	&	\\
$^1A_1 (\pi \ra \pis)$		&5.53&5.52&5.51& 					&5.52&5.50	&5.49&5.47&5.54	&5.52	&	\\
$^1B_2 (n \ra 3p)$		&5.83&5.81&5.83& 					&5.86&5.84	&5.93&5.90&5.99	&5.88	&	\\
$^3A_2 (n \ra \pis)$		&3.33&3.31&	   & 3.29$\pm$0.03		&3.34&3.32	&3.30&	   &3.38	&3.20	&	\\
$^3B_1 (n \ra \pis)$		&3.34&3.33&	   & 					&3.30&3.30	&3.31&3.32&3.40	&3.30	&	\\
$^3B_2 (\pi \ra \pis)$		&4.01&4.00&	   & 4.03$\pm$0.03		&4.03&4.02	&4.02&	   &4.17	&3.86	&	\\
$^3A_1 (\pi \ra \pis)$		&4.06&4.04&	   &					&4.09&4.07	&4.03&	   &4.13	&3.99	&	\\
\hline
		 \mc{12}{c}{Methylenecyclopropene}\\
		& \mc{4}{c}{\Pop} & \mc{2}{c}{\AVDZ}& \mc{3}{c}{\AVTZ}  &  \mc{3}{c}{Litt.}\\
State 	& {\CCT} & {\CCSDT} & {\CCSDTQ}  & {\FCI} & {\CCT}  &  {\CCSDT}& {\CCT} & {\CCSDT}   & {\NEV}& Th.$^c$ & Th.$^d$ & Exp.$^e$\\
\hline
$^1B_2 (\pi \ra \pis)$		&4.38&4.37&4.34& 4.32$\pm$0.03		&4.32&4.31	&4.31&4.31  &4.37	&4.13&4.36	&4.01\\
$^1B_1 (\pi \ra 3s)$		&5.65&5.66&5.66& 					&5.35&5.35	&5.44&5.44  &5.49	&5.32&5.44	&5.12\\
$^1A_2 (\pi \ra 3p)$		&5.97&5.98&5.98& 5.92$\pm$0.10		&5.86&5.88	&5.95&5.96  &6.00	&5.83&		&\\
$^1A_1(\pi \ra  \pis)$$^f$	&6.17&6.18&6.17& 6.20$\pm$0.01		&6.15&6.15	&6.13&6.13  &6.36	&	&6.13	&6.02\\
$^3B_2 (\pi \ra \pis)$		&3.50&3.50&       & 3.44$\pm$0.06		&3.49&3.49$^g$&3.50&3.49 &3.66	&3.24&		&\\
$^3A_1 (\pi \ra \pis)$		&4.74&4.74&       & 4.67$\pm$0.10		&4.74&4.74$^g$&4.74&        &4.87	&4.52&		&\\
\hline
\end{tabular}
 \end{footnotesize}
\vspace{-0.5 cm}
\begin{flushleft}
\begin{footnotesize}
$^a${{\CASPT} results from Ref.~\citenum{Ser02};}
$^b${Electron impact experiment from Ref.~\citenum{Har74}. Note that the $5.5$ eV peak was assigned differently in the original paper, and we follow here the analysis of Serrano-Andr\'es, \cite{Ser02}
whereas the $6.1$ eV assignment was ``supposed'' in the original paper; experimental $\lambda_{\mathrm{max}}$ have been measured at $3.62$ eV and $6.52$ eV for the $^1B_1$ ($n \ra \pis$) and
$^1B_2$ ($\pi \ra \pis$) transitions, respectively; \cite{Bre72}}
$^c${{\CASPT} results from Refs.~\citenum{Mer96} and \citenum{Roo96};}
$^d${{\CCT} results from Ref.~\citenum{Sch17};}
$^e${$\lambda_{\mathrm{max}}$ in pentane at $-78^o$C from Ref.~\citenum{Sta84};}
$^f${Significant state mixing with the $^1A_1$($\pi \ra  3p$) transition, yielding unambiguous attribution difficult;}
$^g${As can be seen in the SI, our {\FCI}/{\AVDZ} estimates are $3.45 \pm 0.04$ and $4.79 \pm 0.02$ eV for the two lowest triplet states of methylenecyclopropene hinting that the CC3 and CCSDT
results might be slightly too low for the second transition. }
\end{footnotesize}
\end{flushleft}
\end{table}

Our results are listed in Tables \ref{Table-2} and S2. As above, considering the {\Pop} basis set, we notice very small differences between {\CCT}, {\CCSDT}, and {\CCSDTQ}, the latter method giving transition energies
systematically falling within the {\FCI} extrapolation incertitude, except in one case (the lowest totally symmetric state of methylenecyclopropene for which the {\CCSDTQ} value is ``off'' by $0.02$ eV  only). Depending on the state, it is
either {\CCT} or {\CCSDT} that is closest to {\CCSDTQ}. In fact, considering the {\CCSDTQ}/{\Pop} data listed in Table \ref{Table-2} as reference, the MAE of {\CCT} and {\CCSDT} are $0.019$ and $0.016$ eV, respectively,
hinting that the improvement brought by the latter, more expensive method is limited for these three compounds.  For the lowest $B_2$ state of methylenecyclopropene, one of the most challenging cases ($\Td = 85\%$),
it is clear from the {\FCI} value that only {\CCSDTQ} is energetically close, the {\CCT} and {\CCSDT} results being slightly too large by $\sim 0.05$ eV. It seems reasonable to believe that the same observation can be made for the corresponding state of
cyclopropenethione, although in this case the FCI error bar is too large, which prevents any definitive conclusion. Interestingly, at the {\CCT} level of theory, the rather small {\Pop} basis set provides data within $0.10$ eV of the CBS limit for $80\%$ of
the transitions.  There are, of course, exceptions to this rule, \eg, the strongly dipole-allowed $^1A_1 (\pi \ra \pis)$ ES of cyclopropenone and the $^1B_1(\pi \ra 3s)$ ES of methylenecyclopropene which are significantly
over blueshifted with the Pople basis set (Table S2).  For cyclopropenone, our {\CCSDT}/{\AVTZ} estimates do agree reasonably well with the {\CASPT} data of Serrano-Andr\'es, except for the $^1B_2 (\pi \ra \pis)$ state
that we locate significantly higher in energy and the three Rydberg states that our CC calculations predict at significantly lower energies. The present {\NEV} results are globally in better agreement with the CC values,
though non-negligible deviations pertain.  Even if comparisons with experiment should be made very cautiously, we note that, for the Rydberg states, the present CC data are clearly more consistent with the electron
impact measurements\cite{Har74} than the original {\CASPT} values. For cyclopropenethione, we typically obtain transition energies in agreement or larger than those obtained with {\CASPT}, \cite{Ser02} though there is no obvious relationship between the
valence/Rydberg nature of the ES and the relative {\CASPT} error.  The average absolute deviation between our  {\NEV}  and  {\CCT}  results is $0.08$ eV only. Finally, in the case of methylenecyclopropene, our values logically agree
very well with the recent estimates of Schwabe and Goerigk, \cite{Sch17} obtained at the {\CCT}/{\AVTZ} level of theory on a different geometry. As anticipated, the available {\CASPT} values, \cite{Mer96,Roo96} determined without IPEA shift, appear too
low as compared to the present {\NEV} and {\CCSDT} values.  For this compound, the available experimental data are based on the wavelength of maximal absorption determined in condensed phase. \cite{Sta84} Hence, only a
qualitative match is reached between theory and experiment.

\subsubsection{Acrolein, butadiene, and glyoxal}

Let us now turn our attention to the excited states of three pseudo-linear $\pi$-conjugated systems that have been the subject to several investigations in the past, namely, acrolein, \cite{Aqu03,Sah06,Car10b,Lea12,Gua13,Mai14,Aza17b,Sch17,Bat17}
butadiene, \cite{Dal04,Bog04,Sah06,Sch08,Sil10c,Li11,Wat12,Dad12,Lea12,Ise12,Ise13,Sch17,Shu17,Sok17,Chi18,Cop18,Tra19,Loo19c} and glyoxal. \cite{Sta97b,Koh03,Hat05c,Sah06,Lea12,Poo14,Sch17,Aza17b,Loo18b}
Among these works, it is worth highlighting the detailed theoretical investigation of Saha, Ehara, and Nakatsuji, who reported a huge number of ES for these three systems using a coherent theoretical protocol based on the symmetry-adapted-cluster
configuration interaction (SAC-CI) method. \cite{Sah06} In the following, these three molecules are considered in their most stable \emph{trans} conformation. Our results are listed in Tables \ref{Table-3} and S3.

\begin{table}[htp]
\caption{\small Vertical transition energies (in eV) of acrolein, butadiene, and glyoxal.}
\label{Table-3}
  \begin{footnotesize}
\begin{tabular}{p{2.9cm}|p{.5cm}p{.9cm}p{1.1cm}p{1.45cm}|p{.5cm}p{1.1cm}|p{.5cm}p{.9cm}p{1.2cm}|p{.6cm}p{.6cm}p{.6cm}}
\hline
		 \mc{12}{c}{Acrolein}\\
		& \mc{4}{c}{\Pop} & \mc{2}{c}{\AVDZ}& \mc{3}{c}{\AVTZ}  &  \mc{3}{c}{Litt.}\\
State 	& {\CCT} & {\CCSDT} & 		 & {\FCI} & {\CCT}  &  {\CCSDT}& {\CCT} & {\CCSDT}  & {\NEV}   & Th.$^a$ & Th.$^b$ & Exp.$^c$  \\
\hline
$^1A'' (n \ra \pis)$			&3.83&3.80&	&3.85$\pm$0.01&3.77&3.74&		3.74&3.73&3.76	&3.63&3.83&3.71	\\
$^1A' (\pi \ra \pis)$			&6.83&6.86&	&6.59$\pm$0.05$^f$&6.67&6.70&	6.65&6.69&6.67	&6.10&6.92&6.41	\\
$^1A'' (n \ra \pis)$			&6.94&6.89&	&			&6.75&6.72&		6.75&	&7.16	&6.26&7.40&		\\
$^1A' (n \ra 3s)$			&7.22&7.23&	&			&6.99&7.00&		7.07&	&7.05	&6.97&7.19&7.08	\\
$^3A'' (n \ra \pis)$			&3.55&3.53&	&3.60$\pm$0.01&3.47&3.45&		3.46&	&3.46	&3.39&3.61&		\\
$^3A' (\pi \ra \pis)$			&3.94&3.95&	&3.98$\pm$0.03&3.95&3.95&		3.94&	&3.95	&3.81&3.87&		\\
$^3A' (\pi \ra \pis)$			&6.25&6.23&	&			&6.22&6.21&		6.19&	&6.23	&	 &6.21&		\\
$^3A'' (n \ra \pis)$			&6.81&6.74&	&			&6.60&	&		6.61&	&6.83	&	 &7.36&		\\
\hline
		 \mc{12}{c}{Butadiene}\\
		& \mc{4}{c}{\Pop} & \mc{2}{c}{\AVDZ}& \mc{3}{c}{\AVTZ}  &  \mc{3}{c}{Litt.}\\
State 	& {\CCT} & {\CCSDT} & {\CCSDTQ}  & {\FCI} & {\CCT}  & {\CCSDT}& {\CCT} & {\CCSDT}  & {\NEV}  & Th.$^b$  & Th.$^d$ & Exp$^e$ \\
\hline
$^1B_u  (\pi \ra \pis)$		&6.41&6.43&6.41&6.41$\pm$0.02		&6.25&6.27&	6.22&6.24	&6.68	&6.33&6.36&5.92\\
$^1B_g (\pi \ra 3s)$			&6.53&6.55&6.54&					&6.26&6.27&	6.33&6.34	&6.44	&6.18&6.32&6.21\\
$^1A_g  (\pi \ra \pis)$		&6.73&6.63&6.56&6.55$\pm$0.04$^f$	&6.68&6.59&	6.67&6.60	&6.70	&6.56&6.60&	    \\
$^1A_u (\pi \ra 3p)$			&6.87&6.89&6.87&					&6.57&6.59&	6.64&6.66	&6.84	&6.45&6.56&6.64\\
$^1A_u (\pi \ra 3p)$			&6.93&6.95&6.94&6.95$\pm$0.01		&6.73&6.74&	6.80&6.81	&7.01	&6.65&6.74&6.80\\
$^1B_u (\pi \ra 3p)$			&7.98&8.00&7.98&					&7.86&7.87&	7.68&	&7.45	&7.08&7.02&7.07\\
$^3B_u (\pi \ra \pis)$			&3.35&3.36&	   &3.37$\pm$0.03		&3.36&3.36&	3.36&	&3.40	&3.20&	  &3.22\\
$^3A_g (\pi \ra \pis)$			&5.22&5.22&       &					&5.21&5.21&	5.20&	&5.30	&5.08&	  &4.91\\
$^3B_g (\pi \ra 3s)$			&6.46&6.47&       &6.40$\pm$0.03		&6.20&6.21&	6.28&	&6.38	&6.14&	  &\\
\hline
		 \mc{12}{c}{Glyoxal}\\
		& \mc{4}{c}{\Pop} & \mc{2}{c}{\AVDZ}& \mc{3}{c}{\AVTZ}  &  \mc{3}{c}{Litt.}\\
State 	& {\CCT} & {\CCSDT} & {\CCSDTQ}  & {\FCI} & {\CCT}  & {\CCSDT}& {\CCT} & {\CCSDT}  & {\NEV}  & Th.$^b$ & Th.$^g$ & Exp.$^h$\\
\hline
$^1A_u (n \ra \pis)$		&2.94&2.94&2.94& 2.93$\pm$0.03		&2.90&2.90&	2.88&2.88	&2.90	&3.10&2.93	&2.8	\\
$^1B_g (n \ra \pis)$		&4.34&4.32&4.31&4.28$\pm$0.06		&4.30&4.28&	4.27&4.25	&4.30	&4.68&4.39	&$\sim$4.4\\
$^1A_g (n,n  \ra \pis,\pis)$&6.74&6.24&5.67&5.60$\pm$0.09$^f$	&6.70&6.22&	6.76&6.35	&5.52	&5.66&		&\\
$^1B_g (n \ra \pis)$		&6.81&6.83&6.79&					&6.59&6.61&	6.58&6.61	&6.64	&7.54&6.63	&7.45\\
$^1B_u (n \ra 3p)$		&7.72&7.74&7.76&					&7.55&7.56&	7.67&7.69	&7.84	&7.83&7.61	&$\sim$7.7\\
$^3A_u (n \ra \pis)$		&2.55&2.55&	   &2.54$\pm$0.04		&2.49&2.49&	2.49&2.49	&2.49	&2.63&		&2.5\\
$^3B_g (n \ra \pis)$		&3.97&3.95&	   &					&3.91&3.90&	3.90&3.89	&3.99	&4.12&		&$\sim$3.8\\
$^3B_u (\pi \ra \pis)$		&5.22&5.20&	   &					&5.20&5.19&	5.17&5.15	&5.17	&5.35&		&$\sim$5.2\\
$^3A_g (\pi \ra \pis)$		&6.35&6.35&	   &					&6.34&6.34&	6.30&6.30	&6.33	&	  &		&\\
\hline
\end{tabular}
 \end{footnotesize}
\vspace{-0.5 cm}
\begin{flushleft}
\begin{footnotesize}
$^a${{\CASPT} results from Ref.~\citenum{Aqu03};}
$^b${SAC-CI results from Ref.~\citenum{Sah06};}
$^c${Vacuum UV spectra from Ref.~\citenum{Wal45}; for the lowest state, the same $3.71$ eV value is reported in Ref.~\citenum{Bec70}.}
$^d${MR-AQCC results from Ref.~\citenum{Dal04}, theoretical best estimates listed for the lowest $B_u$ and $A_g$ states;}
$^e${Electron impact experiment from Refs.~\citenum{Fli78} and \citenum{Doe81} for the singlet states and from Ref.~\citenum{Mos73} for the two lowest triplet transitions;
note that for the lowest $B_u$ state, there is a vibrational structure with peaks at $5.76$, $5.92$, and $6.05$ eV;}
$^f${From Ref.~\citenum{Loo19c};}
$^g${{\CCT} results from Ref.~\citenum{Sch17};}
$^h${Electron impact experiment from Ref.~\citenum{Ver80} except for the second $^1B_g$ ES for which the value is from another work (see Ref.~\citenum{Rob85b});  note that
for the lowest $^1B_g$ ($^1B_u$) ES, a range of $4.2$--$4.5$ ($7.4$--$7.9$) eV is given in Ref.~\citenum{Ver80}. }
\end{footnotesize}
\end{flushleft}
\end{table}

Acrolein, due to its lower symmetry and high density of ES with mixed characters, is challenging from a theoretical point of view, and {\CCSDTQ} calculations were technically impossible despite all our efforts.  For the lowest $n \ra \pis$
transitions of both spin symmetry, the {\FCI} estimates come with a tiny error bar, and it is obvious that the CC excitation energies are slightly too low, especially with {\CCSDT}. Nevertheless, at the exception of the second singlet
and triplet  $A''$ ES, the {\CCT} and {\CCSDT} transition energies are within $\pm 0.03$ eV of each other. These  $A''$ ES are also the only two transitions for which the discrepancies between  {\CCT} and {\NEV} exceed $0.20$ eV.
This hints at a good accuracy for all other transitions. This statement is additionally supported by the fact that the present CC values are nearly systematically bracketed by previous {\CASPT} (lower bound)\cite{Aqu03} and SAC-CI (upper bound)\cite{Sah06} results,
consistently with the typical error sign of these two models. For the two lowest triplet states, the present {\CCT}/{\AVTZ} values are also within $\pm 0.05$ eV of recent MRCI estimates ($3.50$ and $3.89$ eV). \cite{Mai14} As
can be seen in Table S3, the {\AVTZ} basis set delivers excitation energies very close to the CBS limit: the largest variation when upgrading from {\AVTZ} to {\AVQZ} ($+0.04$ eV) is obtained for the second $^1A'$ Rydberg ES. As experimental data
are limited to measured UV spectra, \cite{Wal45,Bec70} one has to be ultra cautious in establishing TBE for acrolein (\emph{vide infra}).

The nature and relative energies of the lowest bright $B_u$ and dark $A_g$ ES of butadiene have puzzled theoretical chemists for many years.  It is beyond the scope of the present study to provide an exhaustive list of previous calculations
and experimental measurements for these two hallmark ES, and we refer the readers to Refs.~\citenum{Wat12} and \citenum{Shu17} for a general and broader overview.  For the $B_u$ transition, we believe
that the most solid TBE is the $6.21$ eV value obtained by Watson and Chan using a computational strategy similar to ours. \cite{Wat12}  Our {\CCSDT}/{\AVTZ} value of $6.24$ eV is obviously compatible with their reference value, and our TBE/CBS
value is actually $6.21$ eV as well (\emph{vide infra}).  For the $A_g$ state, we believe that our previous basis set corrected {\FCI} estimate of $6.50$ eV \cite{Loo19c} remains the most accurate available to date.  These two values are slightly lower
than the semi-stochastic heath-bath CI data obtained by Chien \emph{et al.} with a double-$\zeta$ basis and a slightly different geometry: $6.45$ and $6.58$ eV for $B_u$ and $A_g$, respectively. \cite{Chi18}  For these two thoroughly studied ES,
one can of course find many other estimates, \eg, at the SAC-CI, \cite{Sah06} {\CCT}, \cite{Sil10c,Sch17} {\CASPT}, \cite{Sil10c} and {\NEV} \cite{Sok17} levels. Globally, for butadiene, we find an excellent coherence between the {\CCT}, {\CCSDT},
and {\CCSDTQ} estimates, that all fall in a $\pm 0.02$ eV window. Unsurprisingly, this does not apply to the already mentioned $^1A_g$ ES that is $0.2$ and $0.1$ eV too high with the two former CC methods, a direct consequence of the large
electronic reorganization taking place during this transition.  For all the other butadiene ES listed in Table \ref{Table-3}, both {\CCT} and {\CCSDT} can be trusted. We also note that the {\NEV} estimates are within $0.1$--$0.2$ eV of the CC values,
except for the lowest $B_u$ ES for which the associated excitation energy is highly dependent on the selected active space (see the SI).  Finally, as can be seen in Table S3, {\AVTZ} produces near-CBS excitation energies for most ES.
However, a significant basis set effect exists for the Rydberg $^1B_u (\pi \ra 3p)$ ES with an energy lowering as large as $-0.12$ eV when going from {\AVTZ} to {\AVQZ}. For the record, we note that the available electron impact data
\cite{Mos73,Fli78,Doe81} provide the very same ES ordering as our calculations.

Globally, the conclusions obtained for acrolein and butadiene pertain for glyoxal, \ie, highly consistent CC estimates, reasonable agreement between {\NEV} and {\CCT} values, and limited basis set effects beyond {\AVTZ}, except for the $^1B_u
(n \ra 3p)$ Rydberg state (see Tables \ref{Table-3} and S3). This Rydberg state also exhibits an unexpectedly large deviation of $0.04$ eV between {\CCT} and {\CCSDTQ}. More interestingly, glyoxal presents a genuine low-lying double ES
of $^1A_g$ symmetry. The corresponding $(n,n)  \ra (\pis,\pis)$ transition is totally unseen by approaches that cannot model double excitations, \eg, TD-DFT, {\CCSD}, or {\AD}. Compared to the {\FCI} values, the {\CCT} and {\CCSDT}
estimates associated with this transition are too large by $\sim 1.0$ and $\sim 0.5$ eV, respectively, whereas both the  {\CCSDTQ} and {\NEV} approaches are much closer, as already mentioned in our previous work. \cite{Loo19c} For the
other transitions, the present {\CCT} estimates are logically consistent with the values of Ref.~\citenum{Sch17} obtained with the same approach on a different geometry, and remain slightly lower than the SAC-CI estimates of Ref.~\citenum{Sah06}.
Once more, the experimental data \cite{Ver80,Rob85b} are unhelpful in view of the targeted accuracy.

\subsubsection{Acetone, cyanoformaldehyde, isobutene, propynal, thioacetone, and thiopropynal}

Let us now turn towards six other compounds with four non-hydrogen atoms. There are several earlier studies reporting estimates of the vertical transition energies for both acetone \cite{Gwa95,Mer96b,Roo96,Wib98,Toz99b,Wib02,Sch08,Sil10c,Car10,Pas12,Ise12,Gua13,Sch17}
and isobutene. \cite{Wib02,Car10,Ise12} To the best of our knowledge, for the four other compounds, the previous computational efforts were mainly focussed on the 0-0 energies of the lowest-lying states. \cite{Koh03,Hat05c,Sen11b,Loo18b,Loo19a}
There are also rather few experimental data available for these six derivatives. \cite{Bir73,Jud83,Bra74,Sta75,Joh79,Jud83,Jud84c,Rob85,Pal87,Kar91b,Xin93}
Our main results are reported in Tables \ref{Table-4} and S4.

\begin{table}[htp]
\caption{\small Vertical transition energies (in eV) of acetone, cyanonformaldehyde, isobutene, propynal, thioacetone, and thiopropynal.}
\label{Table-4}
  \begin{footnotesize}
\begin{tabular}{l|p{.5cm}p{.9cm}p{1.1cm}p{1.4cm}|p{.5cm}p{1.1cm}|p{.5cm}p{.9cm}p{1.2cm}|p{.6cm}p{.6cm}p{.6cm}}
\hline
		 \mc{12}{c}{Acetone}\\
		& \mc{4}{c}{\Pop} & \mc{2}{c}{\AVDZ}& \mc{3}{c}{\AVTZ}  &  \mc{3}{c}{Litt.}\\
State 	& {\CCT} & {\CCSDT} &  {\CCSDTQ} & {\FCI} & {\CCT}  &  {\CCSDT}& {\CCT} & {\CCSDT}  & {\NEV}   & Th.$^a$ & Th.$^b$ & Exp.$^c$  \\
\hline
$^1A_2 (n \ra \pis)$			&4.55&4.52&4.53&4.60$\pm$0.05		&4.50&4.48&	4.48&4.46&4.48	&4.18&4.18&4.48\\
$^1B_2 (n \ra 3s)$			&6.65&6.64&6.68&					&6.31&6.30&	6.43&6.42&6.81	&6.58&6.58&6.36\\
$^1A_2 (n \ra 3p)$			&7.83&7.83&7.87&					&7.37&7.36&	7.45&7.43	&7.65	&7.34&7.34&7.36\\
$^1A_1 (n \ra 3p)$			&7.81&7.81&7.84&					&7.39&7.38&	7.48&7.48&7.75	&7.26&7.26&7.41\\
$^1B_2 (n \ra 3p)$			&7.87&7.87&7.91&					&7.56&7.55&	7.59&7.58	&7.91	&7.48&7.48&7.45\\
$^3A_2 (n \ra \pis)$			&4.21&4.19&       &4.18$\pm$0.04		&4.16&4.14&	4.15&	&4.20	&3.90&3.90&4.15\\
$^3A_1 (\pi \ra \pis)$			&6.32&6.30&       &					&6.31&6.28&	6.28&	&6.28	&5.98&5.98&\\
\hline
		 \mc{12}{c}{Cyanoformaldehyde}\\
		& \mc{4}{c}{\Pop} & \mc{2}{c}{\AVDZ}& \mc{3}{c}{\AVTZ}  & \mc{1}{c}{Litt.}\\\
State 	& {\CCT} & {\CCSDT} & 	 & {\FCI} & {\CCT}  & {\CCSDT}& {\CCT} & {\CCSDT}  & {\NEV}  &  Exp$^d$ \\
\hline
$^1A'' (n \ra \pis)$			&3.91&3.89&	&3.92$\pm$0.02		&3.86&3.84&3.83	&3.81&3.98&	3.26\\
$^1A'' (\pi \ra \pis)$			&6.64&6.67&	&6.60$\pm$0.07		&6.51&6.54&6.42	&6.46&6.44&	\\
$^3A'' (n \ra \pis)$			&3.53&3.51&	&3.48$\pm$0.06		&3.47&3.45&3.46	&	&3.58&	\\
$^3A' (\pi \ra \pis)$			&5.07&5.07&	&					&5.03&5.03&5.01	&	&5.35&	\\
\hline
		 \mc{12}{c}{Isobutene}\\
		& \mc{4}{c}{\Pop} & \mc{2}{c}{\AVDZ}& \mc{3}{c}{\AVTZ}  & \mc{3}{c}{Litt.}\\\
State 	& {\CCT} & {\CCSDT} & 	 & {\FCI} & {\CCT}  & {\CCSDT}& {\CCT} & {\CCSDT}  & {\NEV}  &  Th.$^e$ & Exp.$^f$ & Exp.$^g$ \\
\hline
$^1B_1 (\pi \ra 3s)$			&6.77&6.77&	&6.78$\pm$0.08		&6.39&6.39&	6.45&6.46&6.63&6.40&6.15&6.17	\\
$^1A_1 (\pi \ra 3p)$			&7.16&7.17&	&7.16$\pm$0.02		&7.00&7.00&	7.00&7.01&7.20&6.96&	     &6.71	\\
$^3A_1 (\pi \ra \pis)$			&4.52&4.53&	&4.56$\pm$0.02		&4.54&4.54&	4.53&	&4.61&	  &4.21  &4.3	\\
\hline
		 \mc{12}{c}{Propynal}\\
		& \mc{4}{c}{\Pop} & \mc{2}{c}{\AVDZ}& \mc{3}{c}{\AVTZ}  & \mc{1}{c}{Litt.} \\
State 	& {\CCT} & {\CCSDT} & 	  & {\FCI} & {\CCT}  & {\CCSDT}& {\CCT} & {\CCSDT}  & {\NEV}  &  Exp$^h$  \\
\hline
$^1A'' (n \ra \pis)$			&3.90&3.87&	&3.84$\pm$0.06		&3.85&3.82&3.82	&3.80&3.95&	3.24\\
$^1A'' (\pi \ra \pis)$			&5.69&5.73&	&5.64$\pm$0.08		&5.59&5.62&5.51	&5.54&5.50&	\\
$^3A'' (n \ra \pis)$			&3.56&3.54&	&3.54$\pm$0.04		&3.50&3.48&3.49	&	&3.59&	2.99\\
$^3A' (\pi \ra \pis)$			&4.46&4.47&	&4.44$\pm$0.08		&4.40&4.44&4.43	&	&4.63&	\\
\hline
		 \mc{12}{c}{Thioacetone}\\
		& \mc{4}{c}{\Pop} & \mc{2}{c}{\AVDZ}& \mc{3}{c}{\AVTZ}  &  \mc{1}{c}{Litt.}\\
State 	& {\CCT} & {\CCSDT} & {\CCSDTQ}  & {\FCI} & {\CCT}  & {\CCSDT}& {\CCT} & {\CCSDT}  & {\NEV}  &  Exp$^i$ \\
\hline
$^1A_2 (n \ra \pis)$			&2.58&2.56&2.56&2.61$\pm$0.05		&2.59&2.57&2.55	&2.53&2.55	&2.33\\
$^1B_2 (n \ra 4s)$			&5.65&5.64&5.66&					&5.44&5.43&5.55	&5.54&5.72	&5.49\\
$^1A_1 (\pi \ra \pis)$			&6.09&6.10&6.07&					&5.97&5.98&5.90	&5.91&6.24	&5.64\\
$^1B_2 (n \ra 4p)$			&6.59&6.59&6.59&					&6.45&6.44&6.51	&	 &6.62	&6.40\\
$^1A_1 (n \ra 4p)$			&6.95&6.95&6.96&					&6.54&6.53&6.61	&6.60&6.52	&6.52\\
$^3A_2 (n \ra \pis)$			&2.36&2.34&       &2.36$\pm$0.00		&2.36&2.35&2.34	&	 &2.32	&2.14\\
$^3A_1 (\pi \ra \pis)$			&3.45&3.45&       &					&3.51&3.50&3.46	&	 &3.48	&\\
\hline
		 \mc{12}{c}{Thiopropynal}\\
		& \mc{4}{c}{\Pop} & \mc{2}{c}{\AVDZ}& \mc{3}{c}{\AVTZ}  &  \mc{1}{c}{Litt.}\\
State 	& {\CCT} & {\CCSDT} &  & {\FCI} & {\CCT}  & {\CCSDT}& {\CCT} & {\CCSDT}  & {\NEV}  &  Exp$^j$ \\
\hline
$^1A''  (n \ra \pis)$			&2.09&2.06&       &2.08$\pm$0.01		&2.09&2.06&2.05	&2.03&2.05	&1.82\\
$^3A''   (n \ra \pis)$			&1.84&1.82&       &					&1.83&1.82&1.81	&	 &1.81	&1.64\\
\hline
\end{tabular}
 \end{footnotesize}
\vspace{-0.5 cm}
\begin{flushleft}
\begin{footnotesize}
$^a${{\CASPT} results from Ref.~\citenum{Mer96b};}
$^b${EOM-CCSD results from Ref.~\citenum{Gwa95};}
$^c${Two lowest singlet states: various experiments summarized in Ref.~\citenum{Rob85}; three next singlet states: REMPI experiments from Ref.~\citenum{Xin93}; lowest triplet: trapped electron measurements from Ref.~\citenum{Sta75};}
$^d${0-0 energy reported in Ref.~\citenum{Kar91b};}
$^e${EOM-CCSD results from Ref.~\citenum{Car10};}
$^f${Energy loss experiment from Ref.~\citenum{Joh79};}
$^g${VUV experiment from Ref.~\citenum{Pal87} (we report the lowest of the $\pi \ra 3p$ state for the $^1A_1$ state)};
$^h${0-0 energies from Refs.~\citenum{Bra74} (singlet) and \citenum{Bir73} (triplet);}
$^i${0-0 energies from Ref.~\citenum{Jud83};}
$^i${0-0 energies from Ref.~\citenum{Jud84c}.}
\end{footnotesize}
\end{flushleft}
\end{table}

For acetone, one should clearly distinguish the valence ES, for which both methodological and basis set effects are small, and the Rydberg transitions that are both very basis set sensitive, and upshifted by ca.~$0.04$ eV with {\CCSDTQ} as
compared to {\CCT} and {\CCSDT}. For this compound, the 1996 {\CASPT} transition energies of Merch\'an and coworkers listed on the right panel of Table \ref{Table-4} are clearly too low, especially for the three valence ES. \cite{Mer96b}
As expected, this error can be partially ascribed to the computational set-up, as the Urban group obtained {\CASPT}  excitation energies of $4.40$, $4.09$ and $6.22$ eV for the $^1A_2$, $^3A_2$, and $^3A_1$ ES, \cite{Pas12} in much
better agreement with ours. Their estimates of the three $n \ra 3p$ transitions, $7.52$, $7.57$, and $7.53$ eV for the $^1A_2$, $^1A_1$, and $^1B_2$ ES, also systematically fall within $0.10$ eV of our current CC values, whereas for these
three ES, the current {\NEV} values are clearly too large.

In contrast to acetone, both valence and Rydberg ES of thioacetone are rather insensitive to the excitation order of the CC expansion as illustrated by the maximal discrepancies of $\pm 0.02$ eV between the {\CCT}/{\Pop} and {\CCSDTQ}/{\Pop} results.
While the lowest $n \ra \pis$ transition of both spin symmetries are rather basis set insensitive, all the other states need quite large one-electron bases to be correctly described (Table S4).  As expected, our theoretical vertical transition energies
show the same ranking but are systematically larger than the available experimental 0-0 energies.

For the isoelectronic isobutene molecule, we have considered two singlet Rydberg and one triplet valence ES. For these three cases, we note, for each basis, a very nice agreement between {\CCT} and {\CCSDT}, the
CC results being also very close to the {\FCI} estimates obtained with the Pople basis set. The similarity with the {\CCSD} results of Caricato and coworkers \cite{Car10} is also very satisfying.

For the three remaining compounds, namely, cyanoformaldehyde, propynal, and thiopropynal, we report low-lying valence transitions with a definite single excitation character. The basis set effects are clearly under control (they are only
significant for the second $^1A''$ ES of cyanoformaldehyde) and we could not detect any variation larger than $0.03$ eV between the {\CCT} and {\CCSDT} values for a given basis, indicating that the CC values are very accurate.
This is further confirmed by the {\FCI} data.

\subsubsection{Intermediate conclusions}
\label{sec-ic}

For the 15 molecules with four non-hydrogen atoms considered here, we find extremely consistent transition energies between CC and {\FCI} estimates in the vast majority of the cases. Importantly, we confirm our previous conclusions obtained on
smaller compounds: \cite{Loo18a} i) {\CCSDTQ} values systematically fall within (or are extremely close to) the {\FCI} error bar, ii) both {\CCT} and {\CCSDT} are also highly trustable when the considered ES does not exhibit a strong double
excitation character.  Indeed, considering the 54 ``single'' ES cases for which {\CCSDTQ} estimates could be obtained (only excluding the lowest $^1A_g$ ES of butadiene and glyoxal), we determined negligible MSE $< 0.01$ eV, tiny
MAE ($0.01$ and $0.02$ eV), and small maximal deviations ($0.05$ and $0.04$ eV) for {\CCT} and {\CCSDT}, respectively.   This clearly indicates that these two approaches provide chemically-accurate estimates (errors below $1$
kcal.mol$^{-1}$ or $0.043$ eV) for most electronic transitions. Interestingly, some of us have shown that {\CCT} also provides chemically-accurate 0-0 energies as compared to experimental values for most valence transitions. \cite{Loo18b,Loo19a,Sue19}
When comparing the {\NEV} and {\CCT} ({\CCSDT}) results obtained with {\AVTZ} for the 91 (65) ES for which comparisons are possible (again excluding only the lowest $^1A_g$ states of butadiene and glyoxal),
one obtains a MSE of $+0.09$ ($+0.09$) eV and a MAE of $0.11$ ($0.12$) eV. This seems to indicate that {\NEV}, as applied here, has a slight tendency to overestimate the transition energies. This contrasts with {\CASPT} that is known to
generally underestimate transition energies, as further illustrated and discussed above and below.

\subsection{Five-membered rings}

We now consider five-membered rings, and, in particular, five common derivatives that have been considered in several previous theoretical studies (\emph{vide infra}): cyclopentadiene, furan, imidazole, pyrrole, and thiophene.
As the most advanced levels of theory employed in the previous section, namely {\CCSDTQ} and {\FCI}, become beyond reach for these compounds (except in very rare occasions), one has to rely on the nature of the ES and the
consistency between results to deduce TBE.

For furan, \textit{ab initio} calculations have been performed with almost every available wavefunction method.\cite{Ser93b,Nak96,Tro97b,Chr98b,Chr98c,Wan00,Gro03,Pas06b,Sch08,She09b,Li10c,Sil10b,Sil10c,Sau11,Hol15,Sch17}
However, the present work is, to the best of our knowledge, the first to disclose {\CCSDT} values as well as {\CCT} energies obtained with a quadruple-$\zeta$ basis set.  Our results for ten low-lying ES states are listed in Tables \ref{Table-5} and S5.
All singlet (triplet) transitions are characterized by $\Td$ values in the $92$--$94\%$ ($97$--$99\%$) range. Consistently, the maximal discrepancy between {\CCT} and {\CCSDT} is small ($0.04$ eV). In addition, there is a decent consistency
between the present data and the {\NEV} results of both Ref.~\citenum{Pas06b} and of the present work, as well as the MR-CC values of Ref.~\citenum{Li10c}. This holds for almost all transitions, but the $^1B_2$ ($\pi \ra 3p$) excitation that we predict
to be significantly higher than in most previous works, even after accounting for the quite large basis set effects ($-0.10$ eV between the {\AVTZ} and {\AVQZ} estimates, see Table S5). We believe that our estimate is the most accurate to date
for this particularly tricky ES. Interestingly, the recent {\AT} values of Ref.~\citenum{Hol15} are consistently smaller by ca.~$-0.2$ eV as compared to {\CCSDT} (see Table \ref{Table-6}), in agreement with the error sign we observed in smaller
compounds for {\AT}. \cite{Loo18a}  Again, we note that the experimental data \cite{Vee76b,Fli76,Rob85b} provide the same state ordering as our calculations.

\begin{table}[htp]
\caption{\small Vertical transition energies (in eV) of furan and pyrrole.}
\label{Table-5}
\begin{footnotesize}
\begin{tabular}{l|p{.5cm}p{1.0cm}|p{.5cm}p{1.0cm}|p{.5cm}p{1.0cm}p{1.2cm}|p{.5cm}p{.5cm}p{.5cm}p{.5cm}p{.5cm}p{.6cm}p{.6cm}}
\hline
		 \mc{15}{c}{Furan}\\
		& \mc{2}{c}{\Pop} & \mc{2}{c}{\AVDZ}& \mc{3}{c}{\AVTZ}	&  \mc{7}{c}{Litt.}\\
State 	& {\CCT} & {\CCSDT} & {\CCT} & {\CCSDT} & {\CCT} & {\CCSDT}   & {\NEV}  	&Th.$^a$	&Th.$^b$	&Th.$^c$	&Th.$^d$	&Th.$^e$ &Exp.$^f$&Exp.$^g$\\
\hline
$^1A_2 (\pi \ra 3s)$		&6.26&6.28 	&6.00&6.00	&6.08&6.09&6.28	&5.92&6.13&5.94&5.91&6.10&5.91 & \\
$^1B_2 (\pi \ra \pis)$		&6.50&6.52	&6.37&6.39	&6.34&6.37&6.20	&6.04&6.42&6.51&6.10&6.42&6.04 & 6.06\\
$^1A_1 (\pi \ra \pis)$		&6.71&6.67	&6.62&6.58	&6.58&6.56&6.77	&6.16&6.71&6.89&6.44&      &         & 6.44 \\
$^1B_1  (\pi \ra 3p)$		&6.76&6.77	&6.55&6.56	&6.63&6.64&6.71	&6.46&6.68&6.46&6.45&6.66&6.47 & \\
$^1A_2  (\pi \ra 3p)$		&6.97&6.99	&6.73&6.74	&6.80&6.81&6.99	&6.59&6.79&6.61&6.60&6.83&6.61 & \\
$^1B_2  (\pi \ra 3p)$		&7.53&7.54	&7.39&7.40	&7.23&	  &7.01	&6.48&6.91&6.87&6.72&7.36&6.75 & \\
$^3B_2 (\pi \ra \pis)$		&4.28&4.28	&4.25&4.23	&4.22&	  &4.42	&3.99&	  &4.26&		&   &       &4.0\\
$^3A_1 (\pi \ra \pis)$		&5.56&5.54	&5.51&5.49	&5.48&	  &5.60	&5.15&	  &5.53&		&   &       &5.2\\
$^3A_2 (\pi \ra 3s)$		&6.18&6.19	&5.94&5.94	&6.02&	  &6.08	&5.86&	  &5.89&		&   &       &\\
$^3B_1 (\pi \ra 3p)$		&6.69&6.71	&6.51&6.51	&6.59&	  &6.68	&6.42&	  &6.41&		&   &       &\\
\hline
		 \mc{15}{c}{Pyrrole}\\
		& \mc{2}{c}{\Pop} & \mc{2}{c}{\AVDZ}& \mc{3}{c}{\AVTZ}	&  \mc{7}{c}{Litt.}\\
State 	& {\CCT} & {\CCSDT} & {\CCT} & {\CCSDT} & {\CCT} & {\CCSDT}   & {\NEV}  	&Th.$^a$	&Th.$^h$	&Th.$^c$	&Th.$^i$	&Th.$^j$ &Exp.$^l$&Exp.$^l$\\
\hline
$^1A_2 (\pi \ra 3s)$		&5.25&5.25	&5.15&5.14	&5.24&5.24&5.51&	5.08&5.45&5.10&5.20&5.27&5.22&\\
$^1B_1 (\pi \ra 3p)$		&5.99&5.98	&5.89&5.87	&5.98&6.00&6.32&	5.85&6.21&5.79&5.95&6.00&	    &\\
$^1A_2 (\pi \ra 3p)$		&6.27&6.27	&5.94&5.93	&6.01&       &6.44&	5.83&6.14&5.81&5.94&7.03&       &5.87\\
$^1B_2 (\pi \ra \pis)$		&6.33&6.33	&6.28&6.28	&6.25&6.26&6.48&	5.92&6.95&5.96&6.04&6.08&	   &5.98\\
$^1A_1 (\pi \ra \pis)$		&6.43&6.40	&6.35&6.32	&6.32&6.30&6.53&	5.92&6.59&6.53&6.37&6.15&      &\\
$^1B_2 (\pi \ra 3p)$		&7.20&7.20	&7.00&7.00	&6.83&       &6.62&	5.78&6.26&6.61&6.57& &\\
$^3B_2 (\pi \ra \pis)$		&4.59&4.58	&4.56&4.54	&4.53&       &4.74&	4.27&       &4.53&       &       &4.21\\
$^3A_2 (\pi \ra 3s)$		&5.22&5.22	&5.12&5.12	&5.21&       &5.49&	5.04&       &5.07&       &       &5.1\\
$^3A_1 (\pi \ra \pis)$		&5.54&5.54	&5.49&5.48	&5.46&       &5.56&	5.16&       &5.53&       &       &\\
$^3B_1 (\pi \ra 3p)$		&5.91&5.90	&5.82&5.81	&5.92&       &6.28&	5.82&       &5.74&       &       &\\
\hline
 \end{tabular}
  \end{footnotesize}
\vspace{-0.5 cm}
\begin{flushleft}
\begin{footnotesize}
$^a${{\CASPT} results from Ref.~\citenum{Ser93b};}
$^b${{\NEV} results from Ref.~\citenum{Pas06b};}
$^c${MR-CC results from Ref.~\citenum{Li10c};}
$^d${{\AT} results from Ref.~\citenum{Hol15};}
$^e${{\CCT} results from Ref.~\citenum{Sch17};}
$^f${Various experiments summarized in Ref.~\citenum{Wan00};}
$^g${Electron impact from Ref.~\citenum{Vee76b}: for the $^1A_1$ state two values ($6.44$ and $6.61$ eV) are reported, whereas for the two lowest triplet states, Two values ($3.99$ eV and $5.22$ eV) can be found in Ref.~\citenum{Fli76};}
$^h${{\NEV} results from Ref.~\citenum{Pas06c};}
$^i${Best estimate from Ref.~\citenum{Chr99}, based on CC calculations;}
$^j${XMS-{\CASPT} results from Ref.~\citenum{Hei19};}
$^k${Electron impact from Refs.~\citenum{Vee76b} and \citenum{Fli76b};}
$^l${Vapour UV spectra from Refs.~\citenum{Pal03b}, \citenum{Hor67}, and \citenum{Bav76}.}
\end{footnotesize}
\end{flushleft}
\end{table}

Like furan, pyrrole has been extensively investigated in the literature using a large panel of theoretical methods. \cite{Ser93b,Nak96,Tro97,Pal98,Chr99,Wan00,Roo02,Pal03b,Pas06c,Sch08,She09b,Li10c,Sil10b,Sil10c,Sau11,Nev14,Sch17,Hei19}
We report six low-lying singlet and four triplet ES in Tables \ref{Table-5} and S5. All these transitions have very large $\Td$ values except for the totally symmetric $\pi \ra \pis$ excitation ($\Td = 86\%$).  For each state, we found
highly consistent {\CCT} and {\CCSDT} results, often significantly larger than older multi-reference estimates, \cite{Ser93b,Roo02,Li10c} but in nice agreement with the very recent XMS-{\CASPT} results of the Gonzalez
group, \cite{Hei19} and the present {\NEV} estimates [at the exception of the $^1A_2 (\pi \ra 3p)$ transition].  The match obtained with the twenty years old extrapolated CC values of Christiansen and coworkers \cite{Chr99} is quite remarkable. The only exceptions are the two
$B_2$ transitions that were reported as significantly mixed in this venerable work. For the lowest singlet ES, the {\FCI}/{\Pop} value is $5.24 \pm 0.02$ eV confirming the performance of both {\CCT} and {\CCSDT} for this transition.
As can be seen in Table S5, {\AVTZ} yields basis-set converged transition energies, except, like in furan, for the Rydberg $^1B_2 (\pi \ra 3p)$ transition that is significantly redshifted ($-0.09$ eV) when pushing to the
quadruple-$\zeta$ basis set. As mentioned in Thiel's work, \cite{Sch08} the experimental spectra of pyrrole are quite broad, and the rare available experiments \cite{Hor67,Bav76,Fli76b,Vee76b,Pal98,Pal03b} can only be considered as general guidelines.

\begin{table}[htp]
\caption{\small Vertical transition energies (in eV) of cyclopentadiene, imidazole, and thiophene.}
\label{Table-6}
\begin{footnotesize}
\begin{tabular}{l|p{.5cm}p{1.0cm}|p{.5cm}p{1.0cm}|p{.5cm}p{1.0cm}p{1.2cm}|p{.5cm}p{.5cm}p{.5cm}p{.5cm}p{.6cm}p{.6cm}p{.6cm}}
\hline
		 \mc{15}{c}{Cyclopentadiene}\\
		& \mc{2}{c}{\Pop} & \mc{2}{c}{\AVDZ}& \mc{3}{c}{\AVTZ}	&  \mc{7}{c}{Litt.}\\
State 	& {\CCT} & {\CCSDT} & {\CCT} & {\CCSDT} & {\CCT} & {\CCSDT}   & {\NEV}  	&Th.$^a$	&Th.$^b$	&Th.$^c$	&Th.$^d$	&Exp.$^e$ &Exp.$^f$&Exp.$^g$\\
\hline
$^1B_2 (\pi \ra \pis)$		&5.79&5.80	&5.59&5.60	&5.54&5.56&5.65	&5.27	&5.54&5.19	&5.58	&5.26	&		&5.20\\
$^1A_2 (\pi \ra 3s)$		&6.08&6.08	&5.70&5.70	&5.77&5.78&5.92	&5.65	&5.58&5.62	&5.79	&5.68	&5.63	&\\
$^1B_1  (\pi \ra 3p)$		&6.57&6.58	&6.34&6.34	&6.40&6.41&6.42	&6.24	&6.17&6.24	&6.43	&		&6.35	&\\
$^1A_2  (\pi \ra 3p)$		&6.67&6.67	&6.39&6.39	&6.45&6.46&6.59	&6.30	&6.21&6.25	&6.47	&		&		&6.26\\
$^1B_2  (\pi \ra 3p)$		&7.06&7.07	&6.55&6.55	&6.56&6.56&6.60	&6.25	&6.22&6.27	&6.58	&		&6.31	&\\
$^1A_1 (\pi \ra \pis)$		&6.67&6.60	&6.59&6.53	&6.57&6.52&6.75	&6.31	&6.76&6.42	&6.65	&		&		&$\sim$6.2\\
$^3B_2 (\pi \ra \pis)$		&3.33&3.33	&3.32&3.31	&3.32& 	  &3.41	&3.15	&3.40&		&		&3.10			&\\
$^3A_1 (\pi \ra \pis)$		&5.16&5.15	&5.14&5.13	&5.12& 	  &5.30	&4.90	&5.18&		&		&$>$4.7	&		&\\
$^3A_2 (\pi \ra 3s)$		&6.01&6.02	&5.65&5.65	&5.73& 	  &5.73	&5.63	&5.56&		&		&\\
$^3B_1 (\pi \ra 3p)$		&6.51&6.52	&6.30&6.30	&6.36& 	  &6.40	&6.25	&6.19&		&		&\\
\hline
		 \mc{15}{c}{Imidazole}\\
		& \mc{2}{c}{\Pop} & \mc{2}{c}{\AVDZ}& \mc{3}{c}{\AVTZ}	&  \mc{7}{c}{Litt.}\\
State 	& {\CCT} & {\CCSDT} & {\CCT} & {\CCSDT} & {\CCT} &  & {\NEV}  	&Th.$^h$	&Th.$^i$	&&&Exp.$^j$\\
\hline
$^1A'' (\pi \ra 3s)$		&5.77&5.77	&5.60&5.60	&5.71&&5.93	&5.71&       &&&$\sim$5.2\\
$^1A' (\pi \ra \pis)$$^k$	&6.51&6.51	&6.43&6.43	&6.41&&6.73	&6.72&6.25&&&$\sim$6.4\\
$^1A'' (n \ra \pis)$		&6.66&6.66	&6.42&6.42	&6.50&&6.96	&6.52&6.65&&& \\
$^1A' (\pi \ra 3p)$$^k$	&7.04&7.02	&6.93&6.89	&6.87&&7.00	&6.49&       &&&\\
$^3A' (\pi \ra \pis)$		&4.83&4.81	&4.78&		&4.75&&4.86	&4.49&4.65&&&\\
$^3A'' (\pi \ra 3s)$		&5.72&5.72	&5.57&5.56	&5.67&&5.91	&5.68&	  &&&\\
$^3A' (\pi \ra \pis)$		&5.88&5.88	&5.78&		&5.74&&5.91	&5.47&5.64&&&\\
$^3A'' (n \ra \pis)$		&6.48&6.46	&6.37&6.35	&6.33&&6.48	&6.07&6.25&&&\\
\hline
		 \mc{15}{c}{Thiophene}\\
		& \mc{2}{c}{\Pop} & \mc{2}{c}{\AVDZ}& \mc{3}{c}{\AVTZ}	&  \mc{7}{c}{Litt.}\\
State 	& {\CCT} & {\CCSDT} & {\CCT} & {\CCSDT} & {\CCT} & {\CCSDT}   & {\NEV}  	&Th.$^l$	&Th.$^m$	&Th.$^n$	&Th.$^o$	&Exp.$^p$ &Exp.$^q$&Exp.$^r$\\
\hline
$^1A_1 (\pi \ra \pis)$		&5.79&5.77	&5.70&5.68	&5.65&5.64&5.84	&5.33&5.41&5.70&5.64&5.16&5.13&5.16\\
$^1B_2 (\pi \ra \pis)$		&6.23&6.24	&6.05&6.06	&5.96&5.98&6.10	&5.72&5.72&6.10&5.97&5.99&5.83&\\
$^1A_2 (\pi \ra 3s)$		&6.26&6.26	&6.07&6.06	&6.14&6.14&6.20	&5.93&5.70&6.05&6.23&       &       &\\
$^1B_1 (\pi \ra 3p)$		&6.18&6.17	&6.19&6.17	&6.14&6.14&6.19	&6.30&5.87&6.30&6.17&       &        &6.71\\
$^1A_2 (\pi \ra 3p)$		&6.32&		&6.33&6.31	&6.25&6.21&6.40	&6.35&6.03&6.28&6.33&       &\\
$^1B_1 (\pi \ra 3s)$		&6.62&6.62	&6.42&6.41	&6.50&6.49&6.71	&6.23&6.12&6.36&6.68&       &        &6.47\\
$^1B_2 (\pi \ra 3p)$$^s$	&7.45&7.44	&7.45&7.44	&7.29&7.29&7.25	&6.56&6.41&6.81&6.97&       &        &6.60\\
$^1A_1 (\pi \ra \pis)$		&7.50&7.46	&7.41&      	&7.35&	 &7.39	&6.69&7.32&7.71&7.74&6.61&\\
$^3B_2 (\pi \ra \pis)$		&3.95&3.94	&3.96&3.94	&3.94&	 &4.13	&3.75&3.94&	   &3.96&&3.74&\\
$^3A_1 (\pi \ra \pis)$		&4.90&4.90	&4.82&4.81	&4.77&	 &4.84	&4.50&4.86&       &4.87&&4.62&\\
$^3B_1 (\pi \ra 3p)$		&6.00&5.98	&6.01&5.99	&5.95&	 &5.98	&5.90&5.94&	   &6.01&\\
$^3A_2 (\pi \ra 3s)$		&6.20&6.20	&6.01&6.00	&6.09&	 &6.14	&5.88&5.75&       &5.83&\\
\hline
 \end{tabular}
  \end{footnotesize}
\vspace{-0.5 cm}
\begin{flushleft}
\begin{footnotesize}
$^a${{\CASPT} results from Ref.~\citenum{Ser93b};}
$^b${SAC-CI results from Ref.~\citenum{Wan00b};}
$^c${MR-MP results from Ref.~\citenum{Nak96};}
$^d${{\CCT} results from Ref.~\citenum{Sch17};}
$^e${Electron impact from Ref.~\citenum{Fru79};}
$^f${Gas phase absorption from Ref.~\citenum{McD91b};}
$^g${Energy loss from Ref.~\citenum{McD85} for the two valence states; two-photon resonant experiment from Ref.~\citenum{Sab92} for the $^1A_2$ Rydberg ES;}
$^h${{\CASPT} results from Ref.~\citenum{Ser96b};}
$^i${{\CCT} results from Ref.~\citenum{Sil10c};}
$^j${Gas-phase experimental estimates from Ref.~\citenum{Dev06};}
$^k${The assignments of these two states as valence and Rydberg is based on the oscillator strength, but both have a partial Rydberg character. The CASSCF spatial extend is in fact larger
for the lowest transition and Roos consequently classified both ES as Rydberg transitions; \cite{Ser96b}}
$^l${{\CASPT} results from Ref.~\citenum{Ser93c};}
$^m${SAC-CI results from Ref.~\citenum{Wan01};}
$^n${CCSDR(3) results from Ref.~\citenum{Pas07};} 
$^o${TBE from Ref.~\citenum{Hol14}, based on EOM-CCSD for singlet and ADC(2) for triplets;}
$^p${0-0 energies from Ref.~\citenum{Dil72};}
$^q${0-0 energies from Ref.~\citenum{Var82} for the singlets and energy loss experiment  from Ref.~\citenum{Hab03} for the triplets;}
$^r${0-0 energies from Ref.~\citenum{Hol14};}
$^s${Non-negligible mixing with a close-lying $(\pi \ra \pis)$ transition of the same symmetry.}
\end{footnotesize}
\end{flushleft}
\end{table}

Although a diverse array of wavefunction studies has been performed on cyclopentadiene  (including {\CASPT}, \cite{Ser93b,Sch08,Sil10c} CC, \cite{Sch08,Sil10b,Sch17} SAC-CI \cite{Wan00b} and various multi-reference
approaches \cite{Nak96,She09b}), this compound has received less attention than other members of the five-membered ring family, namely furan and pyrrole (\emph{vide infra}). This is probably due to the presence of the
methylene group that renders computations significantly more expensive. Most transitions listed in Tables \ref{Table-6} and S6 are characterized by $\Td$ exceeding  $93\%$, the only exception being the $^1A_1 (\pi \ra \pis)$
excitation that has a similar nature as the lowest $A_g$ state of butadiene ($\Td = 79\%$). Consistently, the {\CCT} and {\CCSDT} results are nearly identical for all ES except for the $^1A_1$ ES. By comparing the results
obtained for this $A_1 (\pi \ra \pis)$ transition to its butadiene counterpart, one can infer that the {\CCSDT} estimate is probably too large by roughly $0.04$--$0.08$ eV, and that the {\NEV} value is unlikely to be accurate enough to
establish a definitive TBE. This statement is also in line with the results of Ref.~\citenum{Loo19c}. For the two $B_2 (\pi \ra \pis)$ transitions, we could obtain {\FCI}/{\Pop} estimates of $5.78 \pm 0.02$ eV (singlet) and $3.33 \pm 0.05$ eV (triplet), the
{\CCT} and {\CCSDT} transition energies falling inside these energetic windows in both cases.  As one can see in Tables \ref{Table-6} and S6, the basis set effects are rather moderate for the electronic transitions of cyclopentadiene,
with no variation larger than $0.10$ eV ($0.02$ eV) between {\AVDZ} and {\AVTZ}  ({\AVTZ} and {\AVQZ}).  When comparing to literature data, our values are unsurprisingly consistent with the {\CCT} values of
Schwabe and Goerigk, \cite{Sch17} and tend to be significantly larger than earlier {\CASPT} \cite{Ser93b,Sil10c} and MR-MP \cite{Nak96} estimates. As expected, a few gas-phase experiments are available as well for this
derivative, \cite{Fru79,McD85,McD91b,Sab92} but they hardly represent grounds for comparison.

Due to its lower symmetry, imidazole has been less investigated, the most advanced studies available probably remain the {\CASPT} work of Serrano-Andr\`es and coworkers from 1996, \cite{Ser96b} and the
basis-set extrapolated {\CCT} results of Silva-Junior \emph{et al.} for the valence transitions from 2010. \cite{Sil10c}  The experimental data in gas-phase are also limited.\cite{Dev06} Our results are displayed in Tables \ref{Table-6} and S6.
The {\CCT} and {\CCSDT} values are quite consistent despite the fact that the $\Td$ values of the two singlet $A'$ states are slightly smaller than $90\%$. These two states have indeed, at least partially, a Rydberg character
(see the footnote in  Table \ref{Table-6}). The agreement between the CC estimates and previous {\CASPT}, \cite{Ser96b} and current {\NEV} energies is reasonable, the latter being systematically larger than their {\CCT}
counterparts.  For the eight transitions considered here, the basis set effects are moderate and {\AVTZ} yield results within $0.03$ eV of their {\AVQZ} counterparts (Table S6 in the SI).

Finally, the ES of thiophene, which is one of the most important building block in organic electronic devices, were the subject of previous theoretical investigations, \cite{Ser93c,Pal99,Wan01,Kle02,Pas07,Hol14} that unveiled a series of
transitions that were not yet characterized in the available measurements. \cite{Dil72,Fli76,Fli76b,Var82,Hab03,Pal99,Hol14} To the best of our knowledge, the present work is the first to report CC calculations obtained with (iterative)
triples and therefore constitutes the most accurate estimates to date. Indeed, all the transitions listed in Tables \ref{Table-6} and S6 are characterized by a largely dominant single excitation character, with $\Td$ above
$90\%$ except for the two $^1A_1$ transitions for which $\Td = 88\%$ and $87\%$, respectively.  The agreement between {\CCT} and {\CCSDT} remains nevertheless excellent for these low-lying totally symmetric transitions.  Thiophene is
also one of these compounds for which the unambiguous characterization of the nature of the ES is difficult, with, \eg, a strong mixing between the second and third singlet ES of $B_2$ symmetry. This makes the assignment of the valence
($\pi \ra \pis$) or Rydberg ($\pi \ra 3p$) character of this transition particularly tricky at the {\CCT} level. We note that contradictory assignments can be found in  the literature. \cite{Ser93c,Wan01,Pas07} As for the
previously discussed isostructural systems, we note that the only ES that undergoes significant basis set effects beyond {\AVTZ} is the Rydberg $^1B_2 (\pi \ra 3p)$ ($-0.09$ eV when upgrading to {\AVQZ}, see Table S6) and that
the {\NEV} estimates tend to be slightly larger than the {\CCT} values. The data of Table \ref{Table-6} are globally in good agreement with the previously reported values with discrepancies that are significant only for the three highest-lying singlet states.

\subsection{Six-membered rings}

Let us now turn towards seven six-membered rings which play a key role in chemistry: benzene, pyrazine, pyridazine, pyridine, pyrimidine, tetrazine, and triazine. To the best of our knowledge, the present work
is the first to propose {\CCSDT} reference energies as well as {\CCT}/{\AVQZ} values for all these compounds.  Of course, these systems have been investigated before, and beyond Thiel's
benchmarks, \cite{Sch08,Sil10b,Sil10c} it is worth pointing out the early investigation of Del Bene and coworkers \cite{Del97b} performed with a CC approach including perturbative corrections for the triples.
Following a theoretically consistent protocol, Nooijen \cite{Noo99} also performed {\STEOM} calculations to study the ES of each of these derivatives. However, these two works only considered singlet ES.

\subsubsection{Benzene, pyrazine, and tetrazine}

These three highly-symmetric systems allow to directly perform {\CCSDT}/{\AVTZ} calculations for singlet states  without the need of basis set extrapolations. Benzene was studied many times
before, \cite{Sob93,Lor95b,Chr96c,Pac96,Del97b,Noo99,Hal02,Li07b,Sch08,Dev08,Sil10b,Sil10c,Sau11,Li11,Lea12,Kan14,Sch17,Dut18,Sha19,Loo19c} and we report in Tables \ref{Table-7} and S7 estimates obtained for
five singlet and three triplet ES, all characterized by $\Td$ exceeding $90\%$ except for the lowest singlet ($86\%$).   As one can see, the two CC approaches are again yielding very consistent transition energies
with variations in the 0.00--0.03 eV range. Besides, {\AVTZ} is essentially providing basis set converged transition energies (Table S7). The present CC estimates are also very consistent with earlier {\CCT} results \cite{Chr96c} and
are compatible with both the very recent RASPT2 \cite{Sha19} and our {\NEV} values. For states of both spin symmetries, the {\CCT} and {\CCSDT} transitions energies are slightly larger than the available electron impact/multi-photon
measurements, \cite{Doe69,Nak80,Joh76,Joh83,Hir91} but do provide energetic gaps between ES very similar to the measured ones.

\begin{table}[htp]
\caption{\small Vertical transition energies (in eV) of benzene.}
\label{Table-7}
\begin{footnotesize}
\begin{tabular}{l|p{.5cm}p{1.0cm}|p{.5cm}p{1.0cm}|p{.5cm}p{1.0cm}p{1.2cm}|p{.5cm}p{.5cm}p{.5cm}p{.5cm}p{.6cm}p{.6cm}}
\hline
		& \mc{2}{c}{\Pop} & \mc{2}{c}{\AVDZ}& \mc{3}{c}{\AVTZ}	&  \mc{6}{c}{Litt.}\\
State 	& {\CCT} & {\CCSDT} & {\CCT} & {\CCSDT} & {\CCT} & {\CCSDT}   & {\NEV}  	&Th.$^a$	&Th.$^b$	&Th.$^c$	&Th.$^d$	&Exp.$^e$ &Exp.$^f$\\
\hline
$^1B_{2u} (\pi \ra \pis)$		&5.13&5.10	&5.11&5.08	&5.09&5.06&5.32	&4.84&5.08&5.06&5.03&	&4.90\\
$^1B_{1u} (\pi \ra \pis)$		&6.68&6.69	&6.50&6.50	&6.44&6.45&6.43	&6.30&6.54&6.22&6.23&	&6.20\\
$^1E_{1g} (\pi \ra 3s)$		&6.75&6.76	&6.46&6.46	&6.52&6.52&6.75	&6.38&6.51&6.42&       &	&6.33\\
$^1A_{2u}  (\pi \ra 3p)$		&7.24&7.25	&7.02&7.02	&7.08&7.08&7.40	&6.86&6.97&7.06&       &	&6.93\\
$^1E_{2u}  (\pi \ra 3p)$		&7.34&7.35	&7.09&7.09	&7.15&7.15&7.45 	&6.91&7.03&7.12&       &	&6.95\\
$^3B_{1u} (\pi \ra \pis)$		&4.18&4.16	&4.19&4.17	&4.18&	  &4.32	&3.89&4.15&3.88&4.11&3.95&\\
$^3E_{1u}(\pi \ra \pis)$		&4.95&4.94	&4.89&4.88	&4.86&	  &4.92	&4.49&4.86&4.72&4.75&4.75&\\
$^3B_{2u} (\pi \ra \pis)$		&6.06&6.06	&5.86&5.86	&5.81&	  &5.51	&5.49&5.88&5.54&5.67&5.60&\\
\hline
 \end{tabular}
  \end{footnotesize}
\vspace{-0.5 cm}
\begin{flushleft}
\begin{footnotesize}
$^a${{\CASPT} results from Ref.~\citenum{Lor95b};}
$^b${{\CCT} results from Ref.~\citenum{Chr96c};}
$^c${SAC-CI results from Ref.~\citenum{Li07b};}
$^d${RASPT2(18,18) results from Ref.~\citenum{Sha19};}
$^e${Electron impact from Ref.~\citenum{Doe69};}
$^f${Jet-cooled experiment from Ref.~\citenum{Hir91} for the two lowest states, multi-photon experiments from Refs.~\citenum{Joh76} and \citenum{Joh83} for the Rydberg states.}
\end{footnotesize}
\end{flushleft}
\end{table}

There are many available studies of the ES of pyrazine, \cite{Ful92,Del97b,Web99,Noo99,Li07b,Sch08,Sau09,Sil10c,Woy10,Car10,Sau11,Lea12,Kan14,Sch17,Dut18} and tetrazine,
\cite{Sta96,Del97b,Rub99,Noo99,Ada00,Noo00,Ang09,Sch08,Sau09,Sil10b,Sil10c,Car10,Lea12,Kan14,Sch17,Dut18,Pas18b} for which the $D_{2h}$ symmetry helps distinguishing the different ES.
Our results are collected in Tables \ref{Table-8} and S8. In pyrazine, all transitions are characterized by $\Td > 85\%$ at the exception of the $^1B_{1g}  (n \ra \pis)$ transition (84\%).
The excitation energies are basically unchanged going from {\CCT} to {\CCSDT} except possibly for the highest-lying singlet state considered here.  Going from triple- to quadruple-$\zeta$ basis, the variations do
not exceed $0.04$ eV, even for the four Rydberg ES treated here. This indicates that one can be highly confident in the present estimates except for the highest-lying singlet ES.
Again, the previous {\CASPT} estimates \cite{Ful92,Web99,Sch08,Sau11} appear to be globally too low, while the (unconventional) CASPT3 results \cite{Woy10} seem too high.
A similar overestimation can be noticed in previous SAC-CI results \cite{Li07b} and our {\NEV} values, the latter showing  a mean absolute deviation of 0.11 eV compared to {\CCT}.
In fact, the most satisfying agreement between the current estimates and previous works is reached with Nooijen's {\STEOM} values (except for the highest ES), \cite{Noo99} and
the recent Schwabe-Goerigk's {\CCT} estimates. \cite{Sch17}  The available experimental data \cite{Bol84,Oku90,Wal91,Ste11c} do not include all theoretically-predicted transitions, 
but provide a similar energetic ranking for both singlets and triplets.

\begin{table}[htp]
\caption{\small Vertical transition energies (in eV) of pyrazine and tetrazine.}
\label{Table-8}
\begin{footnotesize}
\vspace{-0.3 cm}
\begin{tabular}{p{2.18cm}|p{.5cm}p{1.cm}|p{.5cm}p{1.cm}|p{.5cm}p{1.0cm}p{1.2cm}|p{.5cm}p{.5cm}p{.5cm}p{.5cm}p{.5cm}p{.6cm}}
\hline
		 \mc{14}{c}{Pyrazine}\\
		& \mc{2}{c}{\Pop} & \mc{2}{c}{\AVDZ}& \mc{3}{c}{\AVTZ}	&  \mc{6}{c}{Litt.}\\
State 	& {\CCT} & {\CCSDT} & {\CCT} & {\CCSDT} & {\CCT} & {\CCSDT}   & {\NEV}  	&Th.$^a$	&Th.$^b$	&Th.$^c$&Th.$^d$	&Exp.$^e$ &Exp.$^f$\\
\hline
$^1B_{3u}  (n \ra \pis)$			&4.28&4.28	&4.19&4.19	&4.14&4.15&4.17	&3.83&4.12&4.25&4.19&3.93\\
$^1A_{u}  (n \ra \pis)$			&5.08&5.08	&4.98&4.98	&4.97&4.98&4.77	&4.36&4.93&5.24&4.93&\\
$^1B_{2u}  (\pi \ra \pis)$			&5.10&5.08	&5.07&5.05	&5.03&5.02&5.32	&4.79&4.75&4.84&5.19&4.8&4.81\\
$^1B_{2g}  (n \ra \pis)$			&5.86&5.85	&5.78&5.77	&5.71&5.71&5.88	&5.50&5.85&6.04&5.81&5.19\\
$^1A_{g}  (n \ra 3s)$				&6.74&6.73	&6.54&6.53	&6.66&6.65&6.70	&       &6.83&7.07&6.46&\\
$^1B_{1g}  (n \ra \pis)$			&6.87&6.87	&6.75&6.75	&6.73&6.74&6.75	&6.26&6.73&       &6.73&&6.10\\
$^1B_{1u}  (\pi \ra \pis)$			&7.10&7.11	&6.92&6.93	&6.86&6.88&6.81	&6.60&6.89&6.68&6.99&6.5&6.51\\
$^1B_{1g}  (\pi \ra 3s)$			&7.36&7.37	&7.13&7.14	&7.20&7.21&7.33	&       &7.31&7.08&       &\\
$^1B_{2u}  (n \ra 3p)$			&7.39&7.39	&7.14&7.13	&7.25&	  &7.25	&       &7.45&7.67&7.06&\\
$^1B_{1u}  (n \ra 3p)$			&7.56&7.55	&7.38&7.37	&7.45&	  &7.42	&7.28&7.50&7.73&7.31&&7.67\\
$^1B_{1u}  (\pi \ra \pis)$			&8.19&8.23	&7.99&8.03	&7.94&	  &8.25	&7.43&7.96&8.24&8.08&\\
$^3B_{3u}  (n \ra \pis)$			&3.68&3.68	&3.60&3.60	&3.59&	  &3.56	&3.16&       &       &        &3.33\\
$^3B_{1u}  (\pi \ra \pis)$			&4.39&4.36	&4.40&4.36	&4.39&	  &4.57	&4.15&       &       &        &4.04\\
$^3B_{2u}  (\pi \ra \pis)$			&4.56&4.55	&4.46&4.45	&4.40&	  &4.42	&4.28&       &       &        &$\sim$4.4\\
$^3A_{u}  (n \ra \pis)$			&5.05&5.05	&4.93&4.93	&4.93&	  &4.75	&4.19&       &       &        &4.2\\
$^3B_{2g}  (n \ra \pis)$			&5.18&5.17	&5.11&5.11	&5.08&	  &5.21	&4.81&       &       &        &4.49\\
$^3B_{1u}  (\pi \ra \pis)$			&5.38&5.37	&5.32&5.31	&5.29&	  &5.35	&4.98&       &       &        &\\
\hline
		 \mc{14}{c}{Tetrazine}\\
		& \mc{2}{c}{\Pop} & \mc{2}{c}{\AVDZ}& \mc{3}{c}{\AVTZ}	&  \mc{6}{c}{Litt.}\\
State 	& {\CCT} & {\CCSDT} & {\CCT} & {\CCSDT} & {\CCT} & {\CCSDT}   & {\NEV}  	&Th.$^g$	&Th.$^h$	&Th.$^i$&Th.$^j$	&Th.$^k$ &Exp.$^l$\\
\hline
$^1B_{3u}  (n \ra \pis)$			&2.53&2.54	&2.49&2.50	&2.46&2.47&2.35	&1.96&2.22&2.01&2.29&2.46&2.35\\
$^1A_{u}  (n \ra \pis)$			&3.75&3.75	&3.69&3.70	&3.67&3.69&3.58	&3.06&3.62&3.09&3.41&3.78&3.6\\
$^1A_{g}  (\mathrm{double})$$^m$	&6.22&5.86	&6.22&5.86	&6.21&5.96&4.61	&4.37&5.06&4.34&4.66&       &\\
$^1B_{1g}  (n \ra \pis)$			&5.01&5.02	&4.97&4.98	&4.91&4.93&4.95	&4.51&4.73&4.47&4.53&4.87&\\
$^1B_{2u}  (\pi \ra \pis)$			&5.29&5.26	&5.27&5.25	&5.23&5.21&5.56	&4.89&4.90&       &5.59&5.08&4.97\\
$^1B_{2g}  (n \ra \pis)$			&5.56&5.52	&5.53&5.50	&5.46&5.45&5.63	&5.05&5.09&4.92&5.59&5.28&\\
$^1A_{u}  (n \ra \pis)$			&5.61&5.61	&5.59&5.59	&5.52&5.53&5.62	&5.28&5.23&5.32&5.95&5.39&5.5\\
$^1B_{3g} (\mathrm{double})$$^m$	&7.64&       	&7.62&   	  	&7.62&	  &6.15	&5.16&6.30&5.26&6.01&       &5.92\\
$^1B_{2g}  (n \ra \pis)$			&6.24&6.22	&6.17&6.16	&6.13&	  &6.13	&5.48&6.16&5.78&6.05&6.16&\\
$^1B_{1g}  (n \ra \pis)$			&7.04&7.04	&6.98&6.98	&6.92&	  &6.76	&5.99&6.73&6.20&6.92&6.80&\\
$^3B_{3u}  (n \ra \pis)$			&1.87&1.88	&1.86&1.86	&1.85&	  &1.73	&1.45&1.71&       &       &1.87&1.7\\
$^3A_{u}  (n \ra \pis)$			&3.48&3.49	&3.43&3.44	&3.44&	  &3.36	&2.81&3.47&       &       &3.49&2.90\\
$^3B_{1g}  (n \ra \pis)$			&4.25&4.25	&4.23&4.23	&4.20&	  &4.24	&3.76&3.97&       &       &4.18&\\
$^3B_{1u}  (\pi \ra \pis)$			&4.54&4.49	&4.54&4.49	&4.54&	  &4.70	&4.25&3.67&       &       &4.36&\\
$^3B_{2u}  (\pi \ra \pis)$			&4.65&4.64	&4.58&4.58	&4.52&	  &4.58	&4.29&4.35&       &       &4.39&\\
$^3B_{2g}  (n \ra \pis)$			&5.11&5.11	&5.09&5.08	&5.05&	  &5.27	&4.67&4.78&       &       &4.89&\\
$^3A_{u}  (n \ra \pis)$			&5.17&5.17	&5.15&5.15	&5.11&	  &5.13	&4.85&4.89&       &       &4.96&\\
$^3B_{3g} (\mathrm{double})$$^m$	&7.35&       	&7.33&    	 	&7.35&	  &5.51	&5.08&        &       &       &	   &\\
$^3B_{1u}  (\pi \ra \pis)$			&5.51&5.50	&5.46&5.46	&5.42&	  &5.56	&5.09&5.31&       &       &5.32&\\

\hline
 \end{tabular}
  \end{footnotesize}
\vspace{-0.5 cm}
\begin{flushleft}
\begin{footnotesize}
$^a${{\CASPT} results from Ref.~\citenum{Web99};}
$^b${{\STEOM} results from Ref.~\citenum{Noo99};}
$^c${SAC-CI results from Ref.~\citenum{Li07b};}
$^d${{\CCT} results from Ref.~\citenum{Sch17};}
$^e${Double resonance dip spectroscopy from Ref.~\citenum{Oku90} ($B_{3u}$ and $B_{2g}$ ES) and EEL from Ref.~\citenum{Wal91} (others);}
$^f${UV max from Ref.~\citenum{Bol84};}
$^g${{\CASPT} results from Ref.~\citenum{Rub99};}
$^h${Ext-{\STEOM} results from Ref.~\citenum{Noo00};}
$^i${GVVPT2 results from Ref.~\citenum{Dev08};}
$^j${{\NEV} results  from Ref.~\citenum{Ang09};}
$^k${{\CCT} results from Ref.~\citenum{Sil10c};}
$^l${From Ref.~\citenum{Pal97}, the singlets are from EEL, except for the $4.97$ and $5.92$ eV values that are from VUV; the triplets are from EEL, additional (unassigned) triplet peaks are found at $4.21$, $4.6$, and $5.2$ eV;}
$^m${all these three doubly ES have a $(n,n \ra \pis, \pis)$ character.}
\end{footnotesize}
\end{flushleft}
\end{table}

For tetrazine, we consider valence ES only, including three transitions exhibiting a true double excitation nature ($\Td < 10\%$). Of course, for these double excitations, {\CCT} and {\CCSDT} cannot be considered as reliable. This is illustrated by 
the large change in excitation energies between these two CC models. The theoretical best estimates are likely obtained with {\NEV}. \cite{Loo19c} For all the other transitions, the $\Td$ values are in the $80$--$90\%$ range for singlets and larger than $95\%$ 
for triplets. Consequently, the {\CCT} and {\CCSDT} results are very consistent, the sole exception being the lowest $^3B_{1u}  (\pi \ra \pis)$ transition for which we note a shift of $-0.05$ eV when upgrading the level of theory to {\CCSDT}.
In all other cases, there is a global consistency between our CC values. Moreover, the basis set effects are very small beyond {\AVTZ} with a maximal variation of $0.02$ eV going to {\AVQZ}  (Table S8). The present values are
almost systematically larger than previous {\CASPT}, \cite{Rub99} {\STEOM}, \cite{Noo00} and GVVPT2 \cite{Dev08} estimates. Our {\NEV} values are also globally consistent with the {\CCT} values with a
maximal discrepancy of 0.22 eV for the ES with a dominant single excitation character. One finds a global agreement with Thiel's {\CCT}/{\AVTZ} values, \cite{Sil10c} although we note variations of approximately $0.20$ eV for 
specific excitations like the $B_{2g}$ transitions. This feature might be due to the use of distinct geometries in the two studies. The experimental EEL values from Palmer's work \cite{Pal97} 
show a reasonable agreement with our estimates.

\subsubsection{Pyridazine, pyridine, pyrimidine, and triazine}

\begin{table}[htp]
\caption{\small Vertical transition energies (in eV) of pyridazine and pyridine.}
\label{Table-9}
\begin{footnotesize}
\begin{tabular}{l|p{.5cm}p{1.0cm}|p{.5cm}p{1.0cm}|p{.5cm}p{1.2cm}|p{.5cm}p{.5cm}p{.5cm}p{.5cm}p{.6cm}p{.6cm}}
\hline
		 \mc{13}{c}{Pyridazine}\\
		& \mc{2}{c}{\Pop} & \mc{2}{c}{\AVDZ}& \mc{2}{c}{\AVTZ}	&  \mc{6}{c}{Litt.}\\
State 	& {\CCT} & {\CCSDT} & {\CCT} & {\CCSDT} & {\CCT} & {\NEV}  	&Th.$^a$	&Th.$^b$	&Th.$^c$&Th.$^d$	&Exp.$^e$ &Exp.$^f$\\
\hline
$^1B_1 (n \ra \pis)$			&3.95&3.95	&3.86&3.86	&3.83&3.80	&3.48&3.76&3.65&3.85&&3.36\\
$^1A_2 (n \ra \pis)$			&4.49&4.48	&4.39&4.39	&4.37&4.40	&3.66&4.46&4.28&4.44&&4.02\\
$^1A_1 (\pi \ra \pis)$			&5.36&5.32	&5.33&5.30	&5.29&5.58	&4.86&4.92&4.86&5.20&5.0&5.01\\
$^1A_2 (n \ra \pis)$			&5.88&5.86	&5.80&5.78	&5.74&5.88	&5.09&5.66&5.52&5.66&&5.61\\
$^1B_2  (n \ra 3s)$			&6.26&6.27	&6.06&6.06	&6.17&6.21	&      	 &6.45&       &       &&      \\
$^1B_1 (n \ra \pis)$			&6.51&6.51	&6.41&6.41	&6.37&6.64	&5.80&6.41&6.20&6.33&&6.00\\
$^1B_2 (\pi \ra \pis)$			&6.96&6.97	&6.79&6.80     	&6.74&7.10	&6.61&6.77&6.44&6.68&&6.50\\
$^3B_1 (n \ra \pis)$			&3.27&3.26	&3.20&3.20	&3.19&3.13	&       &       &       &       &&3.06\\
$^3A_2 (n \ra \pis)$			&4.19&4.19	&4.11&4.11	&4.11&4.14	&       &       &       &       &&3.55\\
$^3B_2 (\pi \ra \pis)$			&4.39&4.36	&4.39&4.35	&4.38&4.49	&       &       &       &       &4.0&4.33\\
$^3A_1 (\pi \ra \pis)$			&4.93&4.94	&4.87&4.86	&4.83&4.94	&       &       &       &       &4.4&4.68\\
\hline
		 \mc{13}{c}{Pyridine}\\
		& \mc{2}{c}{\Pop} & \mc{2}{c}{\AVDZ}& \mc{2}{c}{\AVTZ}	&  \mc{6}{c}{Litt.}\\
State 	& {\CCT} & {\CCSDT} & {\CCT} & {\CCSDT} & {\CCT}   & {\NEV}  	&Th.$^g$	&Th.$^b$	&Th.$^c$&Th.$^d$	&Exp.$^h$ &Exp.$^i$ \\
\hline
$^1B_1 (n \ra \pis)$			&5.12&5.10	&5.01&5.00	&4.96&5.15	&4.91&4.90&4.80&4.95&5.24&4.78\\
$^1B_2 (\pi \ra \pis)$			&5.23&5.20	&5.21&5.18	&5.17&5.31	&4.84&4.82&4.81&5.12&4.99&4.99\\
$^1A_2 (n \ra \pis)$			&5.55&5.54	&5.41&5.41	&5.40&5.29	&5.17&5.31&5.24&5.41&5.43&5.40\\
$^1A_1 (\pi \ra \pis)$			&6.84&6.84      	&6.64&6.63	&6.63&6.69	&6.42&6.62&6.36&6.60&6.38&      \\
$^1A_1 (n \ra 3s)$			&6.92&6.92	&6.71&6.71	&6.76&6.99	&6.70&6.96&6.64&       &6.28&6.25\\
$^1A_2 (\pi \ra 3s)$			&6.98&6.99	&6.74&6.75	&6.81&6.86	&6.75&6.90&6.53&       &       &      \\
$^1B_2 (\pi \ra \pis)$$^j$		&7.50&7.52	&7.40&7.42 	&7.38&7.83	&7.48&7.29&7.14&7.33&7.22&7.20\\
$^1B_1 (\pi \ra 3p)$			&7.54&7.55	&7.32&7.32	&7.38&7.45	&7.25&7.42&7.10&       &       &      \\
$^1A_1 (\pi \ra \pis)$			&7.56&      	&7.34&7.34	 &7.39&6.97	&7.23&7.37&7.26&7.39&7.22&6.39\\
$^3A_1 (\pi \ra \pis)$			&4.33&4.31	&4.34&4.31	&4.33&4.60	&4.05&       &       &4.28&       &3.86\\
$^3B_1 (n \ra \pis)$			&4.57&4.56	&4.47&4.47	&4.46&4.58	&4.41&       &       &4.42&       &4.12\\
$^3B_2 (\pi \ra \pis)$			&4.92&4.91	&4.83&4.83	&4.79&4.88	&4.56&       &       &4.72&       &4.47\\
$^3A_1 (\pi \ra \pis)$			&5.14&5.13	&5.08&		&5.05&5.19	&4.73&       &       &4.96&       &      \\
$^3A_2 (n \ra \pis)$			&5.51&5.49	&5.37&5.36	&5.35&5.33	&5.10&       &       &5.53&       &5.40\\
$^3B_2 (\pi \ra \pis)$			&6.46&6.45	&6.30&6.29      	&6.25&6.29	&6.02&       &       &6.22&       &6.09\\
\hline
 \end{tabular}
  \end{footnotesize}
\vspace{-0.3 cm}
\begin{flushleft}
\begin{footnotesize}
$^a${{\CASPT} results from Ref.~\citenum{Ful92};}
$^b${{\STEOM} results from Ref.~\citenum{Noo99};}
$^c${EOM-CCSD({$\tilde{{T}}$}) from Ref.~\citenum{Del97b};}
$^d${CC3-ext.~from Ref.~\citenum{Sil10c};}
$^e${EEL from Ref.~\citenum{Pal91};}
$^f${EEL from Ref.~\citenum{Lin19};}
$^g${{\CASPT}  from Ref.~\citenum{Lor95};}
$^h${EEL from Ref.~\citenum{Wal90};}
$^i${EEL from Ref.~\citenum{Lin16};}
$^j${Significant state mixing with a close-lying Rydberg transition rendering unambiguous attribution difficult. At the {\CCT}/{\AVDZ} level, the Rydberg state is at $7.26$ eV and
has a small $f$, so attribution is rather clear. However, at the {\CCT}/{\AVTZ} level, the two $B_2$ transitions are at $7.35$ and $7.38$ eV (hence strongly mixed), so that the attribution has been 
made using the $f$ of $0.174$ and $0.319$, respectively.}
\end{footnotesize}
\end{flushleft}
\end{table}

Those four azabenzenes with $C_{2v}$ or $D_{3h}$ spatial symmetry are also popular molecules in terms of ES calculations.  \cite{Pal91,Ful92,Wal92,Lor95,Del97b,Noo97,Noo99,Fis00,Cai00b,Wan01b,Sch08,Sil10b,Sil10c,Car10,Sau11,Lea12,Kan14,Sch17,Dut18}
Our results for pyridazine and pyridine are gathered in Tables \ref{Table-9} and S9. For the former compound, the available wavefunction results \cite{Pal91,Ful92,Del97b,Noo99,Fis00,Sch08,Sil10b,Sil10c,Kan14,Sch17,Dut18}
focussed on singlet transitions, at the exception of rather old MRCI, \cite{Pal91} and {\CASPT} investigations. \cite{Fis00}  Again, the $\Td$ values are larger than $85\%$ ($95\%$) for the singlet (triplet) transitions,
and the only state for which there is a variation larger than $0.03$ eV between the {\CCT}/{\AVDZ} and {\CCSDT}/{\AVDZ} energies is the $^3B_2 (\pi \ra \pis)$ transition. As in the previous six-membered cycles, the basis set
effects are rather small and {\AVTZ} provides values close to the CBS limit for the considered transitions. For the singlet valence ES, we find again a rather good match with the results of previous {\STEOM} \cite{Noo99} and 
CC \cite{Del97b,Sil10c} calculations. Yet again, these values are significantly higher than the {\CASPT} estimates reported in Refs.~\citenum{Ful92} and \citenum{Sil10c}. For the triplets, the present data represent the most 
accurate results published to date. Our {\NEV} values are very close to their {\CCT} analogues for the lowest-lying singlet and triplet, but positively deviate for the higher-lying ES.  Interestingly, beyond the popular twenty-year old 
reference measurements, \cite{Inn88,Pal91} there is a very recent experimental EEL analysis for pyridazine, \cite{Lin19} that locates almost all ES.  The transition energies reported in this very recent work are systematically smaller than our 
CC estimates by approximately $-0.20$ eV. Nonetheless, this study provides exactly the same ES ranking as our theoretical protocol.

Pyridine, the hallmark heterocycle, has been more scrutinized than pyridazine and many wavefunction approaches have been applied to estimate its ES energies. \cite{Ful92,Lor95,Del97b,Noo97,Noo99,Cai00b,Wan01b,Sch08,Sil10b,Sil10c,Car10,Sau11,Kan14,Sch17,Dut18} 
Besides, two detailed EEL experiments are also available for pyridine. \cite{Wal90,Lin16} The general trends described above for other six-membered cycles do pertain with: i) large $\Td$ values and consistency between 
{\CCT} and {\CCSDT} estimates for all transitions listed in Table \ref{Table-9}; ii) small basis set effects beyond {\AVTZ} even for the Rydberg transitions; iii) qualitative agreement with past CC results; iv) {\NEV} transitions energies
that are, on average, larger than their CC counterparts; and v) same ES ranking as in the most recent measurements. \cite{Lin16} Beyond these aspects, it is worth mentioning that the second $^1B_2 (\pi \ra \pis)$ ES is strongly mixed with a nearby
Rydberg transition that is separated by only $0.03$ eV at the {\CCT}/{\AVTZ} level. This obviously makes the analysis particularly challenging for that specific transition.

The results obtained for both pyrimidine and triazine are listed in Tables \ref{Table-10} and S10. Because the former derivative can be viewed as the smallest model of DNA bases, previous theoretical  \cite{Ful92,Del97b,Ser97b,Noo99,Ohr01,Fis03b,Li07b,Sch08,Sil10b,Sil10c,Car10,Sau11,Kan14,Sch17,Dut18}
and experimental \cite{Bol84,Pal90,Lin15} studies are rather extensive. For triazine, which belongs to a non-abelian point group, theoretical studies are scarcer,
\cite{Ful92,Wal92,Pal95,Del97b,Noo99,Oli05,Sch08,Sil10b,Sil10c,Kan14,Sch17,Dut18} especially for the triplets,  \cite{Wal92,Pal95,Oli05}  whereas the experimental data are also limited. \cite{Bol84,Wal92} As in pyridazine and pyridine, all the ES listed in Table \ref{Table-10} 
show $\Td$ values larger than $85\%$ for singlets and $95\%$ for triplets, so that {\CCT} and {\CCSDT} are highly coherent, except maybe for the $^3A_1 (\pi \ra \pis)$ transitions in pyrimidine. The basis set effects are also small, with no variation larger than 
$0.10$ ($0.03$) eV between double- and triple-$\zeta$ (triple- and quadruple-$\zeta$) for valence transitions and only slightly larger variations for the two Rydberg transitions (+$0.04$ eV between {\AVTZ} and {\AVQZ}). For both compounds, the current 
values are almost systematically larger than previously published data, with our {\CCT} values being typically bracketed by the published {\CASPT} and our {\NEV} estimates. For the triplets of triazine, the three lowest ES previously estimated by {\CASPT} 
\cite{Oli05} are too low by  roughly half an eV.

\begin{table}[htp]
\caption{\small Vertical transition energies (in eV) of pyrimidine and triazine.}
\label{Table-10}
\begin{footnotesize}
\begin{tabular}{l|p{.5cm}p{1.0cm}|p{.5cm}p{1.0cm}|p{.5cm}p{1.0cm}p{1.2cm}|p{.5cm}p{.5cm}p{.5cm}p{.5cm}p{.6cm}p{.6cm}}
\hline
		 \mc{14}{c}{Pyrimidine}\\
		& \mc{2}{c}{\Pop} & \mc{2}{c}{\AVDZ}& \mc{3}{c}{\AVTZ}	&  \mc{6}{c}{Litt.}\\
State 	& {\CCT} & {\CCSDT} & {\CCT} & {\CCSDT} & {\CCT} & 	 & {\NEV}  	&Th.$^a$	&Th.$^b$	&Th.$^c$&Th.$^d$	&Exp.$^e$ &Exp.$^f$\\
\hline
$^1B_1 (n \ra \pis)$			&4.58&4.57	&4.48&4.48	&4.44&&4.55	&4.26&4.40&4.32&4.24&	4.2	&4.18	\\
$^1A_2 (n \ra \pis)$			&4.99&4.99	&4.89&4.88	&4.86&&4.84	&4.49&4.72&4.74&4.74&		&4.69	\\
$^1B_2 (\pi \ra \pis)$			&5.47&5.44	&5.44&5.41	&5.41&&5.53	&5.47&5.04&5.29&5.01&	5.12	&5.18	\\
$^1A_2 (n \ra \pis)$			&6.07&6.06	&5.98&5.97	&5.93&&6.02	&	 &5.94&5.98&5.84&	6.05	&5.67	\\
$^1B_1 (n \ra \pis)$			&6.39&		&6.29&6.29	&6.26&&6.40	&6.03&6.18&6.35&6.11&		&6.02	\\
$^1B_2  (n \ra 3s)$			&6.81&6.80	&6.61&6.59	&6.72&&6.77	&	 &6.85&6.84&6.57&		&    	  	\\
$^1A_1 (\pi \ra \pis)$			&7.08&7.09	&6.93&6.94	&6.87&&7.11	&7.10&6.87&6.86&6.57&	6.7	&6.69	\\
$^3B_1 (n \ra \pis)$			&4.20&4.20	&4.12&4.11	&4.10&&4.17	&3.81&	  &4.11&	   &		&3.85	\\
$^3A_1 (\pi \ra \pis)$			&4.55&4.52	&4.56&4.52	&4.55&&4.67	&4.35&	  &4.39&	   &		&4.42	\\
$^3A_2 (n \ra \pis)$			&4.77&4.76	&4.67&4.67	&4.66&&4.72	&4.24&	  &4.71&	   &		&4.18	\\
$^3B_2 (\pi \ra \pis)$			&5.08&5.08	&5.00&5.00	&4.96&&5.01	&4.83&	  &4.81&	   &		&4.93	\\
\hline
		 \mc{14}{c}{Triazine}\\
		& \mc{2}{c}{\Pop} & \mc{2}{c}{\AVDZ}& \mc{3}{c}{\AVTZ}	&  \mc{6}{c}{Litt.}\\
State 	& {\CCT} & {\CCSDT} & {\CCT} & {\CCSDT} & {\CCT} & {\CCSDT}   & {\NEV}  	&Th.$^g$	&Th.$^b$	&Th.$^d$&Th.$^h$	&Exp.$^e$ \\
\hline
$^1A_1'' (n \ra \pis)$			&4.85&4.84	&4.76&4.74	&4.73&4.72&4.61	&4.11&4.58&4.49&4.70	&\\
$^1A_2'' (n \ra \pis)$			&4.84&4.84	&4.78&4.78	&4.74&4.75&4.89	&4.30&4.74&4.54&4.71	&4.59\\
$^1E'' (n \ra \pis)$			&4.89&4.89	&4.82&4.81	&4.78&4.78&4.88	&4.32&4.69&4.56&4.75	&3.97\\
$^1A_2' (\pi \ra \pis)$		&5.84&5.80	&5.81&5.78	&5.78&5.75&5.95	&5.59&5.35&5.36&5.71	&5.70\\
$^1A_1' (\pi \ra \pis)$		&7.45&7.45	&7.31&7.31	&7.24&7.24&7.30	&	 &7.21&6.90&7.18	&6.86\\
$^1E' (n \ra 3s)$			&7.44&7.41	&7.24&7.21	&7.35&7.32&7.45	&	 &7.38&7.16& 		& \\
$^1E'' (n \ra \pis)$			&7.89&7.86	&7.82&7.80	&7.79&7.78&7.98	&	 &       &7.78&7.78	&\\
$^1E' (\pi \ra \pis)$			&8.12&8.13	&7.97&		&7.92&7.94&8.34 	&	 &7.82&7.72&7.84	&7.76\\
$^3A_2'' (n \ra \pis)$			&4.40&4.40	&4.35&4.35	&4.33&       &4.51	&3.87&\\
$^3E'' (n \ra \pis)$			&4.59&4.59	&4.52&4.52	&4.51&       &4.61	&4.04&\\
$^3A_1'' (n \ra \pis)$			&4.87&		&4.78&4.76	&4.75&       &4.71	&4.15&\\
$^3A_1' (\pi \ra \pis)$		&4.88&4.85	&4.88&4.85	&4.88&       &5.05	&	 &\\
$^3E' (\pi \ra \pis)$			&5.70&5.68	&5.64&		&5.61&       &5.73	&	 &\\
$^3A_2' (\pi \ra \pis)$		&6.85&6.84	&6.69&6.68	&6.63&       &6.36	&4.76&\\
\hline
 \end{tabular}
  \end{footnotesize}
\vspace{-0.4 cm}
\begin{flushleft}
\begin{footnotesize}
$^a${{\CASPT} results from Ref.~\citenum{Fis03b};}
$^b${{\STEOM} results from Ref.~\citenum{Noo99};}
$^c${SAC-CI results from Ref.~\citenum{Li07b};}
$^d${EOM-CCSD({$\tilde{{T}}$}) results from Ref.~\citenum{Del97b};}
$^e${UV max from Ref.~\citenum{Bol84};}
$^f${EEL from Ref.~\citenum{Lin15};}
$^g${{\CASPT} results from Ref.~\citenum{Oli05};}
$^h${CC3-ext.~results from Ref.~\citenum{Sil10c}.}
\end{footnotesize}
\end{flushleft}
\end{table}

\section{Theoretical Best Estimates}

Table \ref{Table-tbe} reports our two sets of TBE: a set obtained with the {\AVTZ} basis set and one set including an additional correction for the one-electron basis set incompleteness error. The details of our protocol employed to generate 
these TBE are also provided in Table \ref{Table-tbe}. For all states with a dominant single-excitation character (that is when $\Td > 80\%$), we rely on CC results using an incremental strategy to generate these TBE. As explained in the footnotes 
of Table \ref{Table-tbe}, this means that we add the basis set correction (i.e., the excitation energy difference between two calculations performed with a large and a small basis set) obtained with a ``lower''  level of theory, e.g., CC3, to correct the
result obtained at a ``higher''  level of theory, e.g., CCSDTQ, but with the smaller basis set. In our previous contribution, \cite{Loo18a} we have extensively tested this protocol for small compounds for which CCSDTQ/\emph{aug}-cc-pVTZ 
calculations were achievable. It turned out that correcting CCSDTQ/6-31+G(d) with CC3 or CCSDT basis set effect was very effective with a MAE of 0.01 eV as compared to the true value. There are only two exceptions for which we eschew to 
use this CC incremental strategy: two ES in acrolein for which nicely converged FCI values indicated non-negligible CCSDT errors. For ES with $\Td$ values between $70\%$ and $80\%$, our previous works indicated that {\CCSDT} tends to 
overshoot the transition energies by roughly $0.05$--$0.10$ eV, and that {\NEV} errors tend to be, on average, slightly larger. \cite{Loo19c}  Therefore, if {\CCSDTQ} or {\FCI} results are not available, it is extremely difficult to make the final call. 
For the other transitions, we relied either on the current or previous FCI data or the {\NEV} values as reference.  The italicized transition energies in Table \ref{Table-tbe} are believed to be (relatively) less accurate. This is the case when: i) 
the {\NEV} result has to be selected; ii) the CC calculations yield quite large changes in excitation energies while incrementing the excitation order by one unit despite large $\Td$; and iii) there is a very strong ES mixing making hard to follow a 
specific transition from one method (or one basis) to another.

To determine the basis set corrections beyond augmented triple-$\zeta$, we use the {\CCT}/{\AVQZ} or {\CCT}/{\AVPZ} results. For several compounds, we also provide in the SI, {\CCT}/d-{\AVQZ} transition energies (\ie, with an 
additional set of diffuse functions). However, we do not consider such values as reference because the addition of a second set of diffuse orbitals only significantly modifies the transition energies while also inducing a stronger ES mixing. 
We also stick to the frozen-core approximation for two reasons: i) the effect of correlating the core electrons is generally negligible (typically $\pm 0.02$ eV) for the compounds under study  (see the SI for examples); and ii) it would be, in principle, 
necessary to add core polarization functions in such a case.

Table \ref{Table-tbe} encompasses 238 ES, each of them obtained, at least, at the {\CCSDT} level. This set can be decomposed as follows: 144 singlet and 94 triplet transitions, or 174 valence (99 $\pi \ra \pis$, 71  $n \ra \pis$ and 4 double excitations) 
and 64 Rydberg transitions. Among these transition energies, fourteen can be considered as ``unsafe'' and are reported in italics accordingly.  This definitely corresponds to a very significant extension of our previous  ES data sets (see Introduction). 
Taken all together, they offer a consistent, diverse and accurate ensemble of transition energies for approximately 350 electronic transitions of various natures in small and medium-sized organic molecules. Table \ref{Table-tbe} also reports 90 oscillator 
strengths, $f$, which makes it, to the best of our knowledge, the largest set of {\CCT}/{\AVTZ} oscillator strengths reported to date, the previous effort being mostly performed at the {\CCT}/TZVP level for Thiel's set. \cite{Kan14} It should also be pointed 
out that all these data are obtained on {\CCT}/{\AVTZ} geometries, consistently with our previous works. \cite{Loo18a,Loo19c}

\renewcommand*{\arraystretch}{.55}
\LTcapwidth=\textwidth

\begin{footnotesize}
\begin{longtable}{p{3.3cm}lcccccc}
\caption{\small TBE values (in eV) for all considered states alongside their corresponding oscillator strength, $f$, and percentage of single excitations, $\Td$, obtained at the \CCT/{\AVTZ} level.
The composite protocol to generate these TBE is also reported (see footnotes).  In the right-most column, we list the TBE values obtained by including an additional correction (obtained at the {\CCT} level) for basis set incompleteness error.
Values displayed in italics are likely to be relatively less accurate. All values are obtained in the frozen-core approximation.} \label{Table-tbe}  \vspace{-0.3 cm}\\
\hline
			&		&	&		&  \mc{2}{l}{TBE/{\AVTZ}}	&  \mc{2}{l}{TBE/CBS} \\
			& State	 & $f$ &  $\Td$ & 	Value	& Protocol$^a$		& Value	& Corr. 	\\
\hline
\endfirsthead
\hline
			&		&	&		&  \mc{2}{l}{TBE/{\AVTZ}}	&  \mc{2}{l}{TBE/CBS} \\
			& State	 & $f$ &  $\Td$ & 	Value	& Protocol $^a$		& Value	& Corr. 	\\
\hline
\endhead
\hline \mc{7}{r}{{Continued on next page}} \\
\endfoot
\hline
\endlastfoot
Acetone			&$^1A_2 (\Val; n \ra \pis)$						&		& 91.1 & 4.47	& B					& 4.48	&  \AVQZ \\
				&$^1B_2 (\mathrm{R}; n \ra 3s)$				&0.000	& 90.5 & 6.46	& B					& 6.51	&  \AVQZ \\
				&$^1A_2 (\mathrm{R}; n \ra 3p)$				&		& 90.9 & 7.47	& B					& 7.44	&  \AVQZ \\
				&$^1A_1 (\mathrm{R}; n \ra 3p)$				&0.004	& 90.6 & 7.51	& B					& 7.55	&  \AVQZ \\
				&$^1B_2 (\mathrm{R}; n \ra 3p)$				&0.029	& 91.2 & 7.62	& B					& 7.63	&  \AVQZ \\
				&$^3A_2 (\Val; n \ra \pis)$						&		& 97.8 & 4.13	& D					& 4.15	&  \AVQZ \\
				&$^3A_1 (\Val; \pi \ra \pis)$					&		& 98.7 & 6.25	& D					& 6.27	&  \AVQZ \\
Acrolein			&$^1A'' (\Val; n \ra \pis)$						&0.000	& 87.6 & 3.78	& G					& 3.79	& \AVQZ \\
				&$^1A' (\Val; \pi \ra \pis)$						&0.344	& 91.2 & 6.69	& {\CCSDT}/\AVTZ		& 6.69	& \AVQZ \\
				&$^1A'' (\Val; n \ra \pis)$						&0.000	& 79.4 & \emph{6.72} &  D			&\emph{6.74} & \AVQZ \\
				&$^1A' (\mathrm{R}; n \ra 3s)$					&0.109	& 89.4 & 7.08	& D					& 7.12	& \AVQZ \\
				&$^3A'' (\Val; n \ra \pis)$						&		& 97.0 & 3.51	& H					& 3.50	& \AVQZ \\
				&$^3A' (\Val; \pi \ra \pis)$						&		& 98.6 & 3.94	& D					& 3.95	& \AVQZ \\
				&$^3A' (\Val; \pi \ra \pis)$						&		& 98.4 & 6.18	& D					& 6.19	& \AVQZ \\
				&$^3A'' (\Val; n \ra \pis)$						&		& 92.7 &  \emph{6.54} & E				& \emph{6.55}& \AVQZ \\
Benzene			&$^1B_{2u} (\Val; \pi \ra \pis)$					&		& 86.3 & 5.06	& {\CCSDT}/\AVTZ		& 5.06	&\AVQZ \\
				&$^1B_{1u} (\Val; \pi \ra \pis)$					&		& 92.9 & 6.45	& {\CCSDT}/\AVTZ		&6.44	&\AVQZ \\
				&$^1E_{1g} (\mathrm{R}; \pi \ra 3s)$				&		& 92.8 & 6.52	& {\CCSDT}/\AVTZ		&6.54	&\AVQZ \\
				&$^1A_{2u}  (\mathrm{R}; \pi \ra 3p)$			&0.066	& 93.4 & 7.08	& {\CCSDT}/\AVTZ		&7.10	&\AVQZ \\
				&$^1E_{2u}  (\mathrm{R}; \pi \ra 3p)$			&		& 92.8 & 7.15	& {\CCSDT}/\AVTZ		&7.16	&\AVQZ \\
				&$^3B_{1u} (\Val; \pi \ra \pis)$					&		& 98.6 & 4.16	& D					&4.17	&\AVQZ \\
				&$^3E_{1u}(\Val; \pi \ra \pis)$					&		& 97.1 & 4.85	& D					&4.86	&\AVQZ \\
				&$^3B_{2u} (\Val; \pi \ra \pis)$					&		& 98.1 & 5.81	& D					& 5.81	&\AVQZ \\
Butadiene			&$^1B_u  (\Val; \pi \ra \pis)$					&0.664	& 93.3 & 6.22	& B					& 6.21	& \AVQZ  \\
				&$^1B_g (\mathrm{R}; \pi \ra 3s)$				&		& 94.1 & 6.33	& B					& 6.35	& \AVQZ \\
				&$^1A_g  (\Val; \pi \ra \pis)$					&		& 75.1 & 6.50	& F					& 6.50	& \AVQZ  \\
				&$^1A_u (\mathrm{R}; \pi \ra 3p)$				&0.001	& 94.1 & 6.64	& B					& 6.66	& \AVQZ  \\
				&$^1A_u (\mathrm{R}; \pi \ra 3p)$				&0.049	& 94.1 & 6.80	& B					& 6.82	& \AVQZ  \\
				&$^1B_u (\mathrm{R}; \pi \ra 3p)$				&0.055	& 93.8 & 7.68	& C					& 7.54	& \AVQZ  \\
				&$^3B_u (\Val; \pi \ra \pis)$					&		& 98.4 & 3.36	& D					& 3.37	& \AVQZ  \\
				&$^3A_g (\Val; \pi \ra \pis)$					&		& 98.7 & 5.20	& D					& 5.21	& \AVQZ  \\
				&$^3B_g (\mathrm{R}; \pi \ra 3s)$				&		& 97.9 & 6.29	& D					& 6.31	& \AVQZ  \\
Cyanoacetylene	&$^1\Sigma^- 	(\Val; \pi \ra \pis)$ 				&		& 94.3 & 5.80	& A					& 5.79	& \AVPZ\\
				&$^1\Delta 	(\Val; \pi \ra \pis)$ 				&		& 94.0 & 6.07	& A					& 6.05	&\AVPZ\\
				&$^3\Sigma^+	 (\Val; \pi \ra \pis)$ 				&		& 98.5 & 4.44	& {\CCSDT}/\AVTZ		& 4.46	&\AVPZ \\
				&$^3\Delta 	(\Val; \pi \ra \pis)$ 				&		& 98.2 & 5.21	& {\CCSDT}/\AVTZ		& 5.21	& \AVPZ\\
				&$^1A'' [\mathrm{F}]	(\Val; \pi \ra \pis)$ 			& 0.004	& 93.6 & 3.54	& A					& 3.54	& \AVQZ \\
Cyanoformaldehyde	&$^1A'' (\Val; n \ra \pis)$						& 0.001	& 89.8 & 3.81	& {\CCSDT}/\AVTZ		& 3.82	& \AVQZ \\
				&$^1A'' (\Val; \pi \ra \pis)$						& 0.000	& 91.9 & 6.46	& {\CCSDT}/\AVTZ		& 6.45	& \AVQZ \\
				&$^3A'' (\Val; n \ra \pis)$						&		& 97.6 & 3.44	& D					& 3.45	& \AVQZ \\
				&$^3A' (\Val; \pi \ra \pis)$						&		& 98.4 & 5.01	& D					& 5.02	& \AVQZ \\
Cyanogen			& $^1\Sigma_u^- (\Val; \pi \ra \pis)$ 				&		& 94.1 & 6.39	& A					& 6.38	& \AVPZ\\
				& $^1\Delta_u (\Val; \pi \ra \pis)$ 				&		& 93.4 & 6.66 	& A					& 6.64	& \AVPZ\\
				& $^3\Sigma_u^+ (\Val; \pi \ra \pis)$ 				&		& 98.5 & 4.91	& B					& 4.93	& \AVPZ \\
				& $^1\Sigma_u^-  [\mathrm{F}]  (\Val; \pi \ra \pis)$	&		& 93.4 & 5.05 	& A					& 5.03	& \AVPZ \\
Cyclopentadiene	&$^1B_2 (\Val; \pi \ra \pis)$	 				&0.084	& 93.8 & 5.56	& {\CCSDT}/\AVTZ		& 5.55	& \AVQZ \\
				&$^1A_2 (\mathrm{R}; \pi \ra 3s)$			 	&		& 94.0 & 5.78	& {\CCSDT}/\AVTZ		& 5.80	& \AVQZ \\
				&$^1B_1  (\mathrm{R}; \pi \ra 3p)$			 	&0.037	& 94.2 & 6.41	& {\CCSDT}/\AVTZ		& 6.42	& \AVQZ \\
				&$^1A_2  (\mathrm{R}; \pi \ra 3p)$			 	&		& 93.8 & 6.46	& {\CCSDT}/\AVTZ		& 6.47 	& \AVQZ \\
				&$^1B_2  (\mathrm{R}; \pi \ra 3p)$			 	&0.046	& 94.2 & 6.56	&{\CCSDT}/\AVTZ		& 6.55	& \AVQZ\\
				&$^1A_1 (\Val; \pi \ra \pis)$		 			&0.001	& 78.9 & \emph{6.52}	& {\CCSDT}/\AVTZ		& \emph{6.52}	& \AVQZ	\\
				&$^3B_2 (\Val; \pi \ra \pis)$		 			&		& 98.4 & 3.31	& D					& 3.31	& \AVQZ \\
				&$^3A_1 (\Val; \pi \ra \pis)$		 			&		& 98.6 & 5.11	& D					& 5.12	& \AVQZ \\
				&$^3A_2 (\mathrm{R}; \pi \ra 3s)$			 	&		& 97.9 & 5.73	& D					& 5.75 	& \AVQZ \\
				&$^3B_1 (\mathrm{R}; \pi \ra 3p)$			 	&		& 97.9 & 6.36	& D					& 6.38	& \AVQZ \\
Cyclopropenone	&$^1B_1 (\Val; n \ra \pis)$						&0.000	& 87.7 & 4.26 	& B					& 4.28	& \AVPZ \\
				&$^1A_2 (\Val; n \ra \pis)$						&		& 91.0 & 5.55	& B					& 5.56	&\AVPZ \\
				&$^1B_2 (\mathrm{R}; n \ra 3s)$				&0.003	& 90.8 & 6.34	& B					& 6.40	& \AVPZ \\
				&$^1B_2 (\Val; \pi \ra \pis$)					&0.047	& 86.5 & 6.54	& B					& 6.56	& \AVPZ\\
				&$^1B_2 (\mathrm{R}; n \ra 3p)$				&0.018	& 91.1 & 6.98	& B					& 7.01	&  \AVPZ \\
				&$^1A_1 (\mathrm{R}; n \ra 3p)$				&0.003	& 91.2 & 7.02	& B					& 7.08	&\AVPZ \\
				&$^1A_1 (\Val; \pi \ra \pis)$					&0.320	& 90.8 & 8.28	& B					& 8.26	&\AVPZ \\
				&$^3B_1 (\Val; n \ra \pis)$						&		& 96.0 & 3.93	& {\CCSDT}/\AVTZ		& 3.96	& \AVPZ \\
				&$^3B_2 (\Val; \pi \ra \pis)$					&		& 97.9 & 4.88	& {\CCSDT}/\AVTZ		& 4.91	& \AVPZ \\
				&$^3A_2 (\Val; n \ra \pis)$						&		& 97.5 & 5.35	& {\CCSDT}/\AVTZ		& 5.37	& \AVPZ \\
				&$^3A_1 (\Val; \pi \ra \pis)$					&		& 98.1 & 6.79	& {\CCSDT}/\AVTZ		& 6.81	& \AVPZ\\
Cyclopropenethione	&$^1A_2 (\Val; n \ra \pis)$						&		& 89.6 & 3.41	& B					& 3.41	& \AVPZ \\
				&$^1B_1 (\Val; n \ra \pis)$						&0.000	& 84.8 & 3.45	& B					& 3.48	& \AVPZ \\
				&$^1B_2 (\Val; \pi \ra \pis)$					&0.007	& 83.0 & 4.60	& B					& 4.62	& \AVPZ \\
				&$^1B_2 (\mathrm{R}; n \ra 3s)$				&0.048	& 91.8 & 5.34	& B					& 5.40	& \AVPZ \\
				&$^1A_1 (\Val; \pi \ra \pis)$					&0.228	& 89.0 & 5.46	& B					& 5.46	& \AVPZ \\
				&$^1B_2 (\mathrm{R}; n \ra 3p)$				&0.084	& 91.3 & 5.92	& B					& 5.94	& \AVPZ \\
				&$^3A_2 (\Val; n \ra \pis)$						&		& 97.2 & 3.28	& D					& 3.28	& \AVPZ \\
				&$^3B_1 (\Val; n \ra \pis)$						&		& 94.5 & 3.32	& {\CCSDT}/\AVTZ		& 3.36	& \AVPZ \\
				&$^3B_2 (\Val; \pi \ra \pis)$					&		& 96.5 & 4.01	& D					& 4.04	& \AVPZ \\
				&$^3A_1 (\Val; \pi \ra \pis)$					&		& 98.2 & 4.01	& D					& 4.01	& \AVPZ \\
Diacetylene		&$^1\Sigma_u^- (\Val; \pi \ra \pis)$				&		& 94.4 & 5.33	& A					& 5.32	& \AVPZ 	\\
				&$^1\Delta_u 	(\Val; \pi \ra \pis)$				&		& 94.1 & 5.61	& A					& 5.60	& \AVPZ 	\\
				&$^3\Sigma_u^+ (\Val; \pi \ra \pis)$				&		& 98.5 & 4.10	& C					& 4.13	& \AVPZ 	\\
				&$^3\Delta_u 	(\Val; \pi \ra \pis)$				&		& 98.2 & 4.78	& B					& 4.78	&\AVPZ 	\\
Furan			&$^1A_2 (\mathrm{R}; \pi \ra 3s)$				&		& 93.8 & 6.09 	&{\CCSDT}/\AVTZ		& 6.11	&\AVQZ \\
				&$^1B_2 (\Val; \pi \ra \pis)$					&0.163	& 93.0 & 6.37 	&{\CCSDT}/\AVTZ		&  6.37	&\AVQZ \\
				&$^1A_1 (\Val; \pi \ra \pis)$					&0.000	& 92.4 & 6.56 	&{\CCSDT}/\AVTZ		& 6.56	&\AVQZ \\
				&$^1B_1  (\mathrm{R}; \pi \ra 3p)$				&0.038	& 93.9 & 6.64	&{\CCSDT}/\AVTZ		&  6.66	&\AVQZ \\
				&$^1A_2  (\mathrm{R}; \pi \ra 3p)$				&		& 93.6 & 6.81 	&{\CCSDT}/\AVTZ		& 6.83	&\AVQZ \\
				&$^1B_2  (\mathrm{R}; \pi \ra 3p)$				&0.008	& 93.5 & 7.24    & D					& 7.14	&\AVQZ \\
				&$^3B_2 (\Val; \pi \ra \pis)$					&		& 98.4 & 4.20    & D					& 4.20	&\AVQZ \\
				&$^3A_1 (\Val; \pi \ra \pis)$					&		& 98.1 & 5.46    & D					& 5.47	&\AVQZ \\
				&$^3A_2 (\mathrm{R}; \pi \ra 3s)$				&		& 97.9 & 6.02    & D					& 6.05	&\AVQZ \\
				&$^3B_1 (\mathrm{R}; \pi \ra 3p)$				&		& 97.9 & 6.59    & D					& 6.61	&\AVQZ \\
Glyoxal			&$^1A_u (\Val; n \ra \pis)$						& 0.000	& 91.0 & 2.88	& B					&  2.88	& \AVPZ \\
				&$^1B_g (\Val; n \ra \pis)$						&		& 88.3 & 4.24	& B					&  4.25	& \AVPZ \\
				&$^1A_g (\Val; n,n  \ra \pis,\pis)$				&		& 0.5   & 5.61	& F					&  5.60	& \AVPZ \\%
				&$^1B_g (\Val; n \ra \pis)$						&		& 83.9 & 6.57 	& B					&  6.58	& \AVPZ \\
				&$^1B_u (\mathrm{R}; n \ra 3p)$				& 0.095	& 91.7 & 7.71	& B					&  7.78	& \AVPZ \\
				&$^3A_u (\Val; n \ra \pis)$						&		& 97.6 & 2.49	& {\CCSDT}/\AVTZ		& 2.50	& \AVPZ \\
				&$^3B_g (\Val; n \ra \pis)$						&		& 97.4 & 3.89	& {\CCSDT}/\AVTZ		& 3.91	& \AVPZ \\
				&$^3B_u (\Val; \pi \ra \pis)$					&		& 98.5 & 5.15	& {\CCSDT}/\AVTZ		& 5.17	& \AVPZ \\
				&$^3A_g (\Val; \pi \ra \pis)$					&		& 98.8 & 6.30	& {\CCSDT}/\AVTZ		& 6.31	& \AVPZ \\
Imidazole			&$^1A'' (\mathrm{R}; \pi \ra 3s)$				& 0.001	& 93.0 & 5.71 	& D					& 5.73	& \AVQZ \\
				&$^1A' (\Val; \pi \ra \pis)$						& 0.124	& 89.6 & 6.41	& D					& 6.41	& \AVQZ \\
				&$^1A'' (\Val; n \ra \pis)$						& 0.028	& 93.6 & 6.50	& D					& 6.53	& \AVQZ \\
				&$^1A' (\mathrm{R}; \pi \ra 3p)$				& 0.035	& 88.9 & \emph{6.83}	& D			&\emph{6.82}	& \AVQZ \\
				&$^3A' (\Val; \pi \ra \pis)$						&		& 98.3 & 4.73	& E					& 4.74	& \AVQZ \\
				&$^3A'' (\mathrm{R}; \pi \ra 3s)$				&		& 97.6 & 5.66	& D					& 5.69	& \AVQZ \\
				&$^3A' (\Val; \pi \ra \pis)$						&		& 97.9 & 5.74	& E					& 5.75	& \AVQZ \\
				&$^3A'' (\Val; n \ra \pis)$						&		& 97.3 & 6.31	& D					& 6.31	& \AVQZ \\
Isobutene			&$^1B_1 (\mathrm{R}; \pi \ra 3s)$				& 0.006	& 94.1 & 6.46	& {\CCSDT}/\AVTZ		& 6.48	& \AVQZ \\
				&$^1A_1 (\mathrm{R}; \pi \ra 3p)$				& 0.228	& 94.2 & 7.01   & {\CCSDT}/\AVTZ		& 7.00	& \AVQZ \\
				&$^3A_1 (\Val; (\pi \ra \pis)$					&		& 98.9 & 4.53   & D					& 4.54	& \AVQZ \\
Methylenecyclopropene&	$^1B_2 (\Val; \pi \ra \pis)$					& 0.011	& 85.4 & 4.28	& B					& 4.29	& \AVPZ \\
				&$^1B_1 (\mathrm{R}; \pi \ra 3s)$				& 0.005	& 93.6 & 5.44	& B					& 5.47	& \AVPZ \\
				&$^1A_2 (\mathrm{R}; \pi \ra 3p)$				&		& 93.3 & 5.96	& B					& 5.98	& \AVPZ \\
				&$^1A_1(\Val; \pi \ra \pis)$					& 0.224	& 92.8 & \emph{6.12}	& B			& \emph{6.03}	& \AVPZ \\
				&$^3B_2 (\Val; \pi \ra \pis)$					&		& 97.2 & 3.49	&  {\CCSDT}/\AVTZ		& 3.50	& \AVPZ \\
				&$^3A_1 (\Val; \pi \ra \pis)$					&		& 98.6 & 4.74	& D					& 4.75	& \AVPZ \\
Propynal			& $^1A'' (\Val; n \ra \pis)$						& 0.000	& 89.0 & 3.80 	& {\CCSDT}/\AVTZ		& 3.81	& \AVQZ \\
				&$^1A'' (\Val; \pi \ra \pis)$						& 0.000	& 92.9 & 5.54	& {\CCSDT}/\AVTZ		& 5.53	& \AVQZ \\
				&$^3A'' (\Val; n \ra \pis)$						&		& 97.4 & 3.47	& D					& 3.48	& \AVQZ \\
				&$^3A' (\Val; \pi \ra \pis)$						&		& 98.3 & 4.47	& D					& 4.48	& \AVQZ \\
Pyrazine			&$^1B_{3u}  (\Val; n \ra \pis)$					&0.006	& 90.1 & 4.15	& {\CCSDT}/\AVTZ		& 4.15	& \AVQZ \\
				&$^1A_{u}  (\Val; n \ra \pis)$					&		& 88.6 & 4.98	& {\CCSDT}/\AVTZ		& 4.99	& \AVQZ \\
				&$^1B_{2u}  (\Val; \pi \ra \pis)$					&0.078	& 86.9 & 5.02	& {\CCSDT}/\AVTZ		& 5.01	& \AVQZ \\
				&$^1B_{2g}  (\Val; n \ra \pis)$					&		& 85.6 & 5.71	& {\CCSDT}/\AVTZ		& 5.71	& \AVQZ \\
				&$^1A_{g}  (\mathrm{R}; n \ra 3s)$				&		& 91.1 & 6.65	& {\CCSDT}/\AVTZ		& 6.69	& \AVQZ \\
				&$^1B_{1g}  (\Val; n \ra \pis)$					&		& 84.2 & 6.74	& {\CCSDT}/\AVTZ		& 6.74	& \AVQZ \\
				&$^1B_{1u}  (\Val; \pi \ra \pis)$					&0.063	& 92.8 & 6.88	& {\CCSDT}/\AVTZ		& 6.87	& \AVQZ \\
				&$^1B_{1g}  (\mathrm{R}; \pi \ra 3s)$			&		& 93.8 & 7.21	& {\CCSDT}/\AVTZ		& 7.24	& \AVQZ \\
				&$^1B_{2u}  (\mathrm{R}; n \ra 3p)$				&0.037	& 90.8 & 7.24	& D					& 7.28	&\AVQZ \\
				&$^1B_{1u}  (\mathrm{R}; n \ra 3p)$				&0.128	& 91.4 & 7.44	& D					& 7.47	&\AVQZ \\
				&$^1B_{1u}  (\Val; \pi \ra \pis)$					&0.285	& 90.5 & \emph{7.98}& D				& \emph{7.97}	&\AVQZ \\
				&$^3B_{3u}  (\Val; n \ra \pis)$					&		& 97.3 & 3.59 	& D					& 3.59	& \AVQZ \\
				&$^3B_{1u}  (\Val; \pi \ra \pis)$					&		& 98.5 & 4.35	& D					& 4.36	& \AVQZ \\
				&$^3B_{2u}  (\Val; (\pi \ra \pis)$					&		& 97.6 & 4.39	& D					& 4.39	& \AVQZ \\
				&$^3A_{u}  (\Val; n \ra \pis)$					&		& 96.1 & 4.93	& D					& 4.94	& \AVQZ \\
				&$^3B_{2g}  (\Val; n \ra \pis)$					&		& 97.0 & 5.08	& D					& 5.09	& \AVQZ \\
				&$^3B_{1u}  (\Val; \pi \ra \pis)$					&		& 97.0 & 5.28	& D					& 5.28 	& \AVQZ \\
Pyridazine			&$^1B_1 (\Val; n \ra \pis)$						&0.005	& 89.0 & 3.83	& D					& 3.83	& \AVQZ \\
				&$^1A_2 (\Val; n \ra \pis)$						&         	& 86.9 & 4.37   & D					&4.38	 & \AVQZ	\\
				&$^1A_1 (\Val; \pi \ra \pis)$					&0.016	& 85.8 & 5.26 	& D					& 5.26	 & \AVQZ \\
				&$^1A_2 (\Val; n \ra \pis)$						&        	& 86.2 & 5.72	& D					&5.72	&  \AVQZ \\
				&$^1B_2  (\mathrm{R}; n \ra 3s)$				&0.001	& 88.5 & 6.17	& D					&6.21	&  \AVQZ \\
				&$^1B_1 (\Val; n \ra \pis)$						&0.004	& 87.0 & 6.37	& D					&6.37	& \AVQZ \\
				&$^1B_2 (\Val; \pi \ra \pis)$					&0.010	& 90.6 & 6.75	& D					&6.74	 &\AVQZ \\
				&$^3B_1 (\Val; n \ra \pis)$						&		& 97.1 & 3.19	& D					& 3.20	& \AVQZ	\\
				&$^3A_2 (\Val; n \ra \pis)$						&		& 96.2 & 4.11	& D					& 4.12	& \AVQZ	\\
				&$^3B_2 (\Val; \pi \ra \pis)$					&		& 98.5 & \emph{4.34}	& D			& 4.35	& \AVQZ	\\
				&$^3A_1 (\Val; \pi \ra \pis)$					&		& 97.3 & 4.82	& D					& 4.81	&  \AVQZ \\
Pyridine			&$^1B_1 (\Val; n \ra \pis)$						& 0.004	& 88.4 & 4.95 	&  D					& 4.95	& \AVQZ \\
				&$^1B_2 (\Val; \pi \ra \pis)$					& 0.028	& 86.5 & 5.14	&  D					& 5.14	&  \AVQZ \\
				&$^1A_2 (\Val; n \ra \pis)$						&		& 87.9 & 5.40	& D					& 5.41	& \AVQZ \\
				&$^1A_1 (\Val; \pi \ra \pis)$					&0.010	& 92.1 & 6.62	& D					& 6.61	& \AVQZ \\
				&$^1A_1 (\mathrm{R}; n \ra 3s)$				&0.011	& 89.7 & 6.76	& D					& 6.80	& \AVQZ \\
				&$^1A_2 (\mathrm{R}; \pi \ra 3s)$				&		& 93.2 & 6.82	& D					& 6.84	 &\AVQZ \\
				&$^1B_2 (\Val; \pi \ra \pis)$					& 0.319	& 90.0  & \emph{7.40}& D				& \emph{7.42}	&\AVQZ \\
				&$^1B_1 (\mathrm{R}; \pi \ra 3p)$				& 0.045	& 93.6 & 7.38	& D 					& 7.40	&\AVQZ \\
				&$^1A_1 (\Val; \pi \ra \pis)$					&0.291	& 90.5 & 7.39	& D					& 7.40	& \AVQZ \\
				&$^3A_1 (\Val; \pi \ra \pis)$					&		& 98.5 & 4.30	& D					&4.31	 & \AVQZ \\
				&$^3B_1 (\Val; n \ra \pis)$						&		& 97.0 & 4.46	& D					&4.47	 &  \AVQZ \\
				&$^3B_2 (\Val; \pi \ra \pis)$					&		& 97.3 & 4.79	& D					&4.79 	&  \AVQZ \\
				&$^3A_1 (\Val; \pi \ra \pis)$					&		& 97.1 & 5.04	& E					&5.04	 & \AVQZ	\\
				&$^3A_2 (\Val; n \ra \pis)$						&		& 95.8 & 5.36	& D					&5.38	 &\AVQZ	\\
				&$^3B_2 (\Val; \pi \ra \pis)$					&		& 97.7 & 6.24	& D					&6.24	& \AVQZ \\
Pyrimidine			&$^1B_1 (\Val; n \ra \pis)$						& 0.005	& 88.6 & 4.44	& D					&4.45	 &\AVQZ\\
				&$^1A_2 (\Val; n \ra \pis)$						&		& 88.5 & 4.85	& D					&4.86	 &\AVQZ\\
				&$^1B_2 (\Val; \pi \ra \pis)$					&0.028	& 86.3 & 5.38	& D					&5.37	&\AVQZ	\\
				&$^1A_2 (\Val; n \ra \pis)$						&		& 86.7 & 5.92	& D					&5.92	&\AVQZ	\\
				&$^1B_1 (\Val; n \ra \pis)$						&0.005	& 86.7 & 6.26	& D					&6.27	 &\AVQZ	\\
				&$^1B_2  (\mathrm{R}; n \ra 3s)$				&0.005	& 90.3 & 6.70	& D					&6.74	& \AVQZ \\
				&$^1A_1 (\Val; \pi \ra \pis)$					&0.036	& 91.5 & 6.88	& D					&6.87	& \AVQZ \\
				&$^3B_1 (\Val; n \ra \pis)$						&		& 96.8 & 4.09	& D					&4.10	& \AVQZ \\
				&$^3A_1 (\Val; \pi \ra \pis)$					&		& 98.3 & \emph{4.51} & D				&\emph{4.52} 	& \AVQZ \\
				&$^3A_2 (\Val; n \ra \pis)$						&		& 96.5 & 4.66	& D					&4.67	& \AVQZ \\
				&$^3B_2 (\Val; \pi \ra \pis)$					&		& 97.4 & 4.96	& D					&4.96	& \AVQZ \\
 Pyrrole			&$^1A_2 (\mathrm{R}; \pi \ra 3s)$				&		& 92.9 & 5.24	& {\CCSDT}/\AVTZ		& 5.27	& \AVQZ \\
				&$^1B_1 (\mathrm{R}; \pi \ra 3p)$				&0.015	& 92.4 & 6.00 	& {\CCSDT}/\AVTZ		& 6.03	& \AVQZ \\
				&$^1A_2 (\mathrm{R}; \pi \ra 3p)$				&		& 93.0 & 6.00	& D					& 6.02	& \AVQZ \\
				&$^1B_2 (\Val; (\pi \ra \pis)$					&0.164	& 92.5 & 6.26 	& {\CCSDT}/\AVTZ		& 6.23	& \AVQZ \\
				&$^1A_1 (\Val; \pi \ra \pis)$					&0.001	& 86.3 & 6.30 	& {\CCSDT}/\AVTZ		& 6.29	& \AVQZ \\
				&$^1B_2 (\mathrm{R}; \pi \ra 3p)$				&0.003	& 92.6 & 6.83	& D					& 6.74	& \AVQZ \\
				&$^3B_2 (\Val; \pi \ra \pis)$					&		& 98.3 & 4.51	& D					& 4.51	& \AVQZ \\
				&$^3A_2 (\mathrm{R}; \pi \ra 3s)$				&		& 97.6 & 5.21	& D					& 5.24 	& \AVQZ \\
				&$^3A_1 (\Val; \pi \ra \pis)$					&		& 97.8 & 5.45	& D					& 5.46	& \AVQZ \\
				&$^3B_1 (\mathrm{R}; \pi \ra 3p)$				&		& 97.4 & 5.91	& D					& 5.94	& \AVQZ \\
Tetrazine			&$^1B_{3u}  (\Val; n \ra \pis)$					& 0.006	& 89.8 & 2.47	& {\CCSDT}/\AVTZ		& 2.46	& \AVQZ \\
				&$^1A_{u}  (\Val; n \ra \pis)$					&		& 87.9 & 3.69	& {\CCSDT}/\AVTZ		& 3.70	& \AVQZ \\
				&$^1A_{g}  (\Val; n,n \ra \pis, \pis)$				&		& 0.7   & \emph{4.61} & {\NEV}/\AVTZ	& \emph{4.59}	& \AVQZ\\%
				&$^1B_{1g}  (\Val; n \ra \pis)$					&		& 83.1 & 4.93	& {\CCSDT}/\AVTZ		& 4.92	& \AVQZ \\
				&$^1B_{2u}  (\Val; \pi \ra \pis)$					& 0.055	& 85.4 & 5.21	& {\CCSDT}/\AVTZ		& 5.20	& \AVQZ \\
				&$^1B_{2g}  (\Val; n \ra \pis)$					&		& 81.7 & 5.45	& {\CCSDT}/\AVTZ		& 5.45	& \AVQZ \\
				&$^1A_{u}  (\Val; n \ra \pis)$					&		& 87.7 & 5.53	& {\CCSDT}/\AVTZ		& 5.53	& \AVQZ \\
				&$^1B_{3g} (\Val; n,n \ra \pis, \pis)$				&		& 0.7   & \emph{6.15} & {\NEV}/\AVTZ	& \emph{6.13}	& \AVQZ\\
				&$^1B_{2g}  (\Val; n \ra \pis)$					&		& 80.2 & 6.12	& D					& 6.12	& \AVQZ \\
				&$^1B_{1g}  (\Val; n \ra \pis)$					&		& 85.1 & 6.91	& D					& 6.91	& \AVQZ \\
				&$^3B_{3u}  (\Val; n \ra \pis)$					&		& 97.1 & 1.85	& D					& 1.86	& \AVQZ \\
				&$^3A_{u}  (\Val; n \ra \pis)$					&		& 96.3 & 3.45	& D					& 3.46	& \AVQZ \\
				&$^3B_{1g}  (\Val; n \ra \pis)$					&		& 97.0 & 4.20	& D					& 4.21	& \AVQZ \\
				&$^3B_{1u}  (\Val; \pi \ra \pis)$					&		& 98.5 & \emph{4.49}	& D			&  \emph{4.49}	& \AVQZ \\
				&$^3B_{2u}  (\Val; \pi \ra \pis)$					&		& 97.5 & 4.52	& D					& 4.52	& \AVQZ \\
				&$^3B_{2g}  (\Val; n \ra \pis)$					&		& 96.4 & 5.04	& D					& 5.04	& \AVQZ \\
				&$^3A_{u}  (\Val; n \ra \pis)$					&		& 96.6 & 5.11	& D					& 5.11	& \AVQZ \\
				&$^3B_{3g}  (\Val; n,n \ra \pis, \pis)$				&		& 5.7   & \emph{5.51} &{\NEV}/\AVTZ	& \emph{5.50}	& \AVQZ\\
				&$^3B_{1u}  (\Val; \pi \ra \pis)$					&		& 96.6 & 5.42	& D					& 5.43	& \AVQZ \\
Thioacetone		&$^1A_2 (\Val; n \ra \pis)$						& 		& 88.9 & 2.53 	& B					&  2.54	&  \AVQZ \\
				&$^1B_2 (\mathrm{R}; n \ra 4s)$				& 0.052	& 91.3 & 5.56 	&  B					&  5.61 	&  \AVQZ \\
				&$^1A_1 (\Val; \pi \ra \pis)$					& 0.242	& 90.6 & 5.88	& B					& 5.86	&  \AVQZ \\
				&$^1B_2 (\mathrm{R}; n \ra 4p)$				& 0.028	& 92.4 & 6.51	& C					& 6.52	&  \AVQZ \\
				&$^1A_1 (\mathrm{R}; n \ra 4p)$				& 0.023	& 91.6 & 6.61	 &B					&  6.64	&  \AVQZ \\
				&$^3A_2 (\Val; n \ra \pis)$						&		& 97.4 & 2.33 	& D					& 2.34	& \AVQZ \\
				&$^3A_1 (\Val; \pi \ra \pis)$					&		& 98.7 & 3.45	& D					& 3.46 	& \AVQZ \\
Thiophene		&$^1A_1 (\Val; \pi \ra \pis)$					&0.070	& 87.6 & 5.64	& {\CCSDT}/\AVTZ		& 5.63	& \AVQZ \\
				&$^1B_2 (\Val; \pi \ra \pis)$					&0.079	& 91.5 & 5.98	& {\CCSDT}/\AVTZ		& 5.96	& \AVQZ \\
				&$^1A_2 (\mathrm{R}; \pi \ra 3s)$				&		& 92.6 & 6.14	& {\CCSDT}/\AVTZ		& 6.16	& \AVQZ \\
				&$^1B_1 (\mathrm{R}; \pi \ra 3p)$				&0.010	& 90.1 & 6.14	& {\CCSDT}/\AVTZ		& 6.11	& \AVQZ \\
				&$^1A_2 (\mathrm{R}; \pi \ra 3p)$				&		& 91.8 & 6.21	& {\CCSDT}/\AVTZ		& 6.18	& \AVQZ \\
				&$^1B_1 (\mathrm{R}; \pi \ra 3s)$				&0.000	& 92.8 & 6.49	& {\CCSDT}/\AVTZ		& 6.52	& \AVQZ \\
				&$^1B_2 (\mathrm{R}; \pi \ra 3p)$				&0.082	& 92.4 & 7.29 	&  {\CCSDT}/\AVTZ		& 7.18	& \AVQZ \\
				&$^1A_1 (\Val; \pi \ra \pis)$					&0.314	& 86.5 & \emph{7.31}& E				& \emph{7.29}	& \AVQZ \\
				&$^3B_2 (\Val; \pi \ra \pis)$					&		& 98.2 & 3.92	& D					& 3.91	& \AVQZ \\
				&$^3A_1 (\Val; \pi \ra \pis)$					&		& 97.7 & 4.76	& D					& 4.76	& \AVQZ \\
				&$^3B_1 (\mathrm{R}; \pi \ra 3p)$				&		& 96.6 & 5.93	& D					& 5.90	& \AVQZ \\
				&$^3A_2 (\mathrm{R}; \pi \ra 3s)$				&		& 97.5 & 6.08	& D					& 5.98	& \AVQZ \\
Thiopropynal		&$^1A''  (\Val; n \ra \pis)$						& 0.000	& 87.5 & 2.03	 & {\CCSDT}/\AVTZ		&  2.04	&  \AVQZ \\
				&$^3A''   (\Val; n \ra \pis)$						&		& 97.2 & 1.80	& D					&  1.81	&  \AVQZ \\
Triazine			&$^1A_1'' (\Val; n \ra \pis)$					&		& 88.3 & 4.72	& {\CCSDT}/\AVTZ		& 4.72	& \AVQZ \\
				&$^1A_2'' (\Val; n \ra \pis)$					&0.014	& 88.3 & 4.75	& {\CCSDT}/\AVTZ		& 4.74	& \AVQZ \\
				&$^1E'' (\Val; n \ra \pis)$						&		& 88.3 & 4.78	& {\CCSDT}/\AVTZ		& 4.78	& \AVQZ \\
				&$^1A_2' (\Val; \pi \ra \pis)$					&		& 85.7 & 5.75	& {\CCSDT}/\AVTZ		&5.75	 &\AVQZ\\
				&$^1A_1' (\Val; \pi \ra \pis)$					&	 	& 90.4 & 7.24	& {\CCSDT}/\AVTZ		& 7.23	& \AVQZ \\
				&$^1E' (\mathrm{R}; n \ra 3s)$					&0.016	& 90.9 & 7.32	& {\CCSDT}/\AVTZ		& 7.36	& \AVQZ \\
				&$^1E'' (\Val; n \ra \pis)$						&		& 82.6 & 7.78	& {\CCSDT}/\AVTZ		& 7.76	& \AVQZ \\
				&$^1E' (\Val; \pi \ra \pis)$						&0.451	& 90.0 & 7.94	& {\CCSDT}/\AVTZ		& 7.93 	& \AVQZ \\
				&$^3A_2'' (\Val; n \ra \pis)$					&		& 96.7 & 4.33	& D					& 4.34	& \AVQZ \\
				&$^3E'' (\Val; n \ra \pis)$						&		& 96.6 & 4.51	& D					& 4.51	& \AVQZ \\
				&$^3A_1'' (\Val; n \ra \pis)$					&	 	& 96.2 & 4.73	& D					& 4.74	& \AVQZ \\
				&$^3A_1' (\Val; \pi \ra \pis)$					&		& 98.2 & 4.85	& D					& 4.86	& \AVQZ \\
				&$^3E' (\Val; \pi \ra \pis)$						&		& 96.9 & 5.59	& E					& 5.59	& \AVQZ \\
				&$^3A_2' (\Val; (\pi \ra \pis)$					&	 	& 97.6 & 6.62	& D					& 6.61	& \AVQZ \\
 \end{longtable}
  \end{footnotesize}
\begin{flushleft}\begin{footnotesize}\begin{singlespace}
  \vspace{-0.6 cm}
$^a${
Protocol A: {\CCSDT}/{\AVTZ} value corrected by the difference between {\CCSDTQ}/{\AVDZ} and {\CCSDT}/{\AVDZ};
Protocol B: {\CCSDT}/{\AVTZ} value corrected by the difference between {\CCSDTQ}/{\Pop} and {\CCSDT}/{\Pop};
Protocol C: {\CCT}/{\AVTZ} value corrected by the difference between {\CCSDTQ}/{\Pop} and {\CCT}/{\Pop};
Protocol D: {\CCT}/{\AVTZ} value corrected by the difference between {\CCSDT}/{\AVDZ} and {\CCT}/{\AVDZ};
Protocol E: {\CCT}/{\AVTZ} value corrected by the difference between {\CCSDT}/{\Pop} and {\CCT}/{\Pop};
Protocol F: {\FCI}/{\AVDZ} value (from Ref.~\citenum{Loo19c}) corrected by the difference between {\CCSDT}/{\AVTZ} and {\CCSDT}/{\AVDZ}.
Protocol G: {\FCI}/{\Pop} value corrected by the difference between {\CCSDT}/{\AVTZ} and {\CCSDT}/{\Pop};
Protocol H: {\FCI}/{\Pop} value corrected by the difference between {\CCT}/{\AVTZ} and {\CCT}/{\Pop}.
}
\end{singlespace}\end{footnotesize}\end{flushleft}

\section{Benchmarks}

\begin{figure}[htp]
  \includegraphics[scale=0.98,viewport=2cm 14.5cm 19cm 27.5cm,clip]{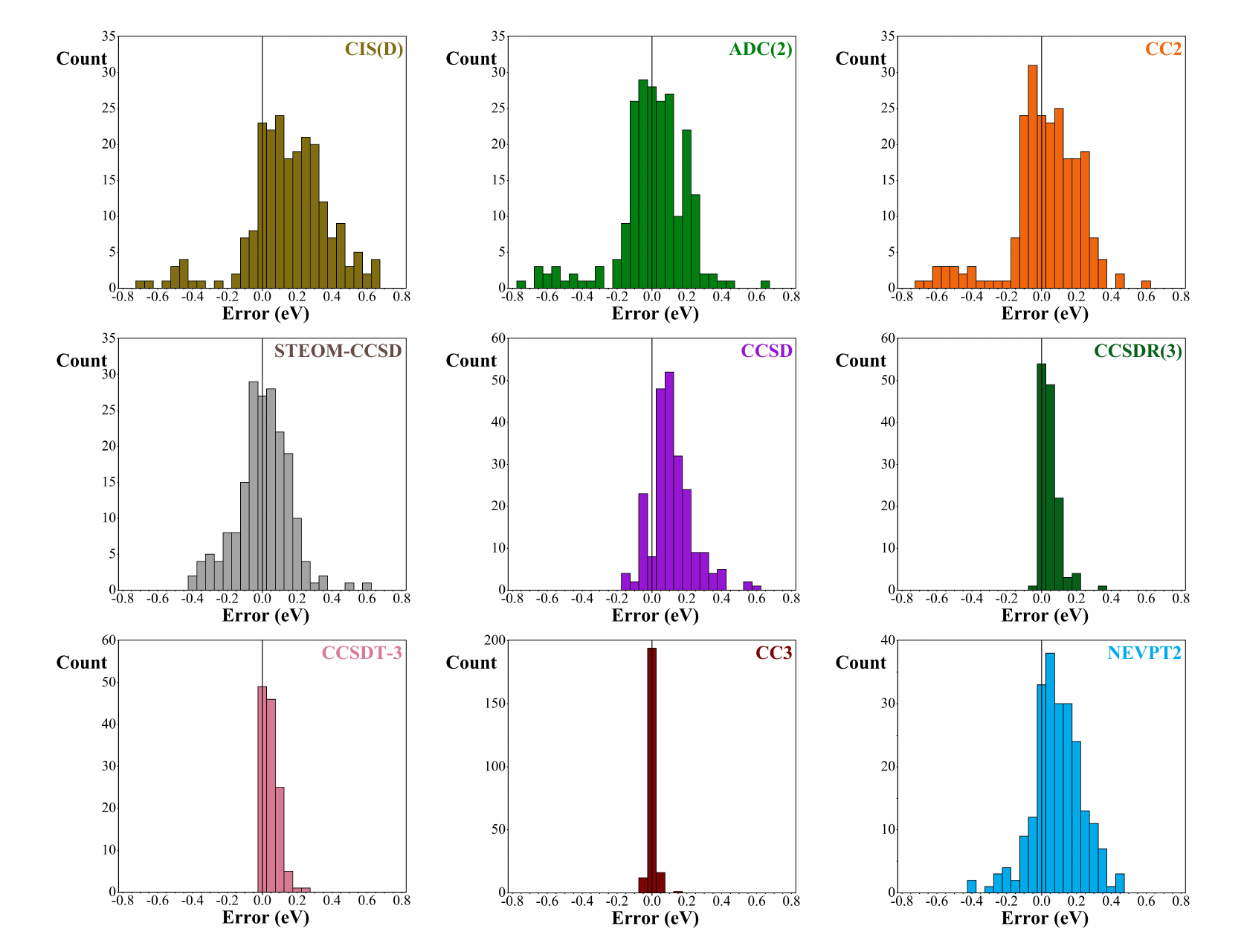}
  \caption{Histograms of the error distribution obtained with various levels of theory, taking the TBE/{\AVTZ} of Table \ref{Table-tbe} as references.
  Note the difference of scaling in the vertical axes.}
  \label{Fig-1}
\end{figure}

Having at hand such a large set of accurate transition energies, it seems natural to pursue previous benchmarking efforts. More specifically, we assess here the performance of eight popular wavefunction approaches, namely, 
CIS(D), {\AD}, {\CCD}, {\STEOM}, {\CCSD}, CCSDR(3), CCSDT-3, and {\CCT}.  The complete list of results can be found in Table S40 of the SI.  To identify the ES for all approaches, we have made, as for the TBE above
choices based on the usual criteria (symmetry, oscillator strength, ordering, and nature of the involved orbitals). Except for a few cases (see above), assignments are unambiguous. In addition, because all tested approaches are single-reference 
methods,  we have removed from the reference set the ``unsafe'' transition energies (in italics in Table \ref{Table-tbe}), as well as the four transitions with a dominant double excitation character (with $\Td < 50\%$ as listed in Table 
\ref{Table-12}). For the latter transitions, only CCSDT-3 and {\CCT} are able to detect their presence, but with, of course, extremely large errors. A comprehensive list of results are collected in Table \ref{Table-12} which, more 
specifically, gathers the MSE, MAE, RMSE, SDE, \MaxP, and \MaxN.  As benchmarks of the {\NEV} method are quite rare, we have also considered the above-listed  {\NEV}  values in our method evaluation. Of course, the results 
of such multi-configurational approaches significantly depend on the active space, but our main purpose is to know what typical error one can expect with such a model when reasonable, yet ``chemically-meaningful'' active spaces are considered.
 Figure \ref{Fig-1} shows  histograms of the error distributions for these nine methods.  Before discussing these, let us stress two obvious biases of this 
molecular set: i) it encompasses only conjugated organic molecules containing 4 to 6 non-hydrogen atoms; and ii) we mainly used {\CCSDTQ} (4 atoms) or {\CCSDT} (5--6 atoms) reference values.  As discussed in 
Section \ref{sec-ic} and in our previous work, \cite{Loo18a} the MAE obtained with these two methods are of the order of $0.01$ and $0.03$ eV, respectively. This means that any statistical quantity smaller than $\sim 0.02$--$0.03$ eV 
is very likely to be irrelevant.

\renewcommand*{\arraystretch}{1.0}
\begin{table}[htp]
\caption{Mean signed error (MSE), mean absolute error (MAE), root-mean square error (RMSE), standard deviation of the errors (SDE), as well as the positive [\MaxP] and negative [\MaxN] maximal errors with respect to the TBE/{\AVTZ} reported in Table \ref{Table-tbe}.
All these statistical quantities are reported in eV and have been obtained with the {\AVTZ} basis set.
``Count'' refers to the number of states.}
\label{Table-12}
\begin{tabular}{lccccccc}
\hline
Method 		& Count 	& MSE 	&MAE 	&RMSE 	&SDE 		&\MaxP 	&\MaxN \\
\hline
CIS(D)		&221&	0.16	&0.23	&0.29	&0.24	&0.96	&-0.69\\
{\AD}			&218&	0.01	&0.14	&0.20	&0.19	&0.64	&-0.73\\
{\CCD}		&223&	0.03	&0.15	&0.21	&0.20	&0.59	&-0.68\\
{\STEOM}		&190&	0.01	&0.12	&0.15	&0.14	&0.59	&-0.42\\
 {\CCSD}		&223&	0.11	&0.13	&0.16	&0.12	&0.62	&-0.16\\
CCSDR(3)	&134&	0.05	&0.05	&0.07	&0.05	&0.36	&-0.03\\
CCSDT-3		&127&	0.05	&0.05	&0.07	&0.04	&0.26	&0.00\\
{\CCT}		&223&	0.00	&0.01	&0.02	&0.02	&0.17	&-0.05\\
{\NEV}		&223	&	0.09	&0.13	&0.17	&0.14	&0.46	&-0.42\\
\hline
 \end{tabular}
 \end{table}

Let us analyze the global performance of all these methods, starting with the most accurate and computationally demanding single-reference models. The relative accuracies of {\CCT} and {\CCSDT}-3 as compared to {\CCSDT} remains an open question in the literature. \cite{Wat13,Dem13}
Indeed, to the best of our knowledge, the only two previous studies discussing this specific aspect are limited to very small compounds. \cite{Kan17,Loo18a} According to the results gathered in Table \ref{Table-12}, it appears that {\CCT} has a slight edge over {\CCSDT}-3, although
{\CCSDT}-3 is closer to {\CCSDT} in terms of formalism.  Indeed, {\CCSDT}-3 seems to provide slightly too large transition energies (MSE of $+0.05$ eV). These conclusions are qualitatively consistent with the analyses performed on smaller derivatives,
\cite{Kan17,Loo18a} but the amplitude of the {\CCSDT}-3 errors is larger in the present set.  Although the performance of {\CCT} might be unduly inflated by the use of {\CCSDT} and {\CCSDTQ} reference values, it is also clear that {\CCT}
 very rarely fails (Figure \ref{Fig-1}). Consequently, {\CCT} transition energies can be viewed as extremely solid references for any transition with a dominant single-excitation character.  This conclusion is again consistent with previous
analyses performed on smaller compounds, \cite{Hat05c,Kan17,Loo18a} as well as with recent comparisons between theoretical and experimental 0-0 energies performed by some of us on medium-sized molecules. \cite{Loo18b,Loo19a,Sue19}
To state it more boldly: it appears likely that {\CCT} is even more accurate than previously thought. In addition, thanks to the exhaustive and detailed comparisons made in the present work, we can safely conclude that {\CCT} regularly outperforms {\CASPT} and {\NEV}, even when
these methods are combined with relatively large active spaces. This statement seems to hold as long as the considered ES does not show a strong multiple excitation character, that is, when $\Td < 70\%$.

The perturbative inclusion of triples as in CCSDR(3) yields a very small MAE ($0.05$ eV) for a much lighter computational cost as compared to {\CCSDT}. Nevertheless, as with {\CCSDT}-3, the CCSDR(3) transition energies have a clear tendency
of being too large, an error sign likely inherited from the parent {\CCSD} model.  The $0.05$ eV MAE for CCSDR(3) is rather similar to the one obtained for smaller compounds when comparing to {\FCI} ($0.04$ eV), \cite{Loo18a} and is also inline with the
2009 benchmark study of Sauer et al. \cite{Sau09}

{\CCSD} provides an interesting case study. The calculated MSE ($+0.11$ eV), indicating an overestimation of the transition energies, fits well with several previous  reports. \cite{Sch08,Car10,Wat13,Kan14,Jac17b,Kan17,Dut18,Jac18a,Loo18a}
It is, nonetheless, larger than the one determined for smaller molecules ($+0.05$ eV), \cite{Loo18a} hinting that the performance of {\CCSD} deteriorates for larger compounds.  Moreover, the {\CCSD} MAE of $0.13$ eV is much smaller than the one reported by
Thiel in his original work ($0.49$ eV) \cite{Sch08} but of the same order of magnitude as in the more recent study of K\'ann\'ar and Szalay performed on Thiel's set ($0.18$ eV for transitions with $\Td > 90\%$). \cite{Kan14} Retrospectively, it is pretty obvious that
Thiel's much larger MAE is very likely due to the {\CASPT} reference values. \cite{Sch08} Indeed, as we have shown several times in the present study, {\CASPT} transitions energies tend to be significantly too low, therefore exacerbating the usual {\CCSD} overestimation.

With a single detailed benchmark study to date, \cite{Dut18} the {\STEOM} approach has received relatively little attention and its overall accuracy still needs to be corroborated. It is noteworthy that {\STEOM} provides a smaller MSE than {\CCSD} and comparable
MAE and RMSE.  The spread of the error is however slightly larger as evidenced by Figure \ref{Fig-1} and the SDE values reported in Table \ref{Table-12}.  These trends are the same as for smaller compounds. \cite{Loo18a} For Thiel's set, Dutta and coworkers
also reported a rather good performance for {\STEOM} with respect to the {\CCT}/TZVP reference data, though a slightly negative MSE is obtained in their case. \cite{Dut18} This could well be due to the different basis set considered in these two studies. It should be
nevertheless stressed that we only consider ``clean'' {\STEOM} results in the present work (see Computational Details), therefore removing several difficult cases that are included in the {\CCSD} statistics, \eg, the $A_g$ excitation in butadiene, which can slightly
bias the relative performance of {\STEOM} and {\CCSD}.

For the three second-order methods, namely CIS(D), {\AD}, and {\CCD}, that are often used as reference to benchmark TD-DFT for ``real-life'' applications, the performance of the former method is clearly worse compared to the latter two which exhibit very
similar statistical behaviors. These trends were also reported in previous works. \cite{Hat05c,Jac18a,Sch08,Sil10c,Win13,Har14,Jac15b,Kan17,Loo18a}  Interestingly, the {\CCD} MAE obtained here ($0.15$ eV) is significantly
smaller than the one we found for smaller compounds ($0.22$ eV). \cite{Loo18a}  Therefore, in contrast to {\CCSD}, {\CCD} performance seems to improve with molecular size. As above, Thiel's original MAE for {\CCD} ($0.29$ eV) was likely too large due
to the selection of {\CASPT} reference values. \cite{Sch08}  As already noticed by Szalay's group, \cite{Kan14,Kan17} although the MSE of {\CCD} is smaller than the one of {\CCSD}, the standard deviation is significantly larger
with the former model, \ie, {\CCD} is less consistent in terms of trends than {\CCSD}.

Finally, one obtains a reasonably tight distribution with {\NEV}, with a net overestimation trend (MSE of $0.09$ eV) and a general behavior that is in fact quite comparable to (STEOM-){\CCSD} in terms of average and maximal deviations. 
Nonetheless, we wish to point out that {\NEV} has the obvious advantage over {\CCSD} to be able to treat accurately ES characterized by a dominant double excitation character. As mentioned above, these were not included in the present 
benchmark set.

In Table \ref{Table-13}, we report a MAE decomposition for different subsets of ES. Note that, due to implementational limitations, only singlet ES could be computed with CCSDR(3) and CCSDT-3 which explains the lack of data for triplet ES.
A few interesting conclusions emerge from these results. First, the errors for singlet and triplet transitions are rather similar with all models, except for {\CCSD} that is significantly more effective for triplets.  Dutta and coworkers observed the same
trend for Thiel's set with MAE of $0.20$ eV and $0.11$ eV for the singlet and triplet ES, respectively. \cite{Dut18} Turning to the comparison between valence and Rydberg states, we find that {\CCD} provides a better description of the former,
whereas {\CCSD} (and higher-order methods) yields the opposite trend. In fact, {\CCD} has the clear tendency to overestimate valence ES energies (MSE of $+0.10$ eV), and to underestimate Rydberg ES energies (MSE of $-0.17$ eV).
{\CCSD} is found to be much more consistent with MSE of $0.12$ and $0.09$ eV, respectively (see the SI).  This relatively poorer performance of {\CCD} as compared to {\CCSD} for Rydberg ES is again perfectly consistent with other benchmarks,
\cite{Kan17,Dut18} although the MAE for {\CCD} ($0.18$ eV) reported in Table \ref{Table-13} remains relatively small as compared to the one given in Ref.~\citenum{Kan17}. We believe that it is likely due to the distinct types of Rydberg states
considered in these two studies. Indeed, we consider here (relatively) low-lying Rydberg transitions in medium-sized molecules, while K\'ann\'ar and Szalay (mostly) investigated higher-lying Rydberg states in smaller compounds.  CIS(D), 
{\AD}, {\CCD}, and {\STEOM} better describe $n\ra\pis$ transitions, whereas {\CCSD} seems more suited for $\pi\ra\pis$ transitions. The variations between the two subsets are probably not significant for the higher-order approaches.  
The former observation agrees well with previous results obtained for smaller compounds \cite{Loo18a} and for Thiel's set, \cite{Sch08,Kan14} whereas the latter, less expected observation is likely dependent on the selected ES subset. \cite{Sch08,Kan17}
Finally, the average errors obtained with {\NEV} are rather uniform for all subsets.

\renewcommand*{\arraystretch}{1.0}
\begin{table}[htp]
\caption{MAE (in eV) obtained with different methods for various classes of excited states.}
\label{Table-13}
\begin{tabular}{lcccccc}
\hline
Method 		& Singlet & Triplet & Valence & Rydberg & $n \ra \pis$ &  $\pi \ra \pis$ 	\\
\hline
CIS(D)		&0.21	&0.25	&0.26	&0.15	&0.22	&0.28\\
{\AD}			&0.15	&0.13	&0.13	&0.17	&0.08	&0.17\\
{\CCD}		&0.16	&0.14	&0.14	&0.18	&0.08	&0.19\\
{\STEOM}		&0.11	&0.13	&0.11	&0.12	&0.08	&0.15\\
 {\CCSD}		&0.16	&0.09	&0.14	&0.09	&0.19	&0.11\\
CCSDR(3)	&0.05	&		&0.07	&0.02	&0.08	&0.06\\
CCSDT-3		&0.05	&		&0.06	&0.03	&0.08	&0.04\\
{\CCT}		&0.01	&0.01	&0.01	&0.01	&0.01	&0.02\\
{\NEV}		&0.15	&0.12	&0.13	&0.15	&0.11	&0.14\\
\hline
 \end{tabular}
 \end{table}

\section{Concluding remarks}

We have computed highly-accurate vertical transition energies for a set of 27 organic molecules containing from 4 to 6 (non-hydrogen) atoms. To this end, we employed several state-of-the-art theoretical models with increasingly large diffuse basis sets.
Most of our theoretical best estimates are based on {\CCSDTQ} (4 atoms) or {\CCSDT} (5 and 6 atoms) excitation energies. For the vast majority of the listed excited states, the present contribution is the very first to disclose (sometimes basis-set extrapolated) {\CCSDT}/{\AVTZ}
and (true) {\CCT}/{\AVQZ} transition energies as well as {\CCT}/{\AVTZ} oscillator strengths for each dipole-allowed transition. Our set contains a total of 238 transition energies and 90 oscillator strengths, with a reasonably good balance between singlet, triplet, valence,
and Rydberg states. Among these 238 transitions, we believe that 224 are ``solid'' TBE, \ie, they are chemically accurate (MAE below $0.043$ eV or $1$ kcal.mol$^{-1}$) for the considered geometry. It allowed us to establish a reasonable error bar for several popular
ES models with lower computational cost: CIS(D), {\AD}, {\CCD}, {\STEOM}, {\CCSD}, CCSDR(3),  CCSDT-3,  {\CCT} and {\NEV}. It turns out that {\CCT} is extremely accurate, and, very likely should be considered as globally more robust and trustworthy than 
{\CASPT} or {\NEV}, except for ES with a predominant double excitation character.  Other methods including corrections for the triples yield a mean absolute deviation around $0.05$ eV, whereas none of the second-order approaches has been found to be chemically accurate with MAE
in the $0.12$--$0.23$ eV range.

Paraphrasing Thiel and coworkers, \cite{Sch08} we hope that this new set of vertical transition energies, combined or not with the ones described in our previous works, \cite{Loo18a,Loo19c} will be useful for the community,
will stimulate further developments and analyses in the field, and will provide new grounds for appraising the \emph{pros} and \emph{cons} of ES models already available or currently under development.  We can
crystal-ball that the emergence of new {\SCI} algorithms optimized for modern supercomputer architectures will likely lead to the revision of some the present TBE, allowing to climb even higher on the accuracy ladder. \cite{Eri17}

\begin{suppinfo}
Basis set and frozen core effects.
Definition of the active spaces for the multi-configurational calculations.
Additional details about the {\SCI} calculations and their extrapolation.
Benchmark data and further statistical analysis.
Geometries.
\end{suppinfo}

\begin{acknowledgement}
P.F.L.~would like to thank \textit{Centre National de la Recherche Scientifique} for funding.
D.J.~acknowledges the \emph{R\'egion des Pays de la Loire} for financial support. This research used resources of  i) the GENCI-TGCC (Grant No.~2019-A0060801738); ii) CCIPL (\emph{Centre de Calcul Intensif des Pays de Loire});
iii) the Troy cluster installed in Nantes; and iv) CALMIP under allocations 2019-18005 (Toulouse).
This work has been supported through the EUR grant NanoX ANR-17-EURE-0009 in the framework of the \textit{``Programme des Investissements d'Avenir''}.
\end{acknowledgement}

\bibliography{biblio-new}

\end{document}